# GROWTH AND CHARACTERIZATION OF MULTI-WALLED CARBON NANOTUBES USING CHEMICAL VAPOR DEPOSITION

*Joydip Sengupta*

# GROWTH AND CHARACTERIZATION OF MULTI-WALLED CARBON NANOTUBES USING CHEMICAL VAPOR DEPOSITION

*Thesis submitted to the*
*Indian Institute of Technology, Kharagpur*
*for award of the degree*

of

Doctor of Philosophy

by

**Joydip Sengupta**

Under the guidance of

**Dr. Chacko Jacob**

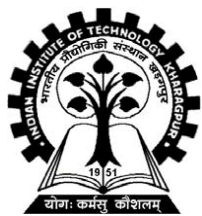

**Materials Science Centre**
**Indian Institute of Technology, Kharagpur**
**January 2010**

# CERTIFICATE OF APPROVAL

Date:…………………

Certified that the thesis entitled "**Growth and characterization of multi-walled carbon nanotubes using chemical vapor deposition**" submitted by **Joydip Sengupta** to the Indian Institute of Technology, Kharagpur, for the award of the degree Doctor of Philosophy has been accepted by the external examiners and that the student has successfully defended the thesis in the viva-voce examination held today.

| | | |
|---|---|---|
| Signature: | Signature: | Signature: |
| Name: Dr. A. Roy | Name: Dr. S. B. Majumder | Name: Dr. S. Das |
| (Member of the DSC) | (Member of the DSC) | (Member of the DSC) |
| Signature: | Signature: | Signature: |
| Name: Dr. C. Jacob | Name: Prof. D. Bahadur | Name: |
| (Supervisor) | (External Examiner) | (Head of the Centre) |

# CERTIFICATE

This is to certify that the thesis entitled "**Growth and characterization of multi-walled carbon nanotubes using chemical vapor deposition**", submitted by **Joydip Sengupta** to Indian Institute of Technology, Kharagpur, is a record of bona fide research work under my supervision and is worthy of consideration for the award of the degree of Doctor of Philosophy of the Institute.

<div style="text-align:right">

Dr. Chacko Jacob  
Supervisor

</div>

# ACKNOWLEDGEMENTS

I would like to take this opportunity to acknowledge several people for their support, collaboration and encouragement during my research tenure at Materials Science Centre, IIT Kharagpur.

First of all, I wish to express my sincere gratitude to my advisor, Dr. C. Jacob. My experience of working as his student is an invaluable treasure. I could not have imagined a better mentor for my Ph.D. than him. Moreover, the family like atmosphere in his group made my graduate life quite enjoyable. Actually, to a student he is much more than an academic advisor. His constant advice, encouragement and invaluable suggestions have not only guided me to find the right way in research but also in personal life.

Thanks are due to Prof. C. K. Das and Prof. B. Adhikari, the former and the present Head of the Centre, for the necessary administrative support. I would like to extend my gratitude to all the faculty members of Materials Science Centre for their favorable assistance. I would also like to convey my deep gratitude to the faculty members of my Doctoral Scrutiny Committee (Dr. A. Roy, Dr. S. Das and Dr. S. B. Majumder) for their invaluable support. I am also grateful to Dr. N. D. Pradeep Singh of the Department of Chemistry for providing the precursor metal complexes for the Mod-PR catalysts. I would also like to express my thanks to Dr. B. Mishra of the Department of Geology & Geophysics and Dr. A. Roy of the Department of Physics and Meteorology for their support to carry out Raman analysis of my samples. I also acknowledge the Administrative staff, Academic staff and the Technical Staff of Materials Science Centre for helping me with their respective fields of expertise.

The Senior Research Fellowship provided by CSIR (Council for Scientific and Industrial Research, India) is gratefully acknowledged. I am also thankful to the authorities of IIT Kharagpur for kindly providing me with other facilities to carry out my research work. Specially, I wish to thank all the staff members of the Central Research Facility for their extreme cooperation in characterizing the samples.

I would also like to thank my fellow group members, who helped to keep me focused and determined through the tough times. In no particular order, thanks are expressed to Aparna Gupta, Samik Pal, Anupama Chanda, Utpal Roy, Jayanta Maity, Sovan Panda, Rakesh Sahoo, Jasbir, Nandan, Harendra, Chittaranjan, Venky, Tarapada, Chandra and of course Biswajit Das along with Tapas Sarkar. I wish to convey my appreciation to all the friends in IIT Kharagpur for providing unforgettable company and moral support during different stages of my Ph.D. tenure.

Finally, I must thank my family for their enormous sacrifice to support me through all these years at IIT Kharagpur. Last, but certainly not least, I need to specially thank my wife, Arpita, for her immense patience and support. Thanks for walking beside me for the last four years, sharing the happy moments and challenges alike. I am grateful being able to say that I have a great family that has always set a good example and provided me with unlimited supported.

# DECLARATION

I certify that
  a. The work contained in the thesis is original and has been done by myself under the general supervision of my supervisor.
  b. The work has not been submitted to any other Institute for any degree or diploma.
  c. I have followed the guidelines provided by the Institute in writing the thesis.
  d. I have conformed to the norms and guidelines given in the Ethical Code of Conduct of the Institute.
  e. Whenever I have used materials (data, theoretical analysis, and text) from other sources, I have given due credit to them by citing them in the text of the thesis and giving their details in the references.
  f. Whenever I have quoted written materials from other sources, I have put them under quotation marks and given due credit to the sources by citing them and giving required details in the references.

<div style="text-align: right;">Joydip Sengupta</div>

# List of Symbols and Abbreviations

$\theta_c$: Chiral angle
$\gamma$: Interfacial energy
$\lambda$ : Radiation wavelength
$\theta$ : Angle of incidence
$d$ : Inter planar spacing
$v$ : Frequency
$\omega$: Wave number (cm$^{-1}$)
AFM: Atomic force microscopy
APCVD: Atmospheric chemical vapor deposition
BSE: Back scattered electrons
CCD: Charged coupled device
CNT: Carbon nanotubes
CVD: Chemical vapor deposition
$E_{laser}$ : Energy of the laser (eV)
EDX: Energy dispersive X-Ray
FEG: Field Emission Gun
FESEM: Field emission scanning electron microscopy
HRTEM: High resolution transmission electron microscopy
$I_D/I_G$: G to D Band Intensity Ratios (Raman)
LPCVD: Low pressure chemical vapor deposition
MOCVD: Metal organic chemical vapor deposition
Mod-PR: Modified photoresist
MWCNT: Multi-walled carbon nanotubes
PECVD: Plasma enhanced chemical vapor deposition
RBM: Radial breathing mode
RMS: Root mean square
SAED: Selected area diffraction pattern
SE: Secondary electrons
SEM: Scanning electron microscopy
SWCNT: Single-walled carbon nanotubes
TEM: Transmission electron microscopy
UV: Ultra violet
VLS: Vapor liquid solid
XRD: X-ray diffraction

**Abstract**

Carbon nanotubes (CNTs) exhibit a unique combination of electronic, thermal, mechanical and chemical properties, which promise a wide range of potential applications in key industrial sectors. An essential step towards the application of nanotubes is a thorough understanding of the effect of process variables on CNT growth and the role of the metal catalyst involved in the synthesis procedure.

In this study, the synthesis of multi-walled carbon nanotubes (MWCNTs) was carried out by chemical vapor deposition (CVD) using propane as the carbon source and Si as the catalyst support. The effect of CVD process variables such as temperature, choice of catalyst, etc on the growth behavior of nanotubes has been examined to understand the catalytic growth of CNTs. The transition metal catalysts, Fe and Ni, were used in both elemental metal form and in a metal complex form. In the case of elemental metal catalysts, the respective metals were deposited over the Si substrate using thermal evaporation following which nanotubes were synthesized by means of CVD. Subsequent studies of the synthesized carbon nanostructures employing elemental metal catalysts revealed a significant influence of the temperature and the catalyst material on the structure of CNTs. The CNTs synthesized using Ni catalyst were bamboo-like whereas the CNTs developed employing Fe catalyst were straight tubes with partial metal filling. Consequently, growth models for the different growth mechanisms have been proposed. Certain limitations of the above process have been overcome by employing spin-coating of a metal complex catalyst material on the Si substrate. The CVD synthesis of nanotubes using metal complex catalysts always resulted in partially catalyst filled CNTs. More importantly, the metal complex catalyst could be easily patterned on the Si substrate using spin-coating and photolithography, which resulted in site selective growth of partially catalyst filled MWCNTs. Since the entire process of site selective growth is suitable for conventional device fabrication, this method is a promising and practical pathway for large-scale fabrication of several magnetic material filled CNT-based devices.

**Keywords**: Multi-walled carbon nanotubes; Lithography; Site selective growth; Chemical vapor deposition

# CONTENTS

# Chapter 1

## Introduction

Careful research on different materials reveals that material properties are not only governed by the atomic composition and the chemical bonding, but also by the dimensions of the material (Buffat, 1976). It has been observed that as the size of a material reduce to nanometer scale dimensions, materials exhibit some remarkable properties, resulting in unique physical and chemical characteristics. Novel synthetic approaches have resulted in the development of materials with reduced dimensions. This technology of synthesizing nanoscale materials and devices was named as "nanotechnology" by Norio Taniguchi in 1974 (Taniguchi, 1974).

Carbon, placed at group 14 (IV A), is one of the most important elements in the periodic table. A carbon atom contains 6 electrons, 2 in the first energy level (1s) and the remaining 4 electrons fill the sp hybrid orbital at the second energy level. Owing to its ability to form $sp^3$, $sp^2$, and sp hybrids and stable multiple pi and sigma bonds, carbon can form 3D (Diamond and Graphite), 2D (Graphene), 1D (Carbon nanotube), and 0D (Fullerene) materials with a wide variety of physical and chemical properties (Figure 1.1).

Among the different forms of carbon, carbon nanotubes (CNTs) represent one of the most exciting research areas in modern science. Nanotubes are nearly one dimensional structures due to their high length to diameter ratio.

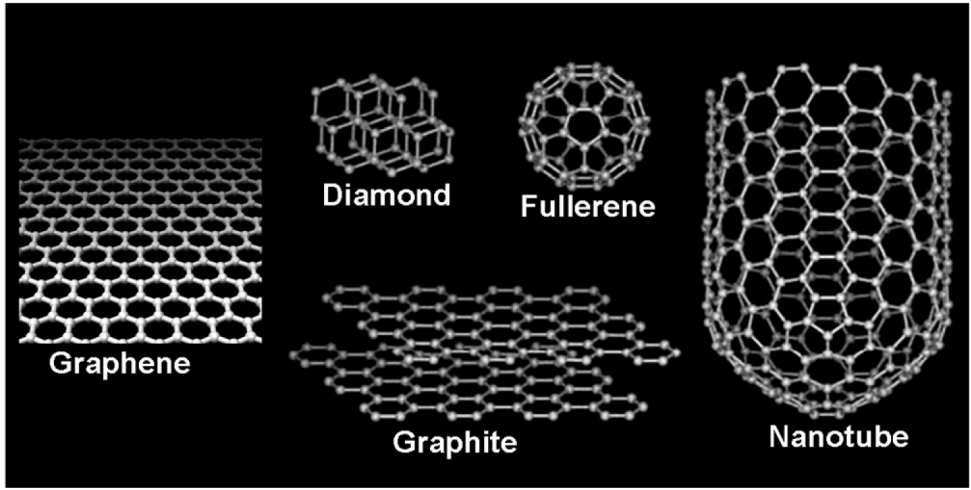

Figure 1.1: Different forms of carbon
(adapted from http://mrsec.wisc.edu/Edetc/IPSE/educators/activities/carbon.html)

CNTs exhibit a unique combination of electronic, thermal, mechanical and chemical properties (Chen et al., 1999; Dai et al., 1996a; Qian et al., 2000; Saito et al., 1997; Tans et al., 1998), which promise a wide range of potential applications in key industrial sectors such as nanoelectronics (Herrero et al., 2006), biotechnology (Martin and Kohli, 2003) and thermal management (Kim et al., 2007).

**1.1 A brief history of carbon nanotubes**

Tubular carbon nanostructures were first observed as early as 1952 by Radushkevich and Lukyanovich (Figure 1.2a), who published this in the Soviet Journal of Physical Chemistry (Radushkevich and Lukyanovich, 1952). Due to the cold war, access to Russian scientific publications for western scientists was not easy. Accordingly this discovery was largely unnoticed. A few years later Oberlin et al. (1976) published clear images of hollow carbon fibres with nanometer scale diameters using a vapour growth technique (Figure 1.2b). However, it was not until nearly two decades later, when Iijima reported the observation of CNTs in Nature (Iijima, 1991) that world-wide interest and excitement was generated. Iijima's work is undoubtedly responsible for the explosion of interest in CNT research in the scientific community which resulted in the rapid development of this field. Iijima clearly observed the so-called multi-walled nanotubes (MWCNTs) while studying the soot made from by-products obtained during the synthesis of fullerenes by the electric arc discharge method. Two years later, single-walled carbon nanotubes (SWCNTs) were discovered independently by Iijima at the NEC Research Laboratory in Japan (Iijima and Ichihashi, 1993) and by Bethune at the IBM Almaden Laboratory in California (Bethune et al., 1993). Since then the field has advanced at a breathtaking pace that is reflected in the number of increased publication (Figure 1.3) along with many unexpected discoveries.

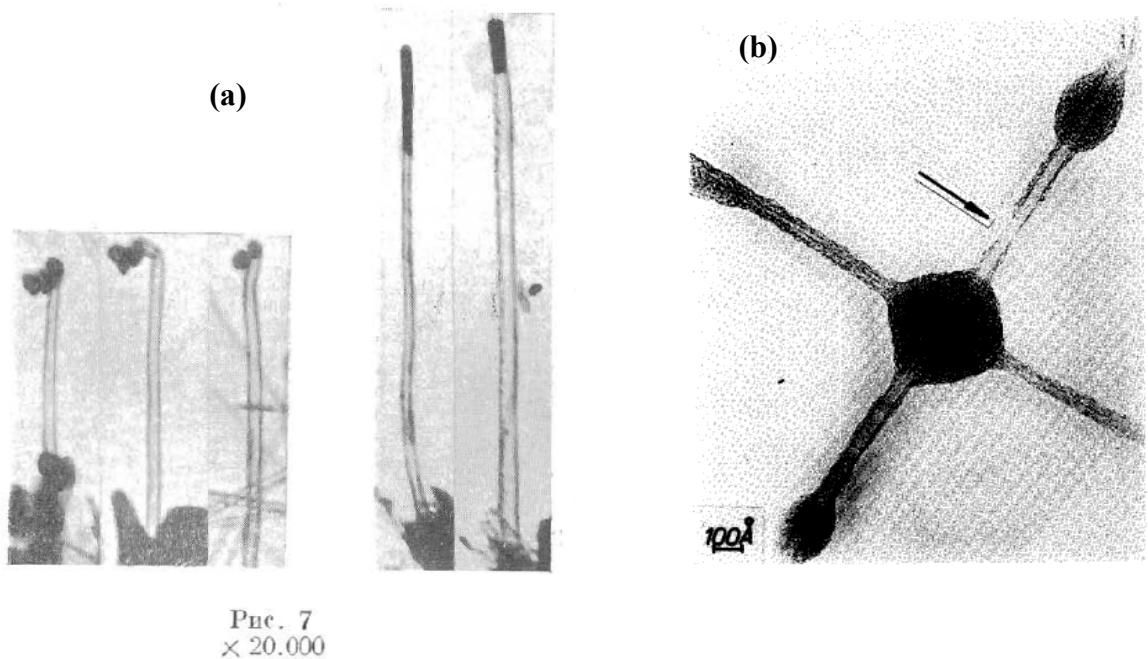

Figure 1.2: Examples of transmission electron microscopy images of carbon nanotubes published by (a) Radushkevich and Lukyanovich (1952) (reprinting permission sought from Nauka Publishers); (b) Oberlin et al. (1976) (reprinted by permission of Elsevier)

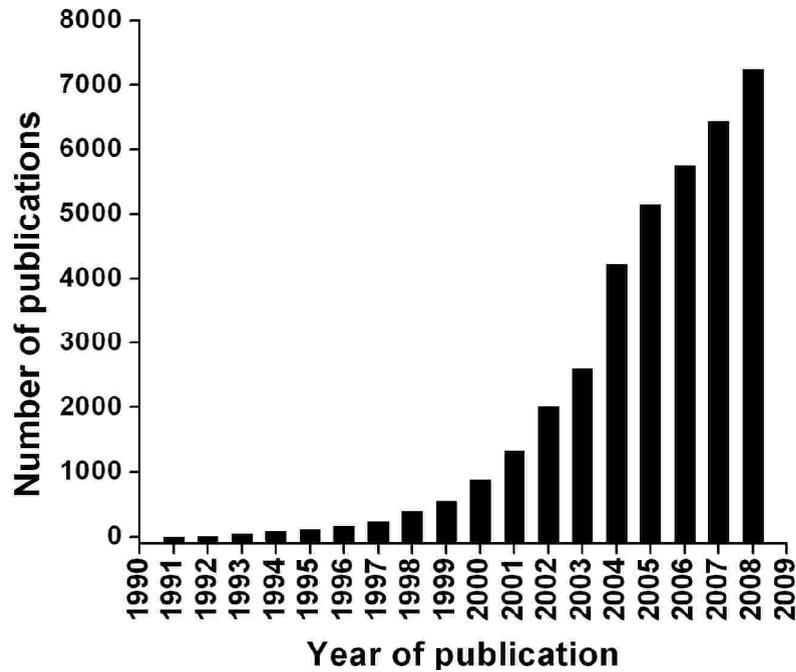

Figure 1.3: The number of papers on carbon nanotubes published annually, from 1991 to 2008 (Data from Scopus, Elsevier)

**1.2 Structure**

The structure of the CNT can be specified in terms of a vector $\vec{C}$. This vector is known as the chiral vector and it is formed by joining two crystallographically equivalent points on the original graphene lattice. $\vec{C}$ can be expressed in terms of a set of two integers (n,m) corresponding to graphite vectors $\vec{a}_1$ and $\vec{a}_2$ (Figure 1.4)

$$\vec{C} = n \cdot \vec{a}_1 + m \cdot \vec{a}_2, \ (0 \leq |m| \leq n) \qquad \ldots(1.1)$$

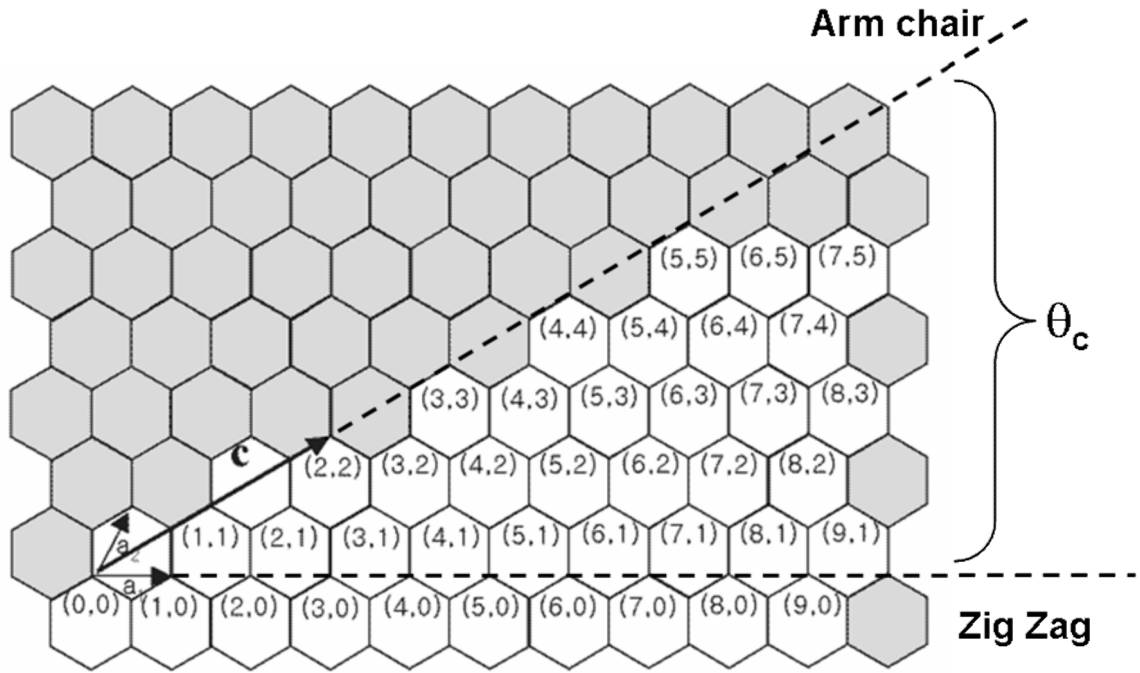

Figure 1.4: Graphene layer defined by unit vectors $\vec{a}_1$ and $\vec{a}_2$ and the chiral angle $\theta_c$, the atoms labeled with (n,m) notation (adapted from Dresselhaus et al., 1996)

The integers n and m are also called indexes and determine the chiral angle

$$\theta_c = \tan^{-1}[\sqrt{3}\{m/(m + 2n)\}] \qquad \ldots(1.2)$$

with respect to the zigzag axis. The direction of the nanotube axis is perpendicular to the chiral vector. Three different types of nanotube structures can be generated by rolling up the graphene sheet into a cylinder along different directions: armchair (n = m, $\theta_c = 30°$), zigzag (m = 0, n > 0, $\theta_c = 0°$), and chiral (0 < |m| < n, 0 < $\theta_c$ < 30°) (Figure 1.5). Armchair CNTs are metallic. Zigzag and chiral nanotubes can be semimetals or semiconductors.

CNTs can be visualized as sheets of carbon atoms rolled up into tubes. There are two types of CNTs: They can consist of one (single-wall) or more (multi-wall) graphitic layers.

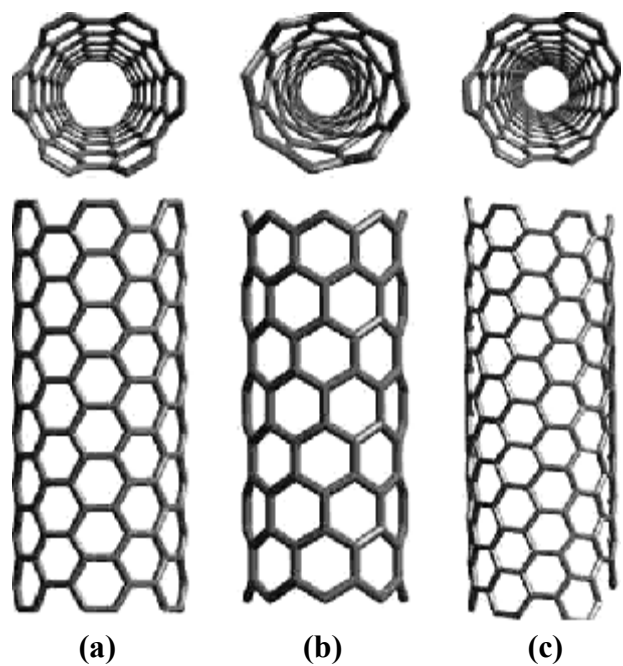

**(a)** **(b)** **(c)**

Figure 1.5: Cross sectional (top) and plan view (bottom) schematics of various types of single-walled carbon nanotubes (a) Armchair nanotube; (b) Zigzag nanotube and (c) Chiral nanotube

The structure of a SWCNT can be thought of as a single atomic layer thick sheet of graphite (called graphene) rolled into a seamless cylinder such that the lattice of carbon atoms remains continuous around the circumference. Depending on the roll-up direction SWCNTs can have different structures namely zigzag, armchair or chiral (Figure 1.5). The term zigzag and armchair refer to the arrangement of hexagons around the circumference of a CNT.

MWCNTs consist of multiple layers of graphitic sheets rolled in on themselves to form a tube shape. The structure of multi-walled nanotube can be explained on the basis of two models– the "Russian doll" model and the "Parchment model" (Figure 1.6).

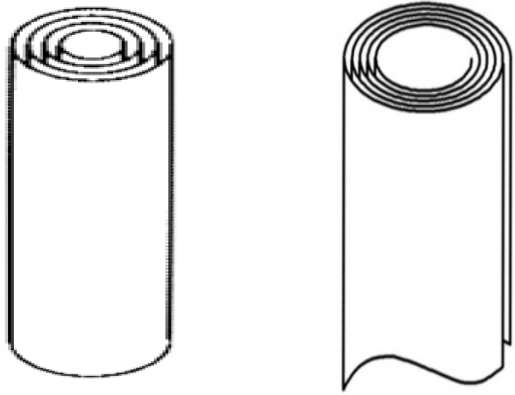

Figure 1.6: Schematics of (a) Russian doll and (b) Swiss roll model of multi-walled carbon nanotube

In the "Russian doll" model, sheets of graphite are arranged in concentric cylinders while in the "Parchment model", a single graphene sheet is rolled in around itself multiple times to form a spiral, like a roll of paper or parchment. The "Parchment model" is often described as the "Swiss roll" model. However, there are reports (Maniwa et al., 2001; Martinez et al., 2007; Zhou et al., 1994) on the structure of MWCNTs which indicate that the structure of MWCNTs may be a mixture of "Russian doll" and "Swiss roll" structures. Evidence has also been found about the presence of zigzag, armchair and chiral tubes within the same MWCNT (Zhang et al., 1993; Zhang et al., 1994a; Zhang et al., 1994b).

**1.3 Properties**

The exotic electronic properties of nanotubes have received a lot of attention from the research community. Nanotubes possess remarkable electronic properties due to their extremely small size and highly symmetric structure. The unparalleled electronic properties, for example the quantum wire like behavior of a SWCNT, SWCNT bundles, and MWCNTs as well as the metallic and semiconducting characteristics of a SWCNT have been established both theoretically and experimentally (Bockrath et al., 1997; Charlier et al., 2007; Chico et al., 1996; Odom et al., 2000; Wildoer et al., 1998). Experimental observation confirms that confinement effect due to the tube circumference is responsible for these phenomena. In the case of electron transport in the metallic nanotubes, as the allowed electronic states are very limited with respect to graphite, therefore the conduction occurs through well separated discrete electron states as in a quantum wire. Consequently, the conduction is of ballistic type, i.e. electrons pass along the nanotubes without any scattering. Effectively, the electrons move without experiencing any resistance and subsequently dissipate no energy. Thereby, metallic nanotubes are capable of conducting high current without producing heat which is one of the highly desirable criteria for the construction of nanoscale circuits. The electron transport in semiconducting nanotubes is much more complicated. However, experimental evidence for extremely high mobilities has been established (Durkop et al., 2004).

Nanotubes also show extraordinary thermal properties. SWCNTs have thermal properties similar to graphite at room and elevated temperatures but unusual behavior at low temperatures because of the effects of phonon quantization (Hone et al., 1999; Yamamoto et al., 2004). This indicates that heat conduction in a nanoscale structure can be quantized since only a small number of phonons are active in the nanoscale device at very low temperatures (Angelescu et al., 1998; Rego and Kirczenow, 1998). Ballistic transport of phonons has also been reported by Brown et al. (2005) using a temperature sensing scanning probe microscope. Theoretical calculations along with experimental measurements reveal that the thermal conductivity at room temperature for SWCNT ropes and MWCNTs may vary between 1800 and 6000 W/(m·K) (Hone et al., 1999; Yi et al., 1999) whereas more than 3000 W/(m·K) is evidenced (Kim et al., 2001) from measurements on a single MWCNT.

Semiconducting SWCNTs have a direct band gap which is the most important criteria for use in optical and optoelectronic applications. Optical properties of nanotubes have been investigated by employing resonant Raman (Rao et al., 1997), fluorescence (Hartschuh et al., 2003), and ultraviolet to the near infrared (UV-VIS-NIR) spectroscopies (Hagen and Hertel, 2003). Furthermore, electrically induced optical emission (Misewich et al., 2003) and photoconductivity (Freitag et al., 2003) measurements have been reported for individual SWCNTs. Theoretical calculations and experimental measurements reveal that there is a difference in the absorption coefficient spectra for nanotubes due to the strong dependence of nonlinear optical properties on the diameter and symmetry of the tubes.

CNTs are also recognized as the stiffest and strongest material known to date featuring high flexibility and large breaking stress with low density. The axial component of sigma bond and $sp^2$ rehybridization in nanotubes is mainly responsible for their unparallel mechanical properties. Several research groups have reported the Young's modulus and tensile strength of SWCNTs and MWCNTs (Lukić et al., 2005; Treacy et al., 1996; Wong et al., 1997; Yu et al., 2000a) based mainly on the techniques of transmission electron microscopy (TEM) (Demczyk et al., 2002), scanning electron microscopy (SEM) (Yu et al., 2000b) and atomic force microscopy (AFM) (Guhados et al., 2007). While there is some scatter in the reported values of Young's modulus and tensile strength observed in these studies, it can be concluded that the Young's modulus of the nanotubes is around 1000 GPa, approximately five times higher than steel while their tensile strength can be up to 50 GPa, around fifty times higher than steel (although higher values for both Young's modulus and tensile strength have been observed in some studies). However, it is important to realize that Young's modulus can vary with tube diameter but stiffness always increases with the increase of the diameter of the tube. The experiment carried out by Falvo et al. (1997) using AFM provides further evidence of extraordinary resilience of CNTs. They found that the CNT could be bent repeatedly through large angles without fracturing.

High reactivity and strong sensitivity of CNTs to chemical or environmental interactions is due to the curvature of the tube. A carbon atom in the sidewall, or in a closed end cap, has three bonds to its nearest neighbor carbon atoms. Carbons at the termination of an open ended tube only have two bonds. Therefore, it is easy to introduce foreign molecules into the structure. Thus, CNTs can be functionalized by the preferential addition of one or more species. Functionalization is a way to enhance properties for specific applications such as for the development of nanotube-based composite (Tasis et al., 2006).

**1.4 Synthesis methods**

There are many methods by which CNTs can be produced, including but not limited to arc discharge, laser ablation and chemical vapor deposition (CVD). These three synthesis method of CNT can be classified into two main categories depending on the growth temperature. High temperature routes are the electric arc method and the laser ablation method whereas medium temperature routes are based on CVD processes. Present work focuses on CVD synthesis since this method can be relatively easily extended to bulk production.

The high temperature process involves sublimation of graphite in an inert atmosphere and condensing the resulting vapor under a high temperature gradient. The difference between the various processes is the method used for subliming graphite. An electric arc formed between two electrodes is used for sublimation of graphite in case of the arc discharge method. An ablation induced by a laser is used for sublimation of graphite in case of the laser ablation technique.

**1.4.1 Arc discharge**

Arc discharge was used by Bacon in 1960 to produce carbon whiskers (Bacon, 1960), around 50 years ago. Later, in 1990, Krätschmer and his co-workers employed the arc discharge technique to produce fullerenes (Krätschmer et al., 1990) while Iijima used arc evaporator for the synthesis of MWCNTs in 1991 (Iijima, 1991). The first synthesis of SWCNTs in 1993 also involved arc evaporation using metal impregnated electrodes. Since then arc evaporation remains an important method for nanotube synthesis. This method creates CNTs through an arc discharge generated between two graphite electrodes placed face to face, separated by approximately 1mm in the airtight chamber (Figure 1.7) that is usually filled with inert gas at low pressure. In general, a low voltage (~ 12 to 25 V) and high current (50 to 120 A) power supply is used for arc discharge.

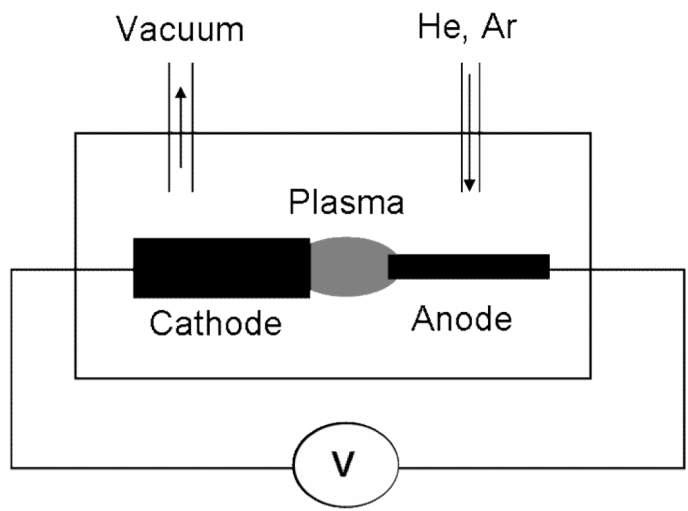

Figure 1.7: Schematic of the arc discharge technique

The electrical discharge creates a high temperature between the two electrodes resulting in the sublimation of the carbon contained in the graphite electrode. During sublimation, the carbon atoms eject from the solid and form plasma. These atoms move toward the colder zones within the chamber, allowing a nanotube deposit to accumulate on the cathode. The type of nanotube that is formed in arc discharge process depends crucially upon the presence of metal catalysts. If small amounts of transition metals such as Fe, Co, Ni, Pt, Pd or Y are introduced in the target graphite, then SWCNTs become the dominant product (Journet et al., 1997; Yudasaka et al., 2002). Radial growth of SWCNTs has also been observed using lanthanum (Saito et al., 1995). In the absence of such metals, the formation of MWCNT is favored. The metal is typically introduced into this reaction via the anode in two predominant methods, drilled hole or uniformly dispersed. In the drilling method, a several centimeter hole is drilled into the end of the anode graphite rod and filled with the catalyst and graphite powder. The other type of anode has the catalyst uniformly dispersed within the rod and is easier to fabricate. Physical conditions of the arc discharge for efficient CNT production include parameters such as potential drop between electrodes, current density in the arc, the inter-electrode spacing, plasma temperature, stability of the plasma formed between the electrodes, inert gas pressure and cooling of electrodes (Ebbesen and Ajayan, 1992; Seraphin et al., 1993). Different type of gases e.g. He, Ar, $CH_4$, $H_2$, $N_2$, $CF_4$ and organic vapors have been used (Ando 1994; Cui et al., 2004; Shimotani et al., 2001; Yokomichi et al., 1998; Zhao et al., 2004). Researchers are trying to improve the method of arc discharge technique in terms of experimental ease, quality of nanotubes and production quantity. To reduce the process cost, several efforts have been made, e.g. under water growth of nanotubes has been demonstrated (Zhu et al., 2002) which does not require vacuum or a water cooled chamber. Similarly, instead of graphite, coal has been used as the electrode material (Qiu et al., 2003). As the growth temperature of the arc discharge method is higher than that of other CNT production methods, therefore the crystallinity and perfection of arc grown CNTs are generally higher than other methods. However, the wide range of CNTs produced in arc discharge method have varying morphologies and suffer from a variety of defects such as amorphous carbon matter deposited on the inside and outside of the CNT walls. Moreover, the CNTs from the arc discharge are often covered with amorphous carbon, which contains metallic particles in the case of metal-carbon co-evaporation. Many purification methods have been developed to purify the nanotubes produced by arc evaporation, e.g. oxidation (Ebbesen et al., 1994), bromination (Chen et al., 1996), chromatography (Duesberg et al., 1998), intercalation (Ikazaki et al., 1994) and combined wet grinding, hydrothermal treatment and oxidation (Sato et al., 2001). In recent years, interest in purifying arc grown tubes has decreased due to the increased focus on in catalytically grown nanotubes. Consequently, there are no completely satisfactory ways of achieving pure arc grown CNTs at present.

**1.4.2 Laser ablation**

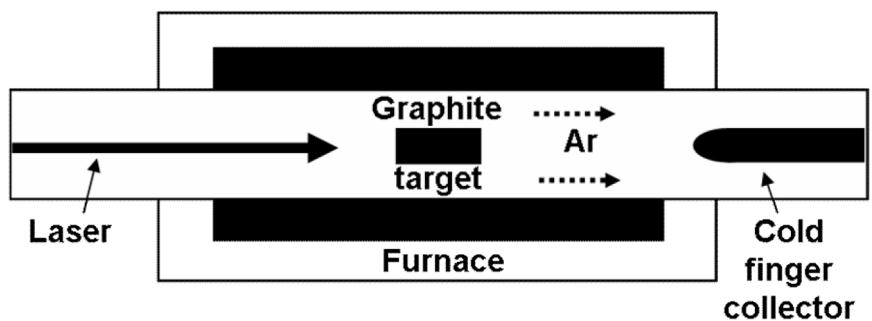

Figure 1.8: Schematic of the laser ablation technique

Historically, production of fullerenes was originally obtained by a laser ablation technique (Kroto et al., 1985). In this process, ablation of a graphite target with a focused laser beam is realized in an inert atmosphere at low pressure (Figure 1.8). Typically a Nd:YAG (Guo et al., 1995a) or $CO_2$ ( Kokai et al., 1999) laser is used as a source. Since the energy density of lasers is much higher than that of other vaporization devices, a laser is suitable for materials with a high boiling temperature such as carbon. Two kinds of methods were developed and they use either a pulsed laser (Guo et al., 1995b) or a continuous laser (Maser et al., 1998). The laser evaporates a solid target of graphite (or graphite and catalyst) into a background gas which is gently flowing through a quartz tube inside a high temperature oven. During the vaporization process the flowing argon gas sweeps the produced soot inside the quartz tube. The nanotubes nucleate in the vapor phase, coalesce, get carried away by the flowing argon and condense downstream on the water-cooled copper collector. The felt-like material, when scraped off the wall, contains MWCNTs or SWCNTs depending on the experimental conditions. As with the electric arc method, MWCNTs are obtained when using a pure graphite target and SWCNTs when the target is a mixture of graphite and metallic catalysts such as Ni-Co or Ni-Y mixtures (Maser et al., 1998). Nanotubes are accompanied by amorphous carbon, metal particles, and carbon onions as in electric arc techniques. Therefore the nanotubes produced by this synthesis method also need extensive purification before use.

Arc discharge and laser vaporization are currently the principal methods for obtaining small quantities of high quality CNTs. However, both methods suffer from drawbacks. The first is that both methods involve high purity graphite rods, consumption of a large amount of energy and low yield of CNTs. So, these techniques are not economically advantageous to scale up and less favorable for nanotube production at an industrial level. The second issue relates to the fact that vaporization methods grow CNTs in highly tangled forms, mixed with unwanted forms of carbon and/or metal species. The CNTs thus produced are difficult to purify, manipulate, and assemble for building nanotube-device architectures for practical applications. Controlled synthesis on substrates with ordered nanotube structures has not been possible by these methods. This need is satisfied by CVD and related techniques.

## 1.4.3 Chemical vapor deposition

There are many types of CVD that can be used in the manufacture of CNTs, but the predominant types are thermal and plasma enhanced which are categorized according to the energy source. When a conventional heat source such as a resistive or inductive heater is used, the technique is called thermal CVD whereas when a plasma source is used to create a glow discharge it is called plasma enhanced CVD (PECVD). This section focuses primarily on thermal CVD as it was employed in the present work.

In the early 1970s, notable work on formation of filamentous carbon by catalytic decomposition of hydrocarbon gases was carried out by several research groups (Baker et al., 1972; Baird et al., 1974; Ruston et al., 1969). Few years later Oberlin et al. (1976) published clear images of hollow carbon filaments using a similar technique. More in-depth studies on the synthesis of carbon filaments were performed in the 1980s. Audier et al. (1981) studied the effect of shape of the catalyst particle on the nanotube growth. Tibbetts (1984) found that it is energetically favorable for the carbon filament to create a hollow core when the outer surface is a curved (0001) basal plane. Baker (1989) reported different growth modes of filamentous carbon and also investigated the influence of the metal-support interaction on them. The observation of high quality MWCNTs by Iijima in 1991 stimulated huge interest in the synthesis of CNTs. In 1993, Yacamán et al. (1993) reported the synthesis of carbon microtubules by catalytic decomposition of acetylene over Fe particles. These carbon microtubules are similar to the helical structure reported by Iijima in 1991. Synthesis of SWCNT using CVD was demonstrated by Dai et al. (1996b) by metal-catalyzed disproportionation of carbon monoxide. Aligned CNT growth was demonstrated in 1996 by a Chinese group who used a technique in which Fe nanoparticles were embedded in mesoporous silica and that was used to catalyze the decomposition of acetylene at 700 ºC (Li et al., 1996). The spatially selective growth of nanotubes was exhibited in 1998 using a patterned Si surface by CVD (Kong et al., 1998). In the same year growth of CNTs using plasma CVD was also achieved using Fe as a catalyst (Ren et al., 1998). However the CNTs produced by CVD process contain residual metal catalyst particle and support material. These impurities can be removed far more easily than in case for arc grown tubes, as shown by a number of groups (Andrews et al., 2001; Chen et al., 2002; Huang et al., 2003; Zhang et al., 2006). It seems that the most successful methods involve high temperature annealing in vacuum. Thus, with the implementation of the CVD process, it became possible to produce a large amount of CNTs directly on the desired substrate with high purity and large yield and at a much lower temperature compared to vaporization methods.

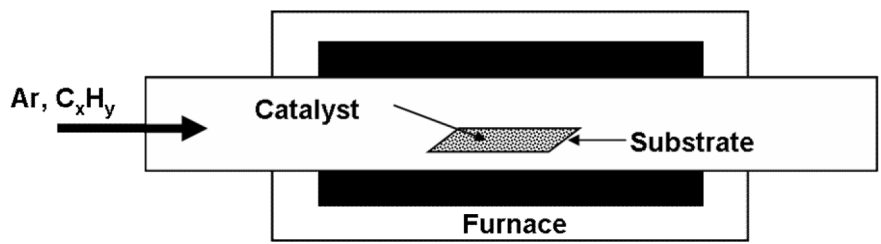

Figure 1.9: Schematic of the chemical vapor deposition technique

A schematic experimental setup for thermal CVD growth is depicted in Figure 1.9. In thermal CVD, an inert gas is used for purging and a hydrocarbon is used as the feedstock. The substrate with catalyst overlayer is placed inside the reactor. A typical growth run involves several steps. Purging of the reactor is done with the inert gas until the reactor reaches the desired growth temperature. Then the inert gas flow is switched to the feedstock for the specified growth period. During the growth period, decomposition of the feedstock gas is catalyzed by catalytic nanoparticles, which also serve as nucleation sites for the initiation of CNT growth. At the end, the feedstock gas flow is switched back to the inert gas and the reactor is allowed to cool down. Materials grown over the catalyst are collected upon cooling the system to room temperature. Care must be taken during cool down since exposure to air at elevated temperatures can cause damage to the CNTs. Several types of hydrocarbon have been used as feedstock e.g. acetylene (Zhong et al., 2009), methane (Homma et al., 2002), propane, etc. These feedstock gases decompose in the presence of catalyst and produce carbon, e.g., $C_3H_8 \rightarrow 2\,C + CH_4 + 2\,H_2$ (Hussain et al., 2008).

**1.4.3.1 Catalyst**

The pathways for the synthesis of CNTs by CVD can be categorized into catalytic and non-catalytic methods (Derycke et al., 2002; Schneider et al., 2008). There are several ways by which catalyst can be prepared for the CNT growth by CVD. A stream of catalyst particles can be injected into the flowing feedstock to produce nanotubes in the gas phase using a floating or unsupported catalyst approach. In contrast, the catalyst can be deposited on the substrate before loading it inside the reactor. This is called supported catalyst approach.

In the unsupported catalyst approach, a volatile compound containing the catalytic element, e.g. iron pentacarbonyl ($Fe(CO)_5$), ferrocene ($Fe(C_5H_5)_2$), or nickelocene ($Ni(C_5H_5)_2$) is used as the catalyst source. The nanotubes form in the vapor phase and condense onto cold surfaces. However, the transition metal sources vaporize at temperatures much lower than that for the gas phase pyrolysis of the carbon sources and a two zone furnace is generally required to produce CNTs by the unsupported catalyst approach. In this method, smaller catalyst clusters tend to evaporate fast and are very unstable. Very large clusters are also not ideal for nanotube growth because they favor graphitic overcoating. Declustering or breakup of large clusters also happens in the reactor. It is the competition between various processes (clustering and evaporation) that creates favorable size clusters (Dateo et al., 2002). Tuning of various parameters e.g. temperature, flow rates of various gases, injection velocity of the catalyst precursor, residence time, etc., is needed to obtain reasonable quantities of nanotubes. Sen et al. (1997) first reported the unsupported catalyst approach and they used ferrocene or nickolecene as a source of the transition metal and benzene as the carbon source. Nikolaev et al. (1999) used CO disproportionation aided by Fe clusters created from $Fe(CO)_5$ to grow CNTs. Cheung et al. (2002) reported the synthesis of nanocluster solutions with distinct and nearly monodisperse diameters of 3.2, 9 and 12.6 nm for three different protective agents used, respectively. Addition of protective agents in the solution prevented the nanoparticles from aggregation. Hence, for large scale continuous production of nanotubes, the floating catalyst approach is suitable. But the major drawback is that CNTs can not be grown with site selectivity. Site selective growth of CNTs is a prerequisite for several device applications of CNTs.

The methods for the preparation of catalyst in case of supported catalyst approach can be divided into two categories: one is solution based catalyst preparation technique and the other is the physical evaporation technique. There are numerous methods for preparing catalysts from solutions, for example sol-gel process (Ermakova et al., 2001; Pan et al., 1999), co-reduction of precursors (Chen et al., 1997), impregnation (Venegoni et al., 2002), reverse micelle method (Ago et al., 2000), spin coating of catalytic solution (Choi et al., 2003) etc. Furthermore, it should be noted that a mixture of transition metal containing compounds is used in the above methods. Usually it is difficult to determine the optimum concentration of each constituent in a trial and error approach because the number of trials is large. Cassell et al. (2001) pioneered a combinatorial optimization process for catalyst discovery for the growth of MWCNTs. This rapid throughput approach coupled with characterization techniques allows development of catalyst libraries with minimal number of growth experiments.

In the physical deposition technique, metal can be evaporated or sputtered to be deposited on the substrate. The metal, if deposited at room temperature, will generally be amorphous and form a nearly smooth film on the surface of the substrate. Upon annealing, the equilibrium shape may be reached. To know this shape, the Young's equation describing a contact between two phases A and B and the ambient atmosphere has to be considered:

$$\gamma_A = \gamma_{AB} + \gamma_B . \cos\theta \qquad ...(1.3)$$

where γ is the corresponding interface energies (Figure 1.10).

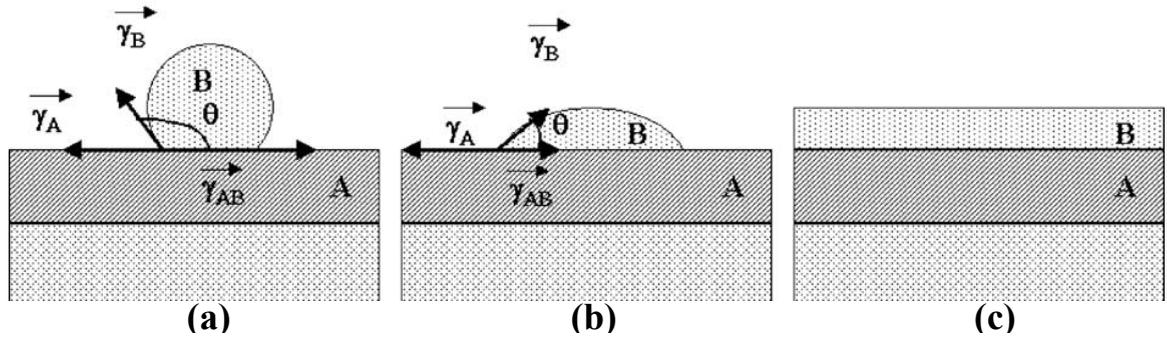

Figure 1.10: Conditions of the surface energies of substrate (A), deposit (B) and interface A–B in determining the type of growth: island or Volmer–Weber growth for (a) non-wetting; (b) wetting and (c) layer-growth

As $\theta \to 0$, $\gamma_A \to \gamma_{AB} + \gamma_B$, in this case, the growth is expected to occur layer by layer. When $\theta \rangle 0$, $\gamma_A \langle \gamma_{AB} + \gamma_B$ and discrete three-dimensional island-like clusters form as shown in Figure 1.10.

It has been observed that the catalyst on the substrate must be in the form of particles instead of smooth, continuous films. The latter do not appear to yield nanotubes. Thus, the property of island growth is required for obtaining nanoparticles on a substrate by physical deposition. After deposition, usually performed at room temperature, annealing allows the atoms to rearrange themselves and reach the energetically most favored configuration. This method has been widely used to obtain nanoparticles of catalyst material to grow CNT. For breaking up the thin film obtained after deposition, many authors reported the use of $NH_3$ (Yoon et al., 2002) or $H_2$ (Wang et al., 2001) during annealing. The surface energies $\gamma_A$ and $\gamma_B$ used above are defined relative to the atmosphere gas and thus a change in the gas can dramatically change the final shape of B on A. Delzeit et al. (2001) have shown that introduction of a metal underlayer (such as Al) can be used instead of any chemical pretreatment steps to create a rough surface. Physical techniques such as electron gun evaporation (Merkulov et al., 2000), thermal evaporation (Chhowalla et al., 2001), pulsed laser deposition (Gao et al., 2003), ion-beam sputtering (Delzeit et al., 2002), and magnetron sputtering (Han et al., 2000) have been successfully used in catalyst preparation.

A wide variety of catalytic species can be used to produce CNTs in CVD growth based on their hydrocarbon decomposition ability, carbon solubility, stability, morphology etc. Different transition metals (e.g. Fe, Ni, Co) and their alloys (e.g. Fe-Mo, Cu-Co, Ni-Ti) have been extensively used to grow CNTs by CVD (Hofmeister et al., 2004; Ižák et al., 2008; Liao et al., 2003; Siegal et al., 2002; Singh et al., 2007; Wang et al., 2001). Noble metals (e.g. Au, Ag) (Takagi et al., 2006), semiconductors (Takagi et al., 2007) and even $SiO_2$ nanoparticles (Huang et al., 2009) behave as catalyst for nanotube growth. In 2006, Deck and Vecchio described a detailed study of the catalytic growth of CNTs using a wide variety of transition metal catalysts (Deck and Vecchio, 2006). They found that Fe, Co and Ni were the only active catalysts, with Cr, Mn, Zn, Cd, Ti, Zr, La, Cu, V and Gd showing no activity. They suggested that the key to catalytic activity is the solubility of the carbon in the metals However, metals like Cu, Pt, Pd, Mn, Mo, Cr, Sn, Au, Mg and Al have been used to grow CNTs as reported by Yuan et al. (2008). These observations indicate that the catalyst–growth dynamics–feedstock picture is not yet complete, and there is much more to explore.

## 1.4.3.2 Growth mechanisms

Thermal CVD relies on thermal decomposition of carbonaceous gas molecules. The general growth process of CNTs by CVD is based on the mechanism proposed by Baker et al. (1972) known as the vapour-liquid-solid (VLS) mechanism (hydrocarbon vapor → metal-carbon liquid → crystalline carbon solid). The VLS synthesis mechanism can be used in the synthesis of many types of one dimensional nanostructures and generally consists of the three primary steps of absorption, saturation, and structure extrusion. The VLS process is energized by heat, which forces the metal catalyst into a molten state and also begins cracking the carbonaceous gases. Once gas decomposition has started, the carbon species begin to diffuse into the catalyst to form a metal-carbon solution. As more and more carbon elements are incorporated into the catalyst, the concentration of carbon exceeds the solubility limit of the catalyst particle and consequently, the metal-carbon solution become supersaturated. When a supersaturated state is reached, carbon precipitates at the surface of the particle in its stable form and leads to the formation of a carbon tube structure. At this juncture two different scenarios are possible (Figure 1.11).

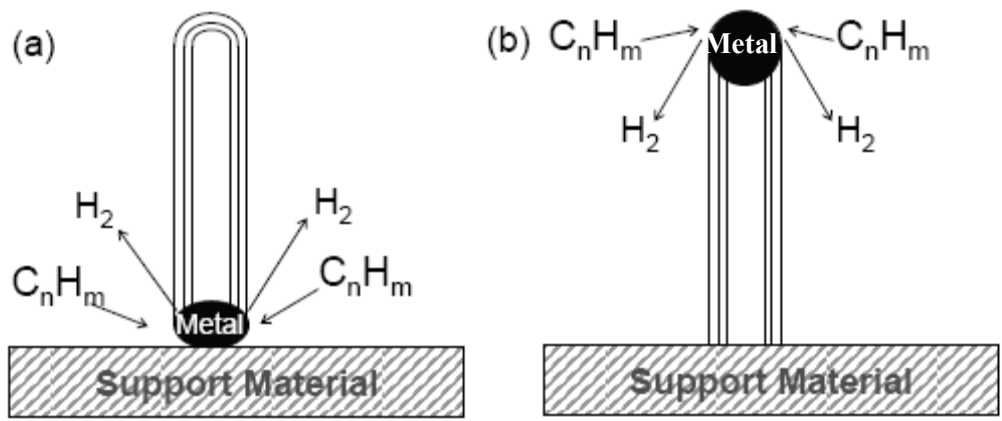

Figure 1.11: Schematic representation of the two typical carbon naotube growth modes, (a) Base growth mode and (b) Tip growth mode

If the particle adherence to the surface is strong, then carbon precipitates from the top surface of the particle and the filament continues to grow with the particle attached to the substrate. This is called the base growth model. In cases where the particle adherence to the surface is weak, then carbon precipitation occurs at the bottom surface of the particle and the growing filament lifts the particle as it grows. In this case, the top end of the filament contains the catalyst particle, and the resultant scenario is called tip growth. Vinciguerra et al. (2003) suggest that tip growth or root growth of CNTs depends solely on the strength of the interaction of the catalyst particle with the substrate. However, the nature of the driving force for carbon diffusion through the catalyst particles is a subject of debate. The driving force could be temperature (Yang and Yang, 1985) or concentration gradient (Nielsen and Trimm, 1977) within the particle.

The key step in temperature-driven carbon diffusion mechanism was believed to be the diffusion of carbon species through the particle from the exposed and hotter front surface on which the exothermic decomposition of hydrocarbons occurs, to the cooler rear surfaces on which carbon is precipitated (endothermic process) from the solid solution (Baker, 1989). The cooler surfaces are generally in contact with the support face. There is considerable experimental evidence to support this mechanism (Massaro and Petersen, 1971). However, a temperature-driven dissolution-precipitation mechanism cannot provide a rational explanation for the endothermic pyrolysis of some hydrocarbons, e.g. methane decomposition.

Concentration-driven carbon diffusion mechanism involves a concentration gradient across the catalyst particle in contact with hydrocarbon on one side and with a graphitic precipitation on the other side. Carbon growth involves a fast gas phase reaction (decomposition of hydrocarbon), carbon atom dissolution in the metal, and carbon precipitation as graphitic structures at the opposite side of the catalyst particle.

So far it has been assumed that the catalytic growth of CNTs involves the volume diffusion of carbon through a catalyst particle. An alternative mechanism based on surface diffusion of carbon around the catalyst particle (Figure 1.12) was put forward by Baird and colleagues in 1974 (Baird et al., 1974) and elaborated by Oberlin in 1976 (Oberlin et al., 1976). In 2004, Helveg et al. reported the study of nanotube growth using a controlled atmosphere TEM (Helveg et al., 2004). This study provides quite compelling evidence for the surface diffusion mechanism.

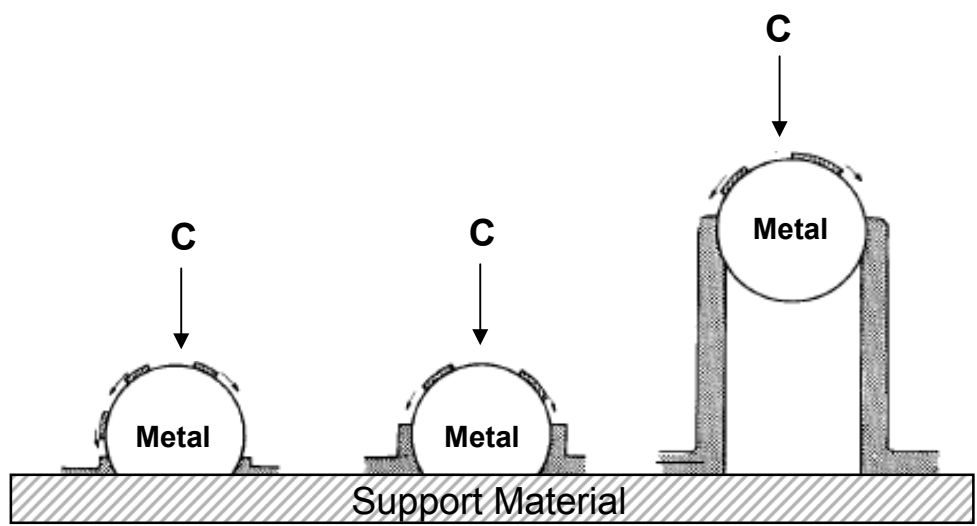

Figure 1.12: An illustration of the growth mechanism based on the surface diffusion of carbon around the metal particle

For a catalyst particle of unchanging size, the growth of CNTs should continue until the hydrocarbon is shut off, either by removing the feedstock from the reaction area or by amorphous or graphitic carbon fully coating the particle, blocking the gas. Additionally, in the case of base growth, growth may slow down or stop due to slow diffusion of hydrocarbons down to the nanoparticle at the bottom of the CNT. If nothing impedes the source of carbon to the nanoparticle, and nothing impedes the nanotube extrusion, the growth should be continuous. In reality, there are competing reactions at the nanoparticle site, such as the formation of graphitic shells and the deposition of amorphous carbon. As a result, in suboptimal growth conditions, amorphous carbon can coat the nanoparticle, preventing feedstock from reacting with the particle and cutting off the carbon source, terminating the growth. Alternatively, if the nanotube exiting the particle encounters an excessive external force, the energy for forming a nanotube might exceed the energy necessary to form a graphitic carbon shell, at which point the nanoparticle will coat itself with a carbon shell, cutting itself off from the carbon feedstock.

### 1.4.3.3 Growth morphologies

Both SWCNT and MWCNT are produced using catalytic CVD. They often have different morphologies depending on the process parameters. Apart from common structures such as zigzag, armchair and chiral nanotubes in case of SWCNT and hollow MWCNT, some special structures like filled and bamboo-like nanotubes are also observed. Their growth mechanisms can be explained with the help of VLS mechanism and are described in detail in Chapter 3. A unique feature of the bamboo-like CNTs is that they offer high surface area, high density of defects and unique inner closed cells, which have extended potential applications to electrochemical bio-sensing (Heng et al., 2005) and hydrogen storage (Jang et al., 2005). The application of bamboo-like CNTs as an electrically conductive additive and anode material for lithium ion batteries has also been reported (Zou et al., 2008).

On the other hand, CNTs filled with magnetic materials have important applications in magnetic force microscopy, magnetoresistance sensors, high density magnetic recording media and biology (Kuo et al., 2003; Monch et al., 2005; Palen 2008; Winkler et al., 2006). They can be also used for microwave absorption (Narayanan et al., 2009) and even in medical sciences for drug delivery (Popov et al., 2007).

**1.5 Motivation, objectives and organization of thesis**

In today's world of nanotechnology, CNTs have become a very promising material and have attracted wide attention both in the research and industrial communities due to their unique structure, novel properties and potential applications. To utilize the unique properties of CNTs in real devices, the nanotubes need to be synthesized over desired substrates. Currently, CVD appears to be the most effective method to achieve this. However, the challenge in the creation of reproducible nanotube-based architectures consisting of arrays of tubes is evidently great. Furthermore, an essential step towards the application of nanotubes is an understanding of nanotube growth and the role of the metal catalyst involved in the synthesis process. Thus, CVD process variables need to be examined intensively to achieve controlled and reproducible growth of nanotubes (Grobert, 2006).

It has been found that the structure of the nanotubes depends on the growth parameters such as reaction temperature, catalyst, reaction gas etc. The growth temperature is especially crucial for selective and controlled growth of CNTs. Unfortunately, despite tremendous progress in synthesizing CNTs, reports on the systematic comparative study of the growth temperature effect on the CNTs are still relatively scarce. Moreover, prior to CNT synthesis, high temperature hydrogen treatment of the catalyst is an important step in order to produce contamination free catalyst and for the removal of the oxides that may exist over the catalyst surface (Takagi et al., 2007). During this procedure, heating above a certain temperature causes catalyst clusters to coalesce and form macroscopic islands. This process is based on cluster diffusion and depends on their density and surface diffusion constant at a given substrate temperature (Jak et al., 2000). These clusters act as a catalyst surface and the cluster size plays a critical role in CNT growth. Therefore, a detailed comparative study on the effects of pre-heating on CNT growth, morphology and microstructure is of importance to achieve a controllable growth of CNTs.

The CVD synthesis of CNTs on plain substrates generally requires catalyst metal deposition over the substrate, which is time consuming and the uniformly deposited area is also finite. These limitations can be overcome by employing spin coating of the catalyst material onto the desired substrate. Moreover, magnetic metal encapsulated CNTs are of great interest due to their distinct magnetic properties compared to the bulk ferromagnetic material (Grobert et al., 1999; Mühl et al., 2003). The cavity of the CNTs can also be used to incorporate metal clusters in order to create novel nanostructured materials with new electronic or magnetic properties as a consequence of the large surface-to-volume ratio of the confined materials and the interaction between the confined materials and the inner walls of nanotube (Karmakar et al., 2005; Solan et al., 2002). Other than the geometrical advantage of a cylindrical shaped nanostructure design, the carbon shells protect the confined metals from oxidation thus ensuring their long-term stability in the core.

For synthesis of metal filled CNTs, several authors have used a floating catalyst method along with dual zone CVD (Deck and Vecchio, 2005; Leonhardt et al., 2003; Müller et al., 2006; Zhang et al., 2002). The obtained CNTs are in bundles grown on the CVD reactor walls or vertically aligned on Si substrates. However, there are many technical barriers to achieve magnetic metal encapsulated CNT-based electronic devices. The placement of CNTs in the desired position is crucial for CNT integration in such devices. From a practical point of view, to achieve CNT-based electronics, the growth and placement of CNTs should be executed in a manner compatible with current microelectronic processes for large scale fabrication. In this regard, photolithography is one of the most important and effective patterning techniques for microfabrication and is comprehensively used in integrated circuit technology. The lithographically patterned growth approach is feasible with discrete catalytic nanoparticles and scalable on large wafers for producing massive arrays of magnetic material filled CNT.

**Objectives of the present investigation**

Taking into account the above scope the following objectives were adopted for the investigation

1. To understand the effect of CVD process variables on CNT growth and improve the quality

   i) Catalyst

       a) Variation in physical form of catalyst (elemental metal and metal complex)

       b) Variation of catalytic element (Fe and Ni)

   ii) Temperature

2. To demonstrate site selective growth of catalyst filled CNTs using a simple process.

**Organization of the thesis**

This thesis focuses on the growth of MWCNTs on Si using the CVD technique in a resistance heated furnace, the characterization of the CNTs using standard techniques and the improvement of the growth quality with site selectivity. The whole thesis is organized into five chapters. The structure of the thesis is as follows: Chapter 1 contains a general introduction to CNT and the current directions in research related to this material. The experimental details of the growth procedures and characterization techniques which have been used in this thesis have been described in Chapter 2. The results of the growth of MWCNTs on Si(111) substrates using elemental metal catalyst by CVD and the influence of different experimental parameters such as growth temperature, pre-heating temperature and nature of catalytic elements on nanotube growth are presented in Chapter 3. Chapter 4 discusses the results obtained from experiments performed using metal complex as catalyst. Lithographically defined site selective growth of partially catalyst filled MWCNTs is also described. Finally, the thesis concludes with a summary in Chapter 5.

# Chapter 2

## Experimental details

The primary objective of this chapter is to present the experimental details of the growth procedure of CNTs and the characterization techniques used to analyze them. The chapter first deals with the different catalyst preparation methods used. Then the CVD setup and growth conditions for CNTs are discussed. After growth, the CNTs were characterized by different techniques. The first step of characterization was to observe the surface morphology of the catalyst film just before CNT growth and atomic force microscopy (AFM) and scanning electron microscopy (SEM) were used for this purpose. SEM was also employed for the analysis of the morphology and number density of the as-grown CNTs, whereas the crystallinity and phases were determined by X-ray diffraction (XRD). High resolution transmission electron microscopy (HRTEM) was applied to analyze the internal structure of the CNTs. Energy dispersive X-ray (EDX) was employed along with SEM and HRTEM for the compositional analysis of the specimens. The degree of graphitization of the CNTs was determined by Raman analysis. A brief description of each experimental technique is discussed in the subsequent sections.

### 2.1 Catalyst preparation

In the catalytic growth of CNTs several semiconductor materials like Ge, Si, SiC, insulator $SiO_2$ and other metals e.g. Mg, Al, Cr, Cu, Mn, Mo, Sn, Pt, Pd, Au, Ag etc. have been utilized. Nevertheless, the transition metals i.e. Fe, Ni and Co are the most widely used catalyst materials for the synthesis of CNTs. From such a large variety of catalyst material, Fe and Ni were chosen for this work. The physical form of the catalysts was also varied, one being the elemental metal form while the other was in a chemically modified form.

### 2.1.1 Deposition of the elemental metal catalysts using thermal evaporation

Deposition of elemental metal catalysts was carried out by thermal evaporation and a brief description of the process is given in the following sections.

### 2.1.1.1 Thermal evaporation

Thermal evaporation is a deposition technique for thin films over a substrate. In evaporation, the substrate is placed inside a vacuum chamber along with the source material to be deposited. During the process the source material is heated up to the point where it starts to evaporate and subsequently, the material vapor condenses in the form of a thin film on the cold substrate surface (Chambers et al., 1998). Usually low pressures are used, about $10^{-6}$ or $10^{-5}$ Torr, to avoid any contamination from the atmosphere. At these low pressures, the mean free path of vapor atoms is much higher than that in normal atmosphere, so these particles travel in straight lines from the evaporation source towards the substrate. In thermal evaporation techniques, different methods can be applied to heat the source material. In e-beam evaporation, an electron beam is aimed at the source material causing local heating and evaporation. In resistive evaporation, a metal filament, containing the source material, is heated electrically with a high current to make the material evaporate.

**Operation:** The source material and the substrate are loaded and the chamber is closed. Then using a mechanical (rotary) pump the chamber is evacuated subsequently down to a pressure of $10^{-3}$ Torr. The diffusion pump is then used to bring the chamber pressure further down to $10^{-6}$ or $10^{-5}$ Torr. During the pumping of chamber by the diffusion pump, the cold trap is cooled by liquid nitrogen to condense any oil that might back-stream from the diffusion pump into the chamber and also to increase the pumping speed (Lafferty, 1998). Upon achieving the desired vacuum in the chamber, the deposition of the source material is performed by passing a high current through the filament holding the source material. The pressure is measured by pressure gages. A schematic diagram of the thermal evaporation system is shown in Figure 2.1.

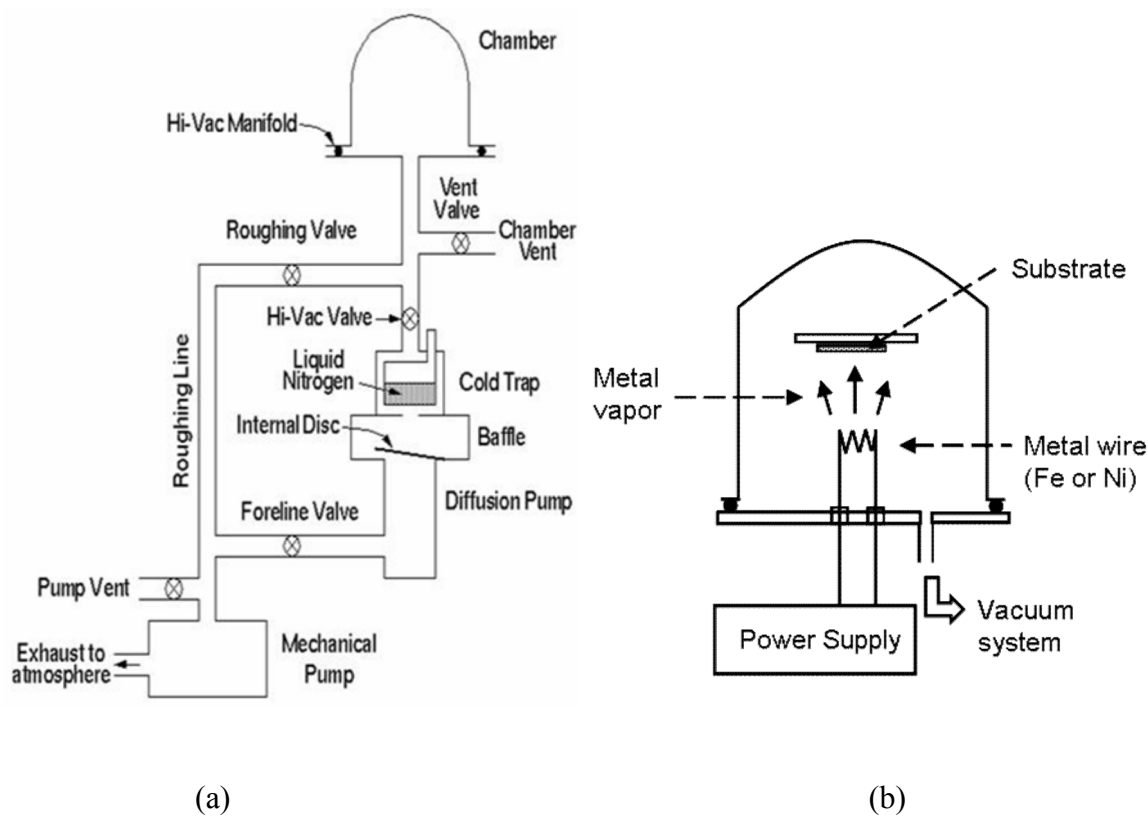

(a)                                          (b)

Figure 2.1: Schematic diagram of vacuum thermal resistive evaporation system (a) The entire setup (without vacuum gages); (b) View of the chamber

## 2.1.1.2 Preparation process

In the present work, for the deposition of catalyst using metal wire, a vacuum system (Hind HiVac, India: Model 12A4D) was used (Figure 2.2). Instead of using a conventional tungsten filament, the respective metal wires [Fe wires (Loba Chemie, India

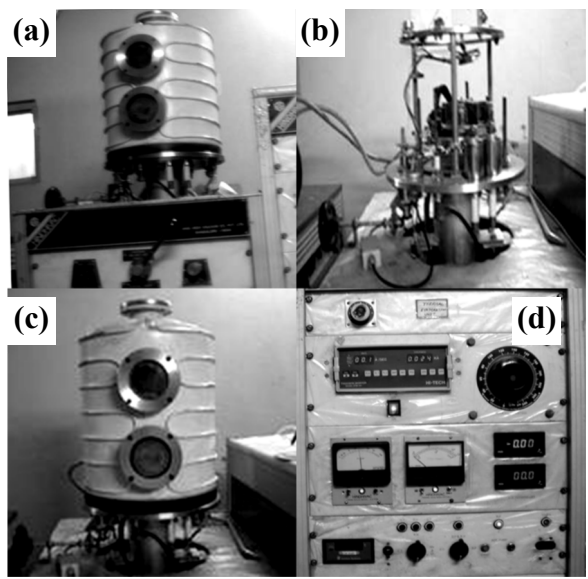

Figure 2.2: (a-d) Photographs of the different parts of the thermal evaporation system

, Purity 99.9%) and Ni wires (Alfa Aesar, Purity 99.9%)] were themselves used as filaments to exclude any contamination from the filament. Prior to the catalyst deposition, the Si(111) substrates were ultrasonically cleaned (Oscar Ultrasonic, India, Model: MIC-101) with acetone (Merck, India) and deionized water (Millipore). The samples were then loaded in the vacuum system and pumped down to a base pressure of $10^{-5}$ Torr. Thereafter, Fe or Ni thin films were deposited by evaporation of the metallic wire.

**2.1.2 Deposition of metal complex catalysts using spin coater**

Other than the elemental metal from, Fe and Ni can also be introduced on the substrate in chemical complex form. Deposition of catalyst can be carried out by mixing the metal complex catalysts with a suitable photoresist and spin coating. Additionally catalyst patterning can also be achieved by simple photolithography. A brief general description of the spin coating and lithography process follows.

**2.1.2.1 Spin Coating**

Spin coating of a fluid on a planar substrate is a common method to produce a thin and relatively uniform film. The process involves deposition of a small puddle of fluid at the center of a substrate and then spinning the substrate at a high speed to spread the fluid evenly over the surface. Fluid flows radially due to the centrifugal force and the excess is ejected off the edge of the substrate. The final film thickness and other properties depend on the nature of the fluid (viscosity, drying rate, percent solids, surface tension, etc.) and the parameters of the spin process (speed, duration etc.) (Bornside et al., 1987). The schematic diagram and photograph of our spin coater system (Headway Research Institute, Garland Texas) are shown in Figure 2.3. The wafer is held to the chuck with vacuum pump and lid is placed over the spinning basin before the spin is initiated.

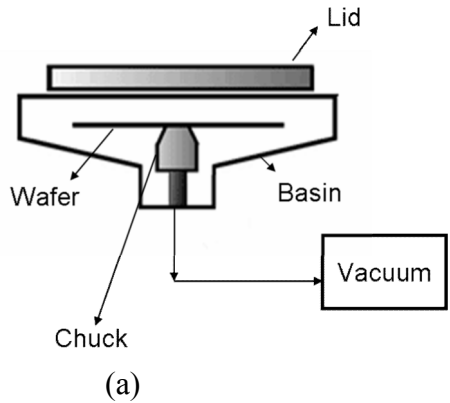 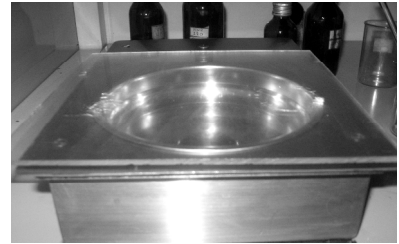

(a)  (b)

Figure 2.3: (a) Schematic diagram of a spin coater system and (b) Photograph of the spin coater

**2.1.2.2 Lithography**

Lithography is the process of transferring geometric shapes on a mask to the surface of a substrate. The substrate is coated with resist (a radiation sensitive material) and then irradiated through a mask. The mask contains clear and opaque features that define the pattern to be created in the resist layer. The resist areas which are exposed to the radiation are either soluble or insoluble in a specific solvent known as developer. When the irradiated (exposed) areas are soluble, a positive image of the mask is produced in the resist. Such a material is therefore termed as a positive resist. If the non-irradiated regions are dissolved by the developer, a negative image of the mask results. Hence that resist is termed a negative resist. This principle is the same for all lithography techniques - only the source used for the irradiation of the resist differs. Generally used radiation sources include UV light, electron beam, X-rays and ion beams. Subsequently, the related resists and lithography methods are also named after the source (Campbell, 2001). A typical photoresist is composed of three types of materials a polymer base, photosensitizer and solvent.

**2.1.2.3 Preparation process**

Iron(III) acetylacetonate (Fe(acac)$_3$) was selected as the precursor for Fe and (N,N'-bis(Salicyliden)-ethylenediiminato) nickel(II) popularly known as Ni(Salen) for Ni. Both the chemicals were in powder form which was necessary for spin coating. For the dispersion of the catalyst powders, a low viscous positive photoresist HPR 504 (Fuji Film) was used. The spin speed curve of HPR 504 is shown in Figure 2.4. HPR 504 uses ethyl lactate as the solvent and has a viscosity of ~ 40 cps.

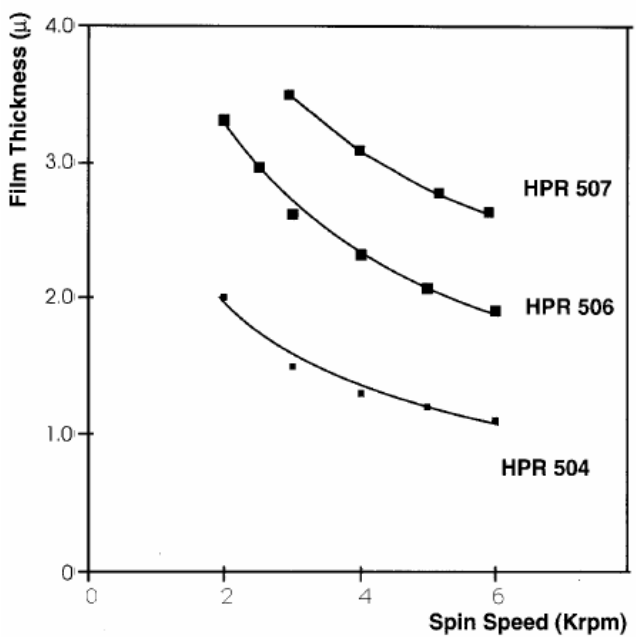

Figure 2.4: Spin speed curve of HPR series

The advantage of using the photoresist as a dispersion medium is that, after addition of metal powder with the photoresist, the modified resist itself plays the role both as a catalyst and also as the medium for patterning.

706 mg of Fe(acac)$_3$ was added to 10 ml of HPR 504 to obtain a modified photoresist (Mod-PR) solution of concentration 0.2 M. The solution was then stirred and sonicated for 30 min to achieve a good dispersion of the metalorganic molecular precursor. This solution was named as Fe-Mod-PR. Prior to spin coating, the Si(111) substrates were ultrasonically cleaned with acetone and deionized water. The solution was then spin coated with a rotation speed of 4000 rpm for 20 s on the Si(111) substrates to get a thin layer of the Fe-Mod-PR. For pattern formation using photolithography, after the spin coating of Fe-Mod-PR, the samples were baked at 90 °C for 15 min followed by an exposure step (contact printing) with a mask aligner to make an array of patterns. For exposure, UV light was used and the exposed specimens were developed in the developer solution for 60 s and then rinsed in distilled water. The samples were annealed in air for 10 min at 200 °C to improve the adhesion of the Fe-Mod-PR to the substrate. For annealing, an open air annealing system connected to a variac (240V, 10A, Type 10D-1P) was used. The temperature of the annealing chamber was controlled by a temperature controller (Maxthermo MC-2438) connected to a K-type thermocouple.

For Ni catalyst, 650 mg of Ni(Salen) was mixed with 10 ml of HPR 504 to obtain a modified photoresist (Mod-PR) solution of concentration 0.2 M. This solution was named as Ni-Mod-PR. The remaining procedure was similar to Fe-Mod-PR and a Ni-Mod-PR pattern was prepared. The samples were then annealed in air for 1 h at 400 °C to improve the adhesion of Ni-Mod-PR to the substrate. The general scheme to develop the catalyst patterns on Si substrates using positive photoresist is shown in Figure 2.5.

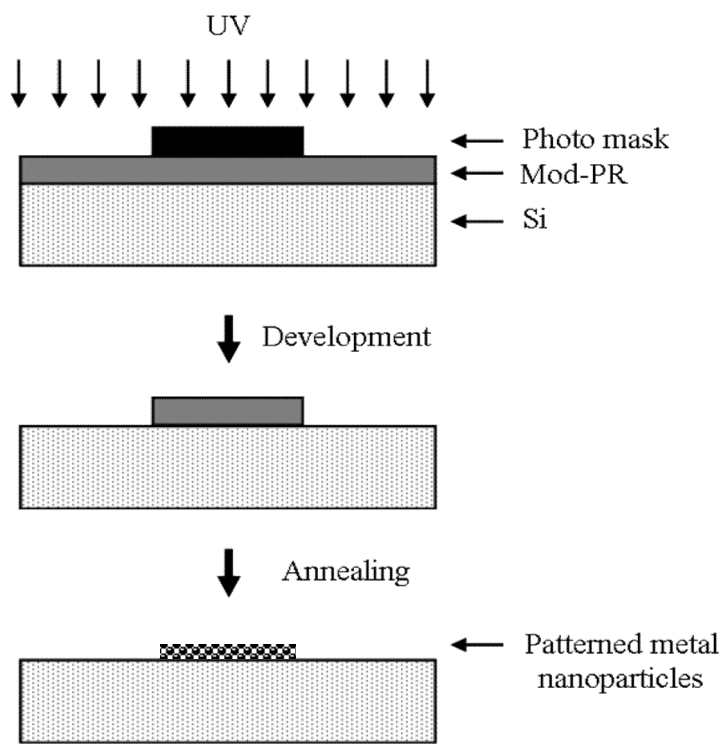

Figure 2.5: Schematic representation of the direct photolithographic route for the generation of Mod-PR pattern

## 2.2 Growth of carbon nanotubes by chemical vapor deposition

Vapor deposition is the transformation of vapors into solids, frequently used to grow solid thin film and powder materials. CVD is a generic name for a group of process that involves deposition of a solid on a heated surface from a chemical reaction in the vapor phase. The process is shown schematically in Figure 2.6.

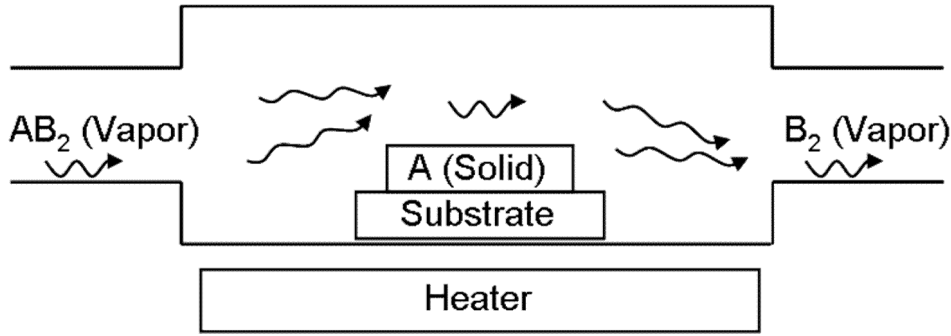

Figure 2.6: Schematic representation of a chemical vapor deposition process

In Figure 2.6, it is observed that the main constituents of the CVD process are a gaseous precursor ($AB_2$), energy (Heater), reactor, solid product (A) and gas phase product ($B_2$). In the CVD process, the substrate is loaded into the chamber using a substrate loading mechanism. A vacuum system is used for the removal of all other gaseous species other than those required for the deposition. Then the precursor gases are delivered into the reaction chamber using a gas delivery system and heated (by the energy source) as it approaches the deposition surface. As the precursor gases pass over or come into contact with a heated substrate, they react or decompose forming a solid phase which is deposited onto the substrate. Finally the volatile by-products are removed from the reaction chamber by an exhaust system (Dobkin and Zuraw, 2003). Although there are several methods for synthesizing CNTs, CVD is most widely used because of its simple construction, easy implementation and potential advantage to produce a large amount of CNTs growing directly on a desired substrate with high purity and large yield.

There are several types of CVD, which are classified based on their heating source, the process pressure, the reactor configuration and other parameters e.g. atmospheric pressure CVD (APCVD), low pressure CVD (LPCVD), metalorganic CVD (MOCVD), plasma enhanced CVD (PECVD) etc.

## 2.2.1 The chemical vapor deposition system

In this work a horizontal APCVD reactor was used for CNT growth. The CVD reactor is designed using a modular approach to allow further modification and enhancement in the future. Care has been taken in selecting components and materials of high purity and quality to minimize contamination. High quality stainless steel tubing (1/4" OD) is used to transport gases. All fittings in the gas lines are compression-type fittings (Swagelok/Excelok). The schematic diagram of the CVD system is shown in Figure 2.7. Figure 2.8 (a) and (b) are photographs of the system. The system consists of a 1 m long and 48 mm diameter quartz reaction tube in conjunction with two water-cooled flanges at the two ends. The outlet of the quartz tube is connected to a rotary pump (max. base pressure ~ $10^{-3}$ Torr). In order to initiate the chemical reaction at a particular temperature (550-1000 °C) followed by the nanotube deposition, a hot-wall resistance heated furnace (ELECTROHEAT EN345T-NC5010, India) with maximum continuous operating temperature of ~ 1250 °C was used. High purity (XL grade) gases (hydrogen ($H_2$), argon (Ar) and propane ($C_3H_8$)) from BOC (India) have been utilized with single stage gas regulators (Apollo, India).

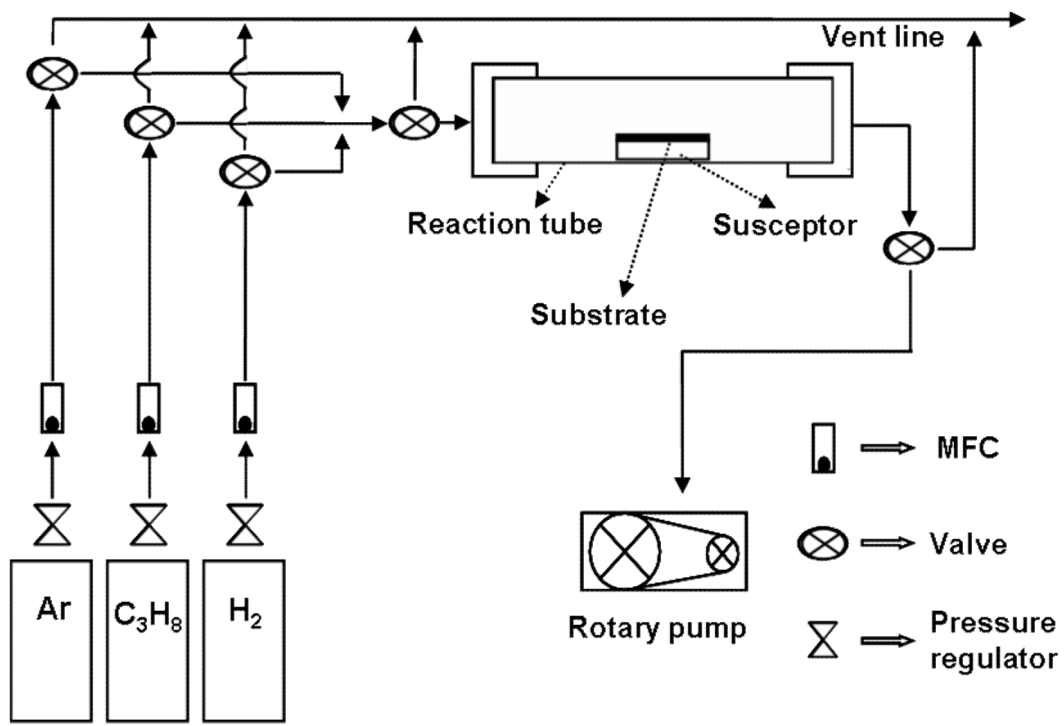

Figure 2.7: Schematic representation of the chemical vapor deposition system

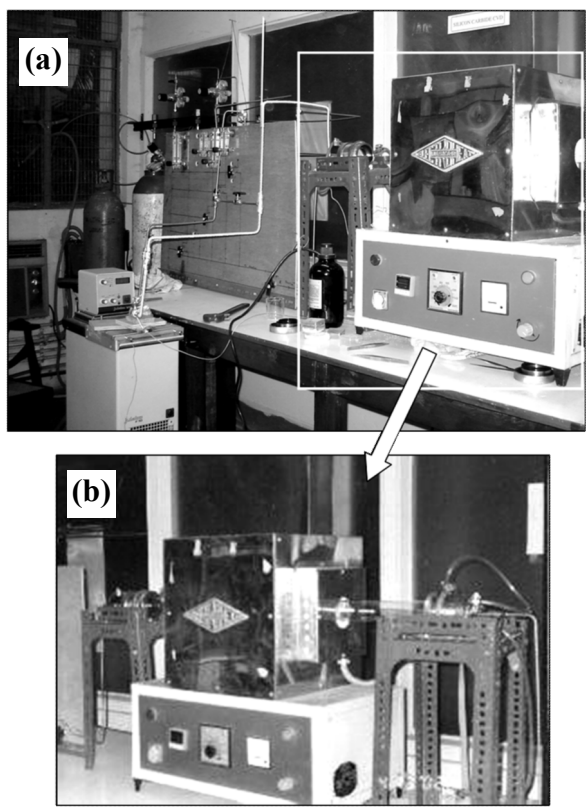

Figure 2.8: (a and b) Photographs of the chemical vapor deposition system

**Operation:** The substrates with a catalyst overlayer (prepared according to section 2.1) were mounted on a SiC-coated graphite susceptor and placed in the center of the reactor tube. The rotary pump was used to remove the air from the reaction chamber and then the chamber was backfilled with flowing argon with a flow rate of 1 slm (standard liters per minute) to atmospheric pressure. Thereafter, the samples were heated in argon up to the pre-heating temperature following which argon was replaced with hydrogen. Subsequently, the samples were annealed in a hydrogen atmosphere with a flow rate of 1 slm for 10 min. Finally, the reactor temperature was brought to growth temperature and the hydrogen was turned off. Consequently, propane was introduced into the gas stream at a flow rate of 200 sccm (standard cubic centimeters per minute) for 1 h for CNT synthesis. All growth was performed at atmospheric pressure and the pre-heating temperature is the same as growth temperature unless otherwise explicitly stated. The temperature of the system was controlled with an accuracy of $\pm 1$ °C using the temperature controller. After the synthesis, propane was turned off and hydrogen, with a flow rate of 1 slm, was introduced into the reactor. The reactor was allowed to cool down to room temperature in hydrogen atmosphere. Finally, the hydrogen flow was cut off and the chamber was purged with argon before it was opened and the sample taken out.

**Yield measurement:** A rough estimate of yield was performed by measuring the weight (in mg) of the sample before and after CNT growth using a micro balance (Sartorius AG Germany, CP225D). Before growth, the area of catalyst deposition over the substrate was measured and the yield expressed in units of mg/(cm$^2$·h).

## 2.3 Characterization methods

### 2.3.1 X-ray diffraction

X-ray diffraction (XRD) is a versatile, non-destructive technique that reveals detailed information about the chemical composition and crystallographic structure of the specimen. X-rays are a relatively short wavelength, high energy electromagnetic radiation and can be produced by rapid deceleration of an electrically charged particle with sufficient kinetic energy. Crystals are ordered, three-dimensional arrangements of atoms with characteristic periodicities. If the spacing between atoms is of the same order as X-ray wavelengths (1-3 Å), then the crystals can diffract the radiation. Diffraction occurs when each object in a periodic array scatters the incident radiation coherently, producing concerted constructive interference at specific angles. The wavelength of the X-rays is determined by the anode material of the X-ray source. Among different X-ray sources (Mo, Cu, Co, Fe etc) the most common radiation used for diffraction studies is the $CuK_\alpha$ radiation with $\lambda = 1.5406$ Å. Whenever the spacing, d, between two parallel planes satisfies the Bragg equation:

$$n\lambda = 2d \sin\theta \qquad \ldots(2.1)$$

(where, $\lambda$ is the wavelength of the incident beam, $\theta$ is the angle between the incident ray and the scattering planes and n is an integer), a diffraction maxima is produced. So, for different values of d spacing, diffraction maxima will form at different angles and the resulting analysis is described graphically as a set of peaks with intensity on the Y-axis and $2\theta$ on the X- axis. Diffraction from different planes of atoms produces a diffraction pattern which contains information about the atomic arrangement within the crystal. The pattern of diffracted X-rays is unique for a particular structure and can be used as a "fingerprint" to identify the material which may also be compared with the ICDD JCPDS data. There are several applications of powder diffraction e.g. phase identification, quantitative phase analysis, crystallite size and residual stress measurement (Suryanarayana and Norton, 1998).

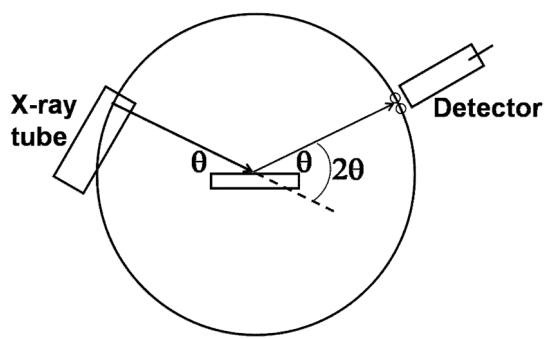

Figure 2.9: The Bragg-Brentano geometry

The X-ray diffractometers typically use the Bragg-Brentano geometry as shown in Figure 2.9. In this arrangement, a point detector and the sample are moved in a certain manner so that the detector is always at 2θ with respect to the incident beam and the sample surface is always at θ to the incident X-ray beam. Divergent X-rays from the source hit the sample at different points on its surface. Following the diffraction process, the X-rays are refocused at the detector slit. In a θ:2θ instrument, the tube is fixed, the sample rotates at θ°/min and the detector rotates at 2θ°/min. In a θ:θ instrument, the sample is fixed and the tube rotates at a rate -θ°/min and the detector rotates at a rate of θ°/min.

In this investigation, XRD was performed using a Philips PW 1710 X-ray diffractometer (Figure 2.10) with CuK$_\alpha$ radiation at an operating voltage of 40 kV and 20 mA current.

The samples were mounted on a holder and the scan rate was 3°/min with a θ:2θ arrangement.

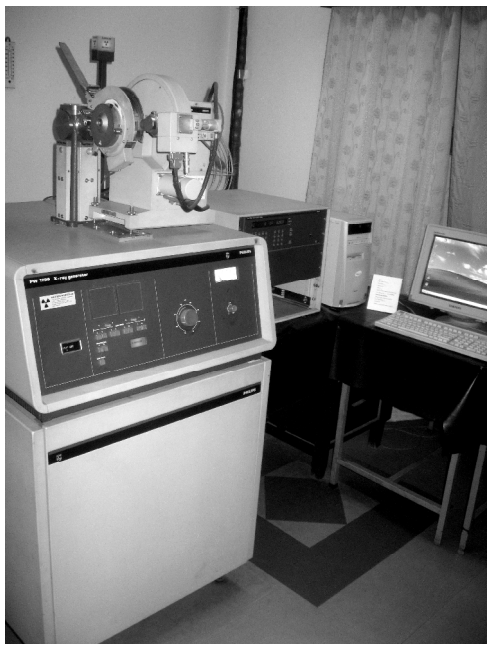

Figure 2.10: Photograph of the Philips PW 1710 X-ray diffractometer

## 2.3.2 Atomic force microscopy

An atomic force microscopy (AFM) measures the topography of a sample by monitoring the inter-atomic forces acting between the scanning probe and the atoms on the surface of the sample. Figure 2.11 shows the schematic diagram of the AFM instrumentation with the major components labeled: Laser (1), Mirror (2), Photodetector (3), Amplifier (4), Register (5), Sample with Scanner (6), Probe (7) and Cantilever (8). In AFM, an atomically sharp tip is brought into close proximity of the sample surface and scanned over the surface with feedback mechanisms to maintain the tip at a constant force (contact mode), or at a constant oscillation amplitude (non-contact & intermittent contact mode) above the sample. A laser is focused onto the back of the reflective cantilever. As the tip scans the surface of the sample, every variation of the surface height varies the force acting on the tip and therefore varies the bending of the cantilever. The laser beam bounce off of the cantilever and produces a quantification of the flexing of the cantilever as variations in the photodetector signal. The photodetector measures the difference in light intensities to evaluate the height of the sample surface at that position. The piezoelectric scanner adjusts Z position (Vertical) of the sample at points along the scan so that the atomic force is kept at a constant reference value. The adjustment made into the Z position reflects the topography of the sample and are used to map the surface.

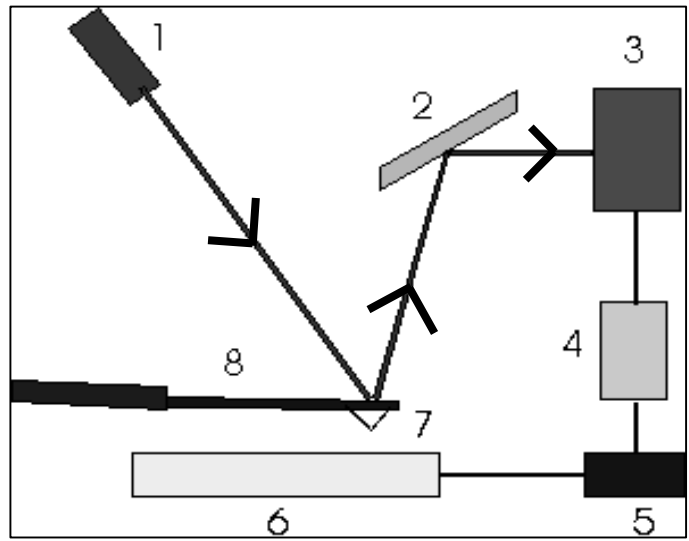

Figure 2.11: Schematic of the atomic force microscope

The actual z-position of the tip is recorded as a function of the lateral x–y position with very high precision. The information collected during a scan of the surface is quantitative in three dimensions which is unique to AFM analysis. It is possible to reconstruct an image of the surface features from different viewpoints. Profiles, or cross-sections, through the surface can be plotted and quantitative distances (horizontal and vertical) can be measured. Different parameters of the surface e.g. average roughness (Ra), RMS roughness ($R_{rms}$) can be computed along lines or over planes using the AFM data. AFM can be performed in all environments; ambient air, liquid and vacuum. There is minimal sample preparation required for AFM and the samples are not damaged as it is a non destructive technique. There are three main operating modes, namely contact mode, intermittent contact mode and non-contact mode. In contact mode the tip and the sample are in close contact whereas in non-contact mode of operation the tip is kept at a distance from the sample so that tip never touches the sample surface. In intermittent contact mode the cantilever is forced to oscillate at its resonant frequency and positioned above the surface such a way that it periodically taps the surface only for a very small fraction of its oscillation period.

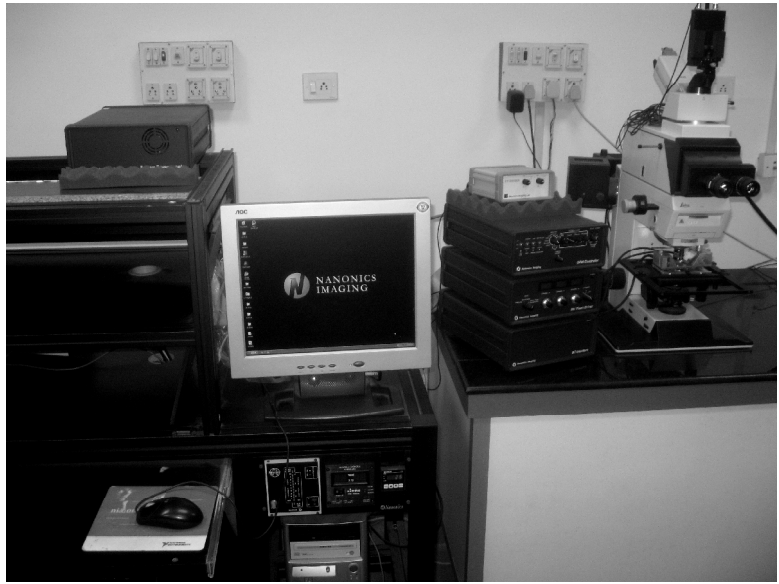

Figure 2.12: Photograph of the atomic force microscope system (Nanonics Multiview 1000$^{TM}$) used in this work

In this research work, a Nanonics Multiview 1000™ system (Figure 2.12) was used in AFM mode with a cantilevered quartz tip. The whole system was placed on an anti-vibration table to minimize vibrations. All the images were taken in intermittent contact mode. The spring constant of the tip was 40 N/m and the resonant frequency was ~ 80 kHz.

**2.3.3 Scanning electron microscopy**

In scanning electron microscopy (SEM), a focused electron beam is rastered across a sample surface to provide high resolution and large-depth-of-field images of the sample surface. As shown in Figure 2.13(a), a beam of electrons (either from a cathode filament or via field emission) is focused (in vacuum) by condenser lenses into a beam with a very small spot size at the surface of the specimen and scanned across it using scan coils. The scan of the sample surface mainly generates secondary electrons (SE) and backscattered electrons (BSE). Secondary electrons consist of low energy electrons originating from specimen due to the inelastic collision of high energy incident electrons. The backscattered electrons consist of high energy electrons originating due to the elastic reflection or back-scattering of incident electrons from the specimen. The emitted electron current is collected and amplified for clear viewing. The intensity of emission of both secondary and backscattered electrons is very sensitive to the angle at which the electron beam strikes the surface, i.e. to topographical features on the specimen. In addition to surface imaging, other information can be gathered from an SEM due to the volume interaction of the electrons in the material which include X-ray mapping, cathodoluminescence etc.

Both conventional and field emission SEMs were used in this work to study the nature of CNTs and to analyze the microstructures in plan view. A Zeiss Supra 40 microscope equipped with a field emission gun (FEG) operating at an acceleration voltage of 5 kV, a working distance of typically 5 mm and in secondary electron (SE) image mode was used. For normal SEM, a Vega Tescan and Zeiss Evo 60 were used. The photograph of a Zeiss Supra 40 is shown in Figure 2.13(b).

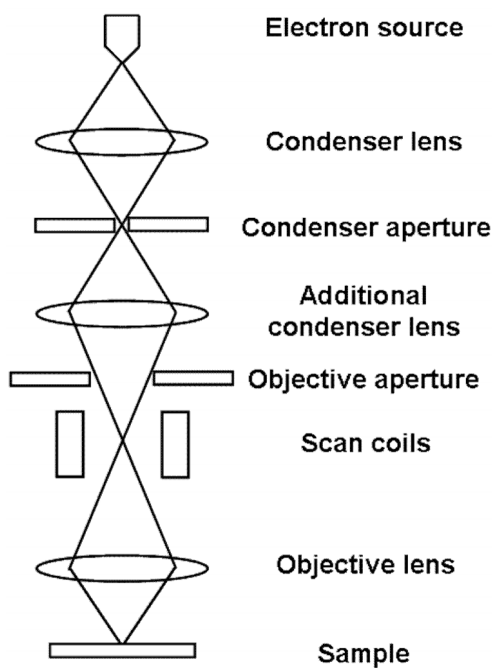
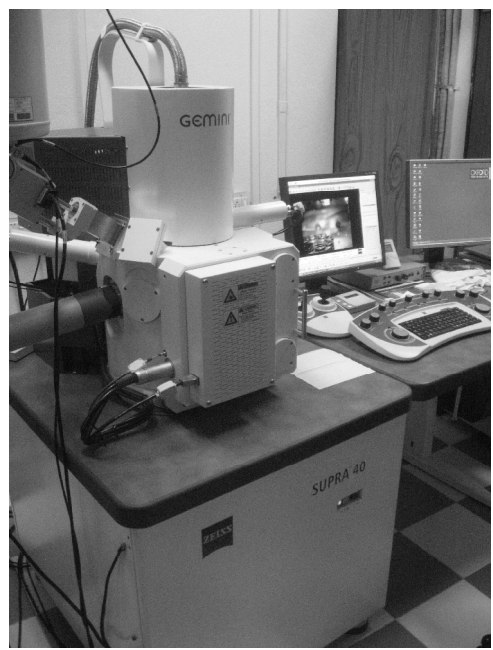

(a) (b)

Figure 2.13: (a) Electron path in a scanning electron microscope; (b) Photograph of the Zeiss Supra 40 scanning electron microscope

### 2.3.4 Transmission electron microscopy

The distinguishing feature of transmission electron microscopy (TEM) is its ability to form images of atomic arrangements at localized regions within materials. As shown in Figure 2.14, the electrons coming from an electron gun travel through vacuum in the column of the microscope and are concentrated on the specimen by a condenser lens. After passing through the specimen, the electrons are focused by the objective lens into a magnified intermediate image. This image is further enlarged by a projector lens and the final image is formed on a fluorescent screen, a photographic film or a CCD (charge-coupled device) camera. The main components of a TEM are vacuum system, specimen stage, electron gun (Tungsten or $LaB_6$), electron lens and apertures. There are several modes of operation of TEM like bright field, dark field, phase contrast, selected area diffraction etc. Among these modes, the bright field and selected area diffraction modes were used for the characterization of CNTs. In bright field imaging, the thicker regions of the sample, or regions with a higher atomic number will appear dark, while regions with no sample in the beam path will appear bright, hence the term "bright field". The bright field image is assumed to be a simple two dimensional projection of the sample. Bright field mode is used to determine the size and shape of nanostructures and at higher magnification it can image the sample with atomic resolution. By adjusting the lenses, a diffraction pattern can also be generated. For thin crystalline samples, this produces an image that consists of a pattern of dots, in the case of a single crystal or a ring pattern, in the case of a polycrystalline sample. A series of diffuse halos is produced for an amorphous material. This image provides information about the space group symmetries in the crystal and the crystal's orientation with respect to the beam path.

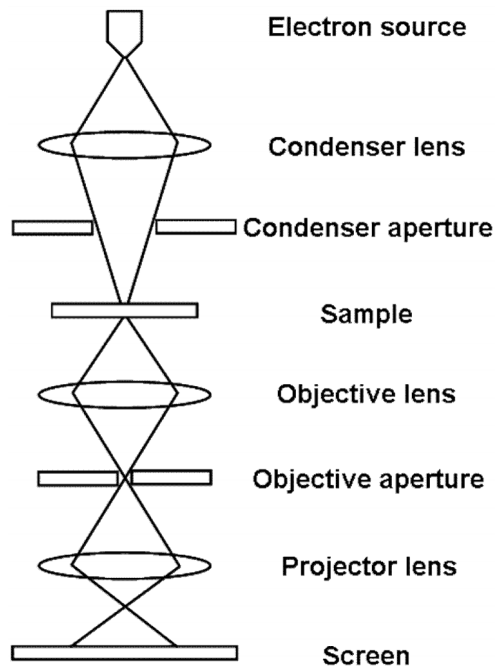

Figure 2.14: Electron path in a transmission electron microscope

In this work a JEOL 2010-High Resolution TEM, operating at 200 keV was used to determine the size, shape, crystallinity and structure of the CNTs and nanoparticles. The CNTs were scraped from the substrate and dispersed in alcohol, followed by ultrasonication for 30 minutes. After that, the samples were placed on a carbon coated copper grid and the alcohol evaporates. A photograph of the system is shown in Figure 2. 15.

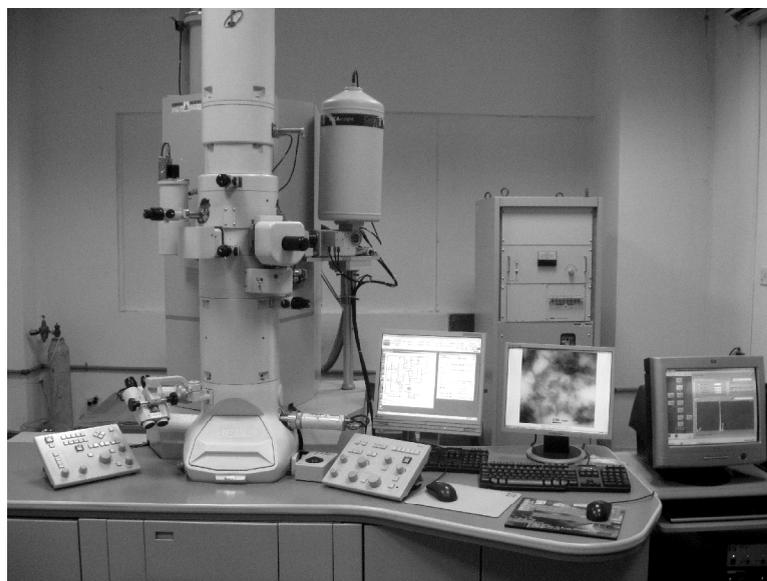
Figure 2.15: Photograph of the JEOL 2010 transmission electron microscope

## 2.3.5 Energy dispersive X-ray spectroscopy

Energy dispersive X-ray spectroscopy (EDX) is a chemical microanalysis technique used for the elemental analysis of a sample and can be coupled with several applications including scanning electron microscopy (SEM) and transmission electron microscopy (TEM). The EDX technique detects X-rays emitted from the sample during bombardment by an electron beam to characterize the elemental composition of the analyzed volume. A high energy beam of charged particles, such as electrons, is focused onto the sample. The incident beam can excite an electron in an inner shell of the specimen and eject it from the shell. The resulting electron vacancies are filled by electrons from a higher state and the difference in energy between the higher energy shell and the lower energy shell may be released in the form of an X-ray. The number and energy of the X-rays emitted from a specimen can be measured by an energy dispersive spectrometer using a lithium drifted silicon (SiLi) detector. The X-rays generated from any particular element are characteristic of that element and can be used to identify the element. The spectrum of X-ray energy versus counts is evaluated for qualitative and quantitative determination of the elements present in the sample volume. Elements of low atomic number are difficult to detect by EDX. In the present study, EDX (Oxford Instruments) was used in conjugation with SEM and TEM for the compositional analysis of CNTs.

## 2.3.6 Raman spectroscopy

Raman spectroscopy is a vibrational spectroscopic technique based on inelastic scattering of incident photons. When the sample is illuminated with a laser beam (a monochromatic light source), some of the photons are absorbed by the material and rest are scattered. The majority of these scattered photons have exactly the same frequency as the incident photons and the process is known as Rayleigh scattering. But a small portion of the scattered photons have frequencies that are shifted up or down in comparison with the original monochromatic frequency upon interaction with the sample caused by some quasi excitations of the medium. This is called the Raman effect. These excitations can be vibrational modes in a molecule, phonons in a crystal, plasmons, single particle electronic excitation, magnons etc (Cardona and Guntherodt, 1982; Hayes and Loudon, 1978). Most of the Raman scattered photons are shifted to the smaller frequency (Stokes shift) but a small portion is shifted to the higher frequency (anti-Stokes shift). Figure 2.16 shows a diagram of Stokes and anti-Stokes Raman scattering. In each case, the incident photon excites an electron into a higher energy level (or virtual state) and then the electron decays back to a lower level, emitting a scattered photon. The resulting frequencies can be expressed as $v = v_p - v_{vib}$ for the Stokes lines and $v = v_p + v_{vib}$ for the anti-Stokes lines with the primary light frequency $v_p$ (Figure 2.16). The difference $v_p - v$ is the Raman shift.

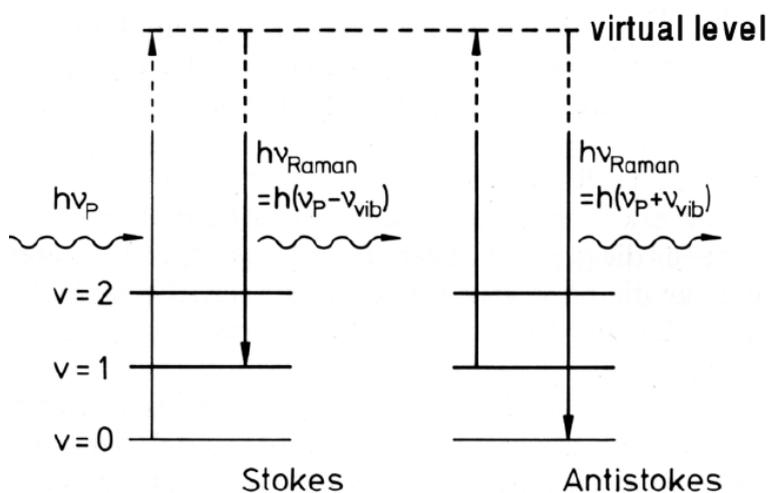

Figure 2.16: Energy-level diagrams of Stokes scattering and anti-Stokes scattering

Raman scattering can be used as an unique tool to characterize carbon materials such as graphite, diamond and CNTs, since the amount of ordering and degree of $sp^2$ and $sp^3$ bonding leaves a unique Raman "fingerprint". Therefore, Raman spectroscopy is used here to identify the CNTs and their degree of graphitization. The most prominent Raman features in CNTs are the radial breathing modes (RBMs), the higher frequency D (disordered) band and the G (graphite) band (Eklund et al., 1995). The G band is the tangential shear mode of carbon atoms that corresponds to the stretching mode in the graphite plane. The D band is related to the presence of defects in the CNT sample; the more defective the sample, the more intense the D band will be. Although the D and G bands are found in graphite, the RBM is specific to CNTs and is representative of the isotropic radial expansion of the tube. Thus the RBM bands are a useful diagnostic tool for confirming the presence of CNTs in a sample. The RBM frequency is inversely proportional to the diameter of the tube (Jinno et al., 2004), making it an important feature in determining the diameter distribution in a sample. The intensity ratio of the D band peak to G band peak is also a direct measure of the quality of the sample. A very small intensity of the D band compared to the G band signifies a high degree of graphitization and good quality of the nanotubes (Saito et al., 1998).

In this work, Raman spectra were recorded in a backscattering configuration using either a 488 nm or 514.5 nm line of Ar ion laser. For the 488 nm line, an assembled Raman spectrometer was used; the spectrometer was equipped with a TRIAX550 single monochromator with a 1200 grooves/ mm holographic grating, a holographic super-notch filter and a Peltier-cooled CCD detector (Figure 2.17). For the 514 nm line, a Renishaw RA1000B LRM Raman analyzer was used. In both the cases, the spectra were calibrated with reference to the Si peak.

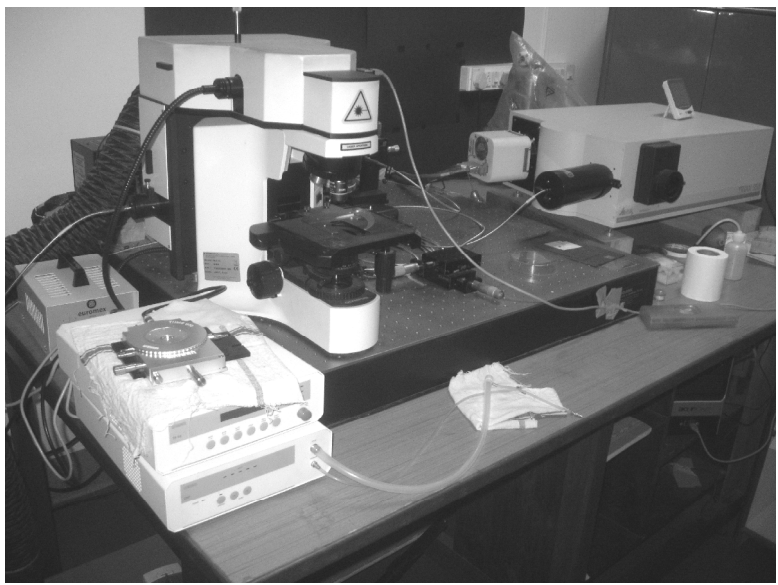

Figure 2.17: The photograph of the Raman Spectrometer with 488 nm Ar ion laser

# Chapter 3

# Growth and characterization of carbon nanotubes by chemical vapor deposition using elemental metal catalysts

The metal clusters acting as a catalyst for CNT growth can be produced in different ways, e.g. by evaporation or sputter deposition of a thin metallic layer on the substrate, by filling of porous materials by metals, by spin coating of solutions containing catalyzing metals or by introduction of organometallic substances into the reactor. Among these catalyst production methods thermal evaporation of elemental metal catalyst and spin coating of an organic complex containing catalyzing metals have been employed for introducing the catalyst over Si(111) substrates. The growth of CNTs using thermally evaporated Fe and Ni catalysts is discussed in this chapter, while the CNT synthesis using spin coated metal complex is discussed in the next chapter.

## 3.1 Synthesis of carbon nanotubes using thermally evaporated elemental Fe metal as catalyst

### 3.1.1 Growth temperature dependence on the Fe catalyzed growth of carbon nanotubes

Here, the results of systematic experiments to study the effect of synthesis temperature on the nanotube growth behavior are reported.

#### 3.1.1.1 Experimental details

Synthesis of CNTs was carried out at 650, 750, 850 and 950 ºC by catalytic decomposition of propane on Fe thin films over Si(111) substrates in a hot-wall horizontal APCVD reactor. The as-grown CNTs were analyzed by atomic force microscopy (AFM), X-ray diffraction (XRD), field emission scanning electron microscopy (FESEM), high resolution transmission electron microscopy (HRTEM), energy dispersive X-ray (EDX) and Raman spectroscopy to understand their crystallinity, morphology and defects.

**3.1.1.2 Results and discussion**

Figure 3.1a shows the AFM image of the surface profile of the as-deposited Fe film on a Si(111) substrate. The AFM image reveals that the initial film consists of tiny clusters with RMS roughness 3.3 nm. This indicates that the catalyst deposition proceeds via island nucleation and coalescence (Tiller, 1991). Figure 3.1(b-e) show the AFM images of the surfaces of Fe films after annealing at 650, 750, 850 and 950 °C in hydrogen. It can be observed that the heat treatment caused the formation of larger islands due to the coalescence of clusters and creates a relatively rough surface. The RMS roughness of the films annealed at 650, 750, 850 and 950 °C are 5.6, 6.7, 9.5 and 10.6 nm, respectively. The coalescence of clusters and formation of macroscopic islands is based on cluster diffusion which terminates when the island shape is of minimum energy for the specific annealing conditions (Jak et al., 2000). The surface diffusion of the clusters increases upon increasing the annealing temperature and the sizes of the islands become larger but their number density decreases as observed in Figure 3.1(b-e) (Müller et al., 2009; Pisana et al., 2007).

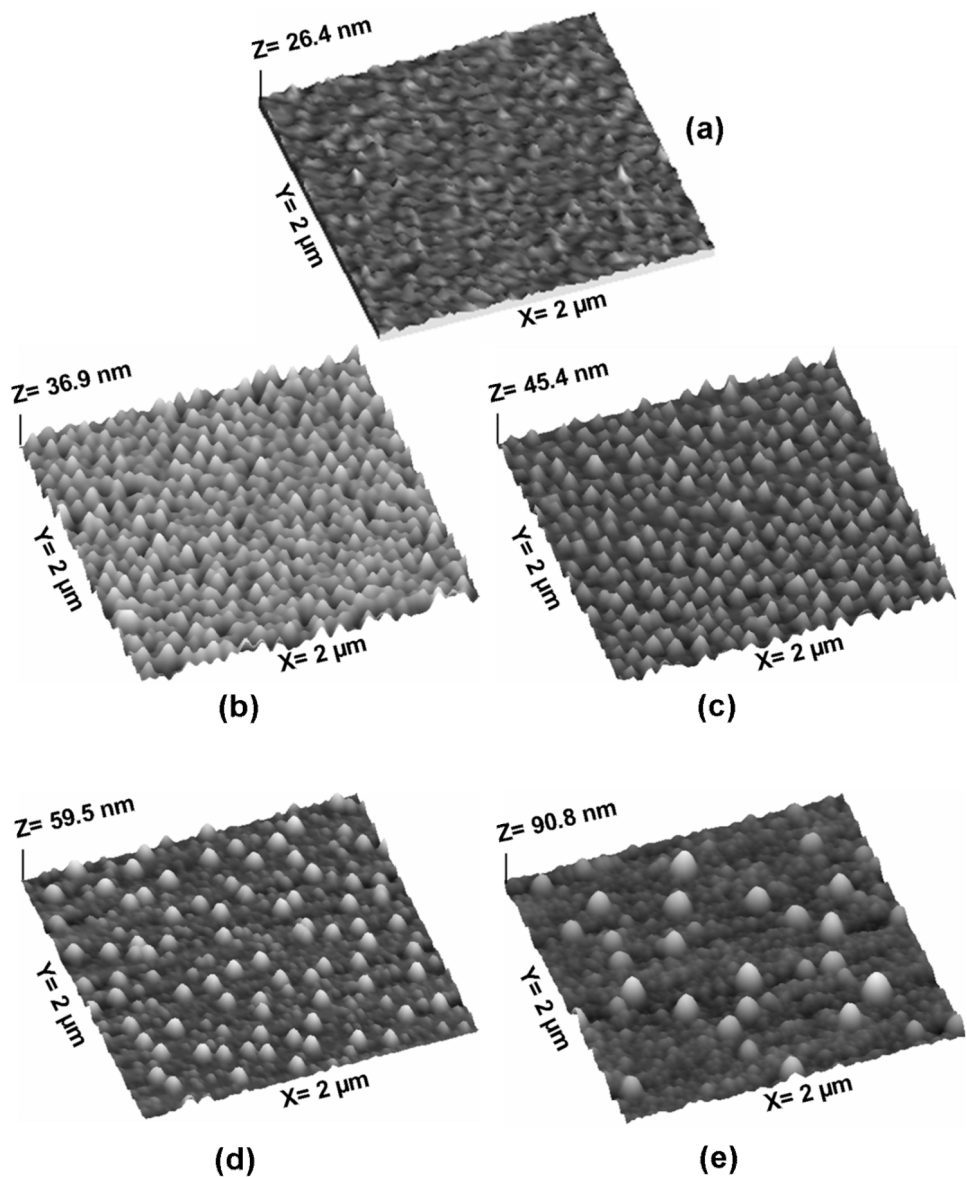

Figure 3.1: Atomic force microscopy images of the surface of the Fe catalyst on the Si substrates (a) The as-deposited catalyst surface; (b-e) The catalyst surface after the thermal treatment at 650, 750, 850 and 950 °C

Figure 3.2(a-d) are the FESEM images of CNTs synthesized at 650, 750, 850 and 950 ºC. The grown CNTs were of high aspect ratio and had randomly oriented spaghetti-like morphology in all the cases. The average diameter of the CNTs increased from ~ 25 nm at 650 ºC to ~ 77 nm at 950 ºC, as summarized in Table 3.1, while the number density of the CNTs decreased.

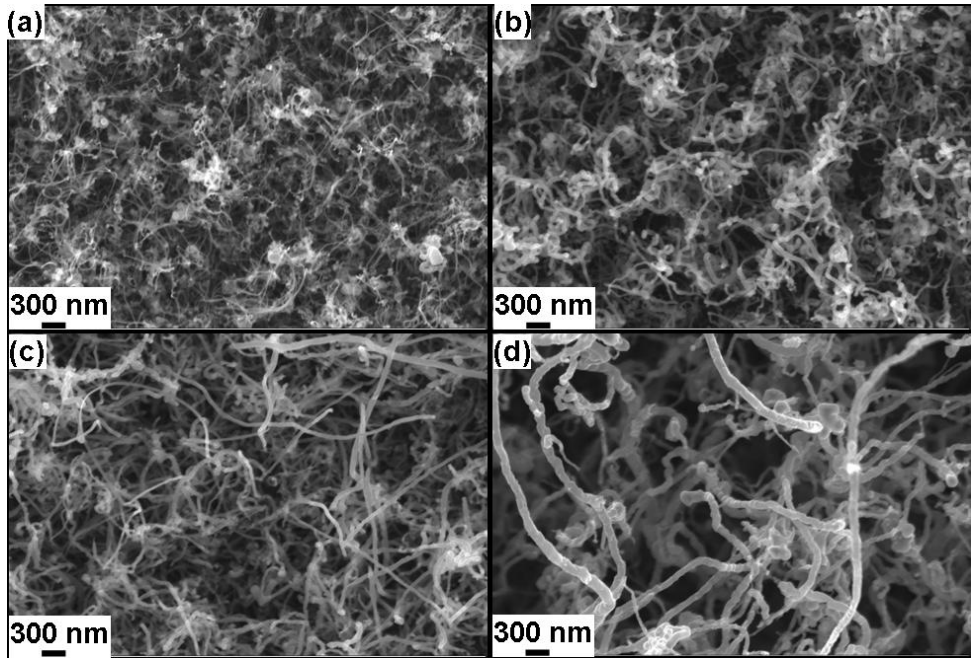

Figure 3.2: Field emission scanning electron microscopy images of carbon nanotubes grown over Fe catalyst at different temperatures (a) 650 °C; (b) 750 °C; (c) 850 °C and (d) 950 °C

Table 3.1: The Diameter Distribution and Averaged Diameter of Carbon Nanotubes Grown on Fe Catalyst using Thermal Chemical Vapor Deposition of Propane at 650, 750, 850 and 950 ºC

| Temperature (°C) | CNT diameter range (nm) | Average CNT diameter (nm) |
|---|---|---|
| 650 | 10-50 | 25 ± 5 |
| 750 | 20-70 | 46 ± 5 |
| 850 | 30-110 | 54 ± 5 |
| 950 | 30-150 | 77 ± 5 |

With the increase in the growth temperature, the migration rate of Fe nanoparticles on the Si surface increases, resulting in significant agglomeration of Fe nanoparticles. Consequently due to this agglomeration, at higher temperatures during the thermal CVD process, the size of the Fe nanoparticles increased but their number density decreased (Yadav et al., 2009). This eventually governs the diameter and number density of the grown nanotubes.

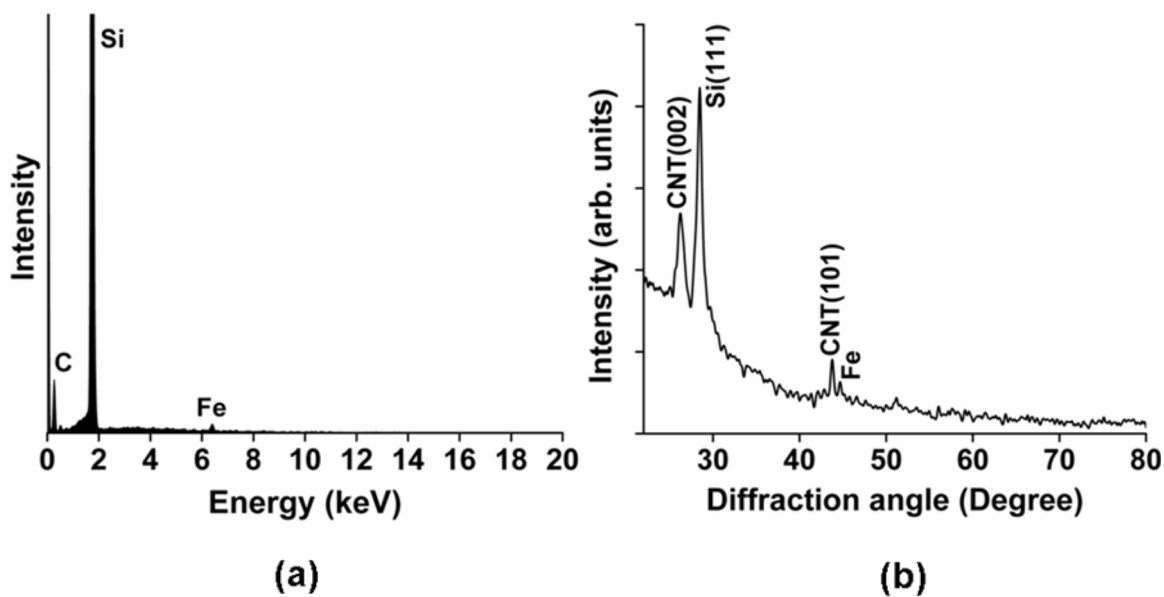

Figure 3.3: (a) Energy dispersive X-ray spectrum obtained from the nanotubes grown at 850 °C; (b) X-ray diffraction spectrum of multi-walled carbon nanotubes grown at 850 °C on Si(111) substrate using elemental Fe metal catalyst

For the compositional analysis of the grown species, EDX and XRD were performed. EDX (Figure 3.3a) shows that the grown material contains only carbon and Fe, while the Si peak is due to the substrate. In the XRD pattern (Figure 3.3b), the peak at 26.2° is the characteristic graphitic peak arising due to the presence of MWCNTs in the sample. The peak near 43.7° is attributed to the (101) plane of the nanotube and the peak at 44.7° is from the Fe catalyst (JCPDS card No. 06-0696). The peak at 28.4°, however, is not from the CNTs and is attributed to (111) plane of the Si substrate. The XRD pattern confirms the presence of MWCNT, Fe and Si which is in agreement with the EDX analysis.

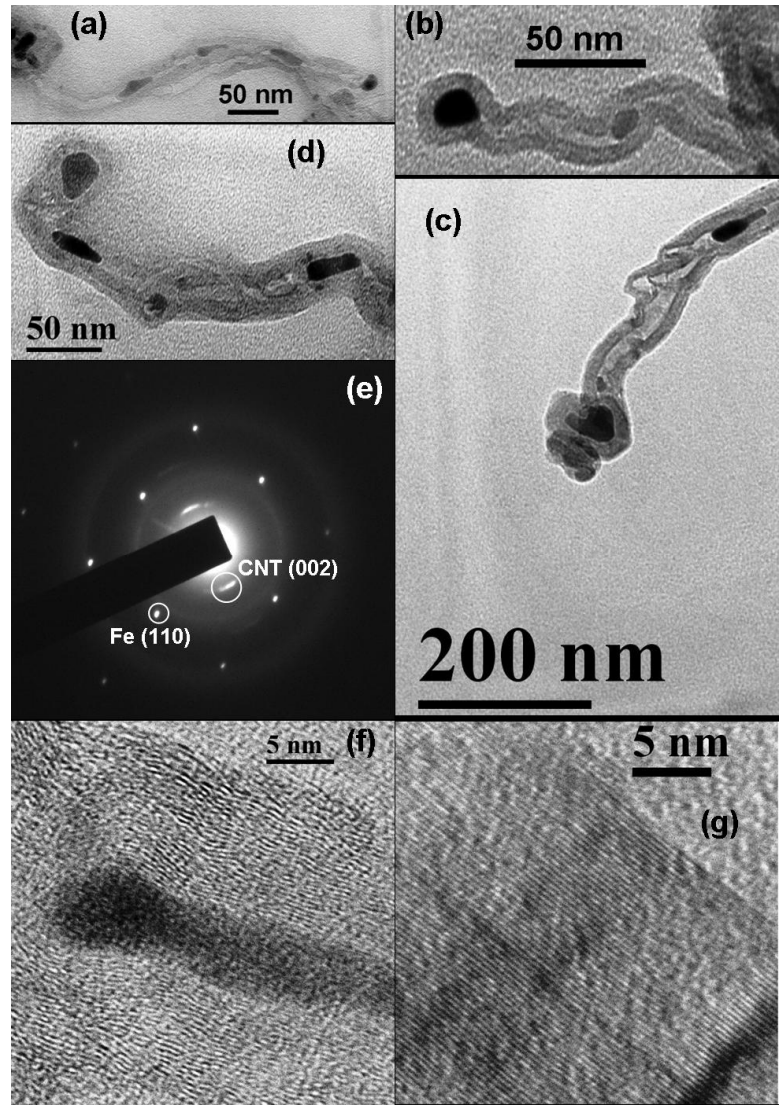

Figure 3.4: High resolution transmission electron microscopy images of carbon nanotubes grown from Fe catalyst at different temperatures (a) 650 °C; (b) 750 °C; (c) 850 °C; (d) 950 °C; (e) Selected area diffraction pattern of an elongated Fe nanoparticle; (f) Lattice image from a carbon nanotube grown at 650 °C; (g) Lattice image from a carbon nanotube grown at 850 °C

Figure 3.4(a-d) are the HRTEM images of CNTs synthesized at 650, 750, 850, and 950 ºC, illustrate that the nanotubes produced at all four temperatures are multi-walled. Elongated nanoparticles encapsulated inside the nanotubes demonstrate the hollow nature of the nanotube deposited using Fe. A selected area electron diffraction (SAED) pattern

of the elongated Fe nanoparticle shows spots due to (110) planes of Fe (Figure 3.4e). The diffraction pattern characteristic of the CNT (002) plane originating from the shielding graphitic layers is also labeled. Typically, the pattern consists of short arcs, rather than dots, because of imperfectly graphitized tubular walls. The SAED pattern indicates the single crystalline nature of Fe particle encapsulated within the nanotube.

To investigate the effect of the growth temperature on the crystallinity of graphite sheets, high resolution images of CNTs were also acquired. The HRTEM image of a CNT grown at 650 ºC (Figure 3.4f) shows the wavy structure of graphitic sheets at short range. It is well known that the wavy structure is caused by defects in the graphite sheet. In contrast, the CNT grown at 850 ºC (Figure 3.4g) have clear well-ordered and straight lattice fringes of graphitic sheets separated by 0.34 nm. HRTEM observations indicate that the crystallinity of CNTs improves as the growth temperature increases.

The room temperature Raman spectra obtained from CNTs that were grown at four different temperatures are shown in Figure 3.5. The Raman spectra show two main bands around 1360 cm$^{-1}$ (D band) and around 1585 cm$^{-1}$ (G band), which indicates the presence of MWCNTs (Sadezky et al., 2005). The strong band around 1585 cm$^{-1}$, which is referred to as the G band, corresponds to the $E_{2g}$ mode i.e. the stretching mode of the C-C bond in the graphite plane and demonstrates the presence of crystalline graphitic carbon. The D band at around 1360 cm$^{-1}$ originates from defects in the curved graphene sheets, tube ends etc. Beside the D and G band, the spectra also show another band at ~ 1620 cm$^{-1}$ (for 650 and 750 ºC) which appears as a shoulder on the G band and is called the D′ band.

This band arises due to the structural disorder. Like the G band of graphite, the D′ band also corresponds to a graphitic lattice mode with $E_{2g}$ symmetry (Sadezky et al., 2005).

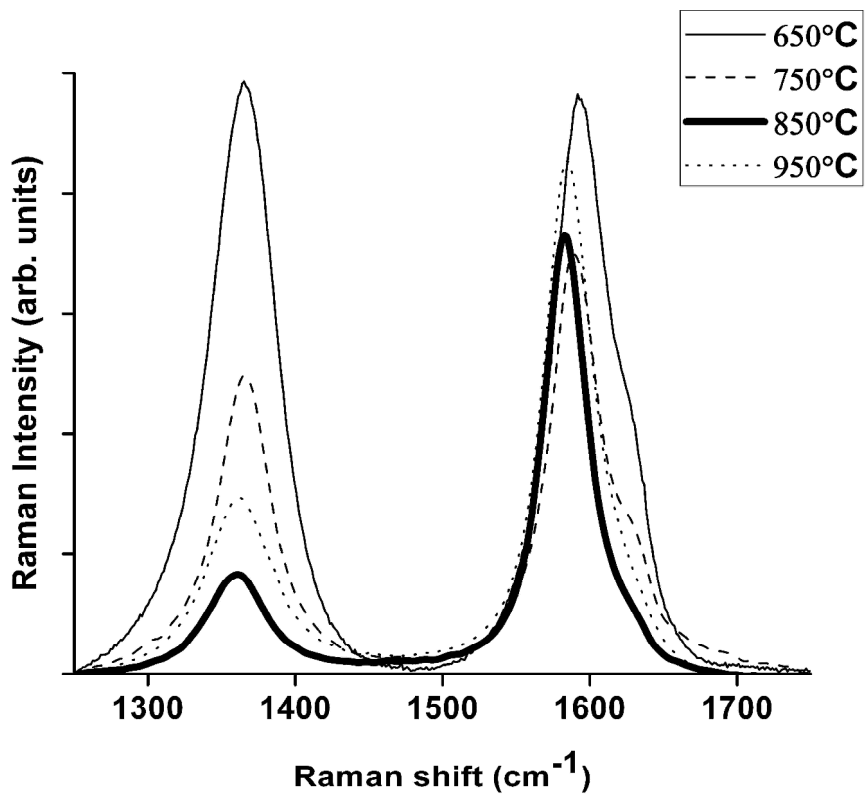

Figure 3.5: Raman spectra (488 nm excitation) of multi-walled carbon nanotubes grown at different temperatures using Fe catalyst

Though the D band and G band are present in all CNT samples, there is a difference in peak positions and also a variation in the intensity ratios of these two bands as a function of growth temperature as summarized in Table 3.2. The observed up-shift of the G band for the nanotubes deposited at 650 and 750 ºC with respect to that at 850 and 950 ºC temperature, can be explained in terms of an overlapping of the G band and D′ band, and a single fit to the G+D′ feature gives a net increase in the G band position (Ferrari and Robertson, 2000). This suggests that the graphitic carbon films deposited at 650 ºC and

Table 3.2: Raman Peak Positions and G to D Band Intensity Ratios for the Carbon Nanotubes Grown at Four Different Temperatures

| Growth temperature (ºC) | D band position (cm$^{-1}$) | G band position (cm$^{-1}$) | $I_D/I_G$ |
|---|---|---|---|
| 650 | 1365.2 | 1591.7 | 1.02 |
| 750 | 1365.1 | 1589.1 | 0.71 |
| 850 | 1361.2 | 1582.7 | 0.23 |
| 950 | 1361.9 | 1584.0 | 0.34 |

750 ºC temperatures consist of more disordered structure. The up-shift of the D band in these films can be interpreted as the signature for the increase of disorder in the graphitic structure in accordance with similar observations from other carbon-based films (Choi et al., 2002). It is well known that the $I_D/I_G$ ratio is dependent on the quality of CNTs and that the D band intensity is also related to the amount of amorphous carbon and defects (Jiao et al., 2009; Tuinstra and Koenig, 1970). The highest value of $I_D/I_G$ obtained for CNTs grown at 650 ºC indicate that the tubes grown at this temperatures are more defective as compared to those grown at higher temperatures. Above 650 ºC, the relative intensity of the D band to the G band decreases with increasing growth temperature (up to 850 ºC) indicating an increasing degree of graphitization in the material. Further increase in the growth temperature results in an increase in the $I_D/I_G$ values. This may be due to the formation of amorphous carbon which results in a slight increase in the D band, thus resulting in higher $I_D/I_G$ values.

In the large number of reports regarding filled CNT synthesis using CVD (Geng and Cong, 2006; Leonhardt et al., 2003; Müller et al., 2006; Zhang et al., 2002) the procedure followed to fill the CNTs is the floating catalyst method whereas the method used here utilizes the fixed catalyst for nanotube growth. On the basis of the aforementioned experimental results and analysis, it can be concluded that the growth is primarily governed by the tip growth mechanism. The following model, schematically shown in Figure 3.6 is suggested to explain the growth mechanism of the nanotubes having a tubular structure with partial Fe filling. The growth process starts with the diffusion of carbon into the metallic Fe nanoparticles. Due to their small diameter, the melting point of these nanoparticles is far below the melting point of the bulk metal (Buffat, 1976; Pan et al., 2004; Thomas and Walker, 1964). This suggests that the nanoparticles are in a liquid state during the decomposition of propane and they can easily change their shape. In particular, the small metal nanoparticles with diameters smaller than the inner tube diameter of the growing CNTs can diffuse into the cavities due to nanocapillarity (Ajayan and Iijima, 1993, Fujita et al., 2003). Constraining forces from the walls of the encapsulating nanotube would confine the catalyst particle to a fixed diameter. This process is shown in Figure 3.6. The diameter restriction provided by the inner nanotube walls also explains why the widths of all the particles observed in the middle of nanotubes matched the inner diameter of those CNTs. As the carbon concentration inside the metallic particle exceeds supersaturation, segregation of graphite layers takes place leading to CNT growth.

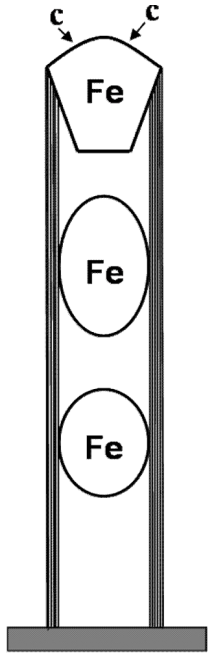

Figure 3.6: The proposed growth model for partially Fe filled multi-walled carbon nanotube

**3.1.1.3 Summary**

To summarize, the Fe catalyzed APCVD growth of partially Fe filled MWCNTs on Si substrates with respect to growth temperature has been investigated. It was observed that the average diameter of CNTs increases with growth temperature from 650 to 950 °C, but their number density decreases. Moreover, the crystallinity of the CNTs improves progressively with increasing growth temperature up to 850 ºC, which is the optimum growth temperature in the present case. The results demonstrate that the number density, diameter and crystallinity of partially Fe filled MWCNTs can be effectively controlled and optimized by adjusting the growth temperature and their growth occurs primarily by the tip growth model based on the capillary action of the liquid-like Fe particle.

## 3.1.2 Pre-heating effect on the Fe catalyzed growth of carbon nanotubes

The surface reconstruction of the Fe catalyst films due to high temperature processing in hydrogen prior to nanotube nucleation and its effect on the growth morphologies of CNTs synthesized using APCVD of propane is discussed in this section.

### 3.1.2.1. Experimental details

APCVD of CNTs was carried out on Si(111) substrates using propane at 850 °C with pre-heating of the Fe catalyst at 850, 900 and 1000 °C. The CNTs synthesized with pre-heating at 850, 900 and 1000 °C were assigned the name a-CNT, b-CNT and c-CNT, respectively. The as-grown CNTs were analyzed using AFM, XRD, SEM, HRTEM, EDX and Raman spectroscopy.

### 3.1.2.2 Results and discussion

Figure 3.7a shows the AFM image of the as-deposited catalyst film over the Si(111) substrate. The AFM image reveals that the initial film consists of Fe clusters instead of a continuous layer. Figure 3.7(b-d) show the intermittent contact mode AFM images of the surfaces of Fe films after being annealed at 850, 900 and 1000 °C in hydrogen for 10 min. Fe islands are formed due to the heat treatment as observed in the AFM images. However, these Fe islands are not uniformly distributed on the substrate and their sizes range from about tens of nanometers to hundreds of nanometers. It is clear that the size and number density distribution of Fe nanoparticles change drastically after annealing at 1000 °C compared to the heat treatment at 850 and 900 °C.

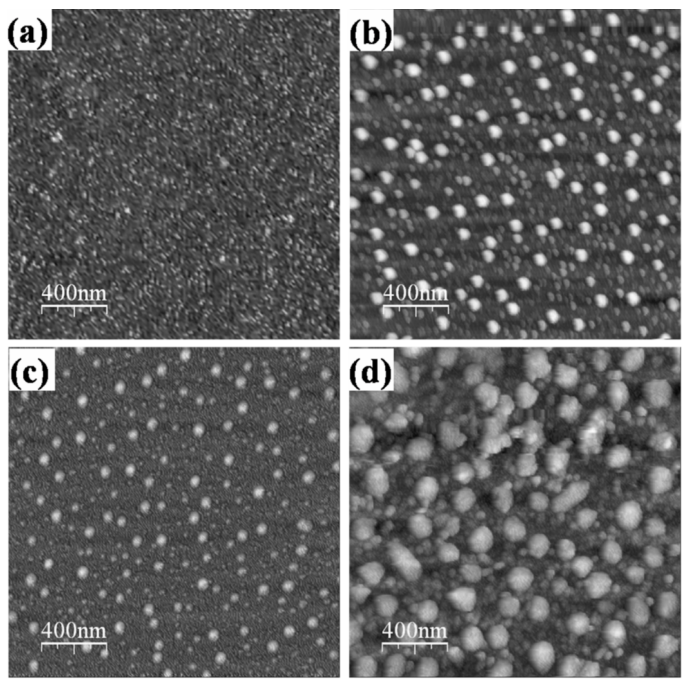

Figure 3.7: Atomic force microscopy images of the surface of the Fe catalyst on the Si substrates (a) The as-deposited catalyst surface; (b-d) The catalyst surface after the pre-heating at 850, 900 and 1000 °C

In order to precisely investigate the effect of annealing temperature on the reconstruction of Fe films, the statistical analysis of the size distribution of the Fe particles obtained after thermal treatment in hydrogen at different temperatures was performed (Figure 3.8a-c). The Fe nanoparticles annealed in hydrogen at 850 °C have a multi-modal distribution and range in about 3 nm to 170 nm. Upon increasing the annealing temperature from 850 to 900 °C, the distribution range of the particle sizes decrease. However, after annealing at 1000 °C the distribution range of particle sizes increases again and the particles have a clear bimodal distribution i.e. below and above 100 nm.

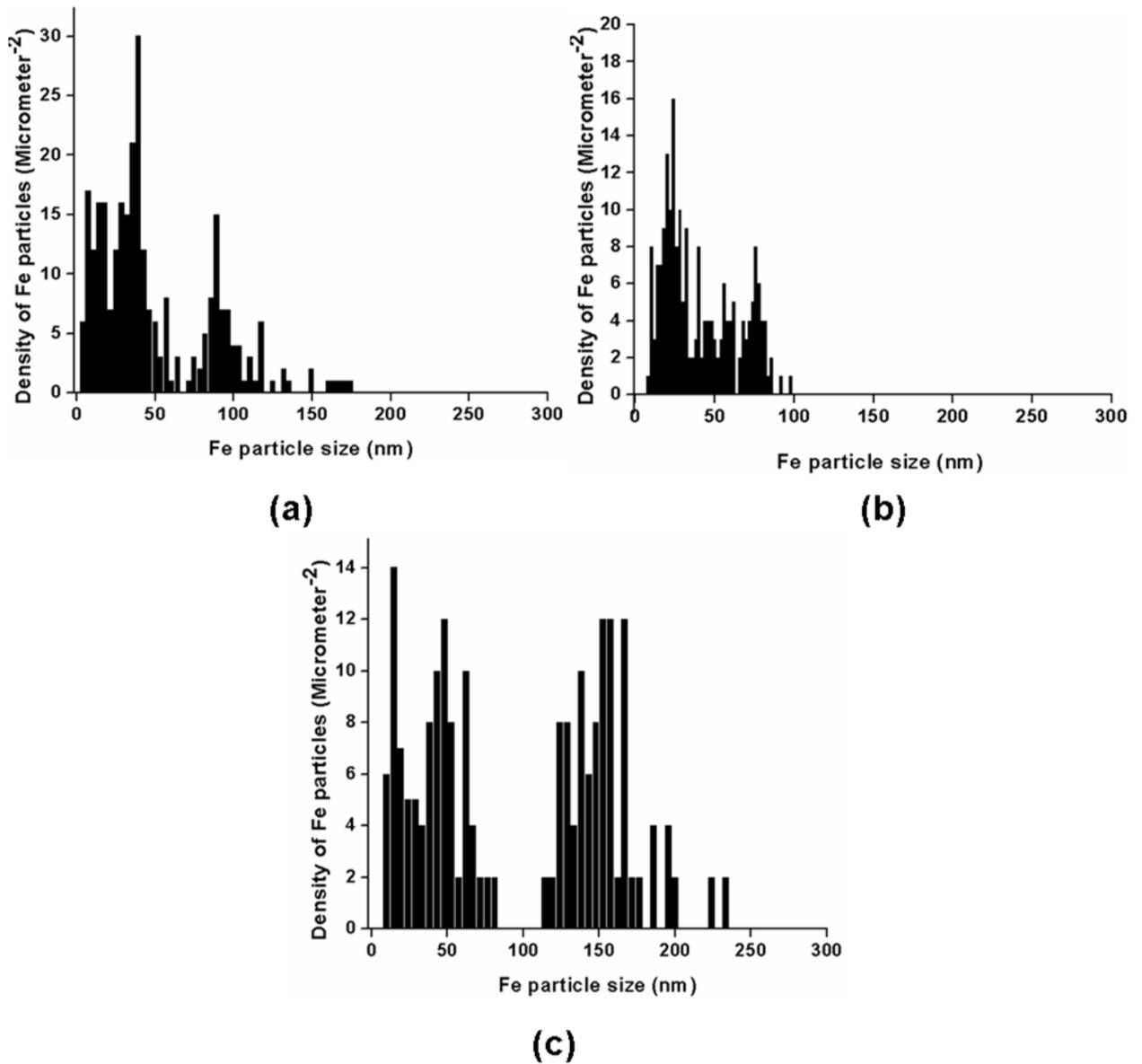

Figure 3.8: The distribution histogram of the Fe particles after pre-heating at (a) 850 °C; (b) 900 °C and (c) 1000 °C

The AFM images and statistical results together demonstrate that the Fe particles are refined and there is a small increase in uniformity upon increasing the annealing temperature from 850 to 900 °C. However, with the heating at 1000 °C, larger particles form and the uniformity decrease again.

Figure 3.9(a-f) are SEM micrographs showing the surface morphologies of the a-CNTs, b-CNTs and c-CNTs. In particular, the growth is very sparse in case of c-CNT (Figure 3.9c); the higher magnification image of c-CNT sample (Figure 3.9f) shows that the nanotubes grow only over the smaller clusters whereas the large clusters are ineffective for nanotube growth. However for the a-CNT and b-CNT films, better results were achieved with a high number density of nanotubes and for b-CNT the measured yield was ~ 0.1 mg/(cm$^2$·h). In many cases, small bright catalyst particles were detected at the tip of the CNTs. This suggests that the tip growth mechanism is likely to be responsible for the nanotube synthesis under the present conditions.

The diffusion model can be used to explain the growth of CNTs on catalyst clusters (Rodriguez, 1993). According to the diffusion model, CNTs grow by diffusion driven precipitation of carbon from the supersaturated catalyst particles. If the cluster size is larger than the diffusion length of the carbon atoms then CNTs cannot grow. In the present case, the catalyst film mostly forms large clusters when pre-heated to 1000 °C whereas, at 850 and 900 °C, relatively smaller clusters are produced.

In case of c-CNTs, the size of most of the clusters is much larger than the diffusion length. Therefore, carbon atoms supplied by decomposition of propane cannot diffuse

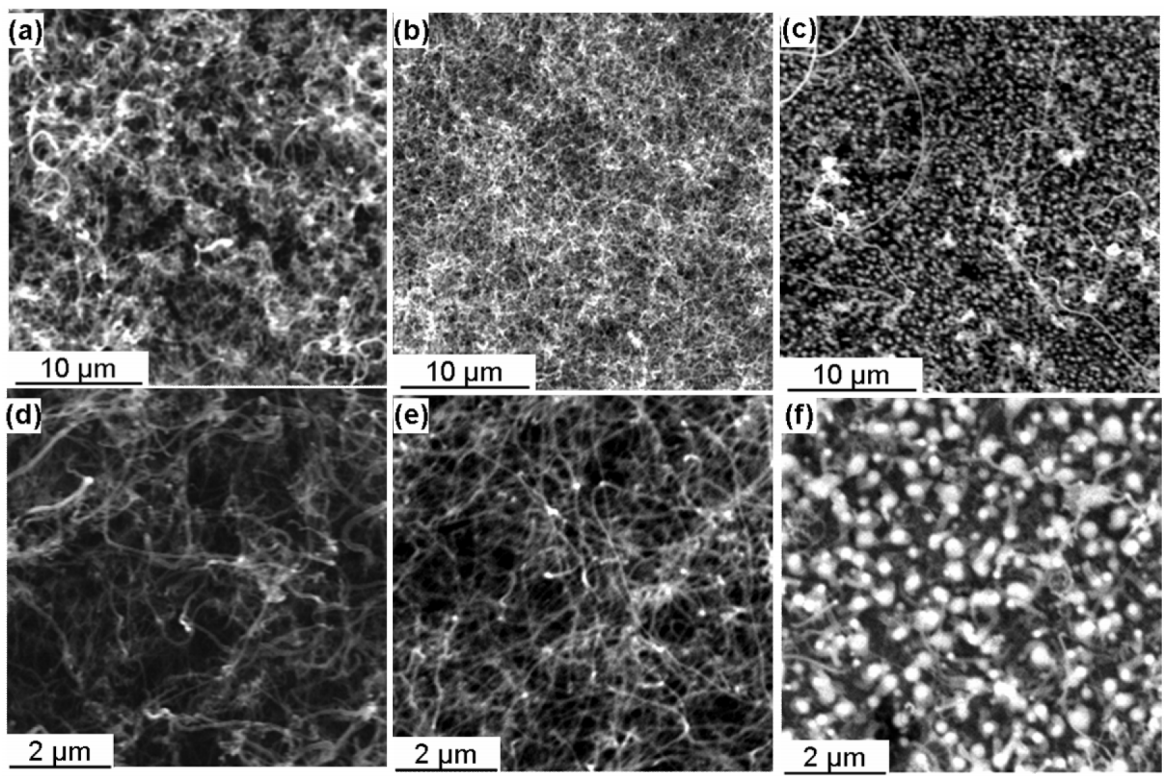

Figure 3.9: Scanning electron microscopy images of carbon nanotube grown from Fe catalyst after pre-heating in hydrogen at different temperatures (a) and (d) 850 °C; (b) and (e) 900 °C; (c) and (f)1000 °C

into the catalyst clusters to form CNTs (Figure 3.9 c and f). Only a limited number of CNTs grow over the smaller clusters. For a-CNT and b-CNT, most of the catalyst clusters are smaller than the diffusion length. Therefore, these clusters can effectively aid CNT growth resulting in a high number density of CNTs.

Figure 3.10-12 show HRTEM images of a-CNT, b-CNT and c-CNT samples, respectively. The lattice images (Figure 3.10b, 3.11b and 3.12b) elucidate the effects of preheating on the degree of graphitization of the a-CNT, b-CNT and c-CNT, respectively. The CNTs exhibit a multi-walled structure for all tubes grown under the three different conditions and their diameters are below 100 nm. A catalytic nanoparticle of nearly 40

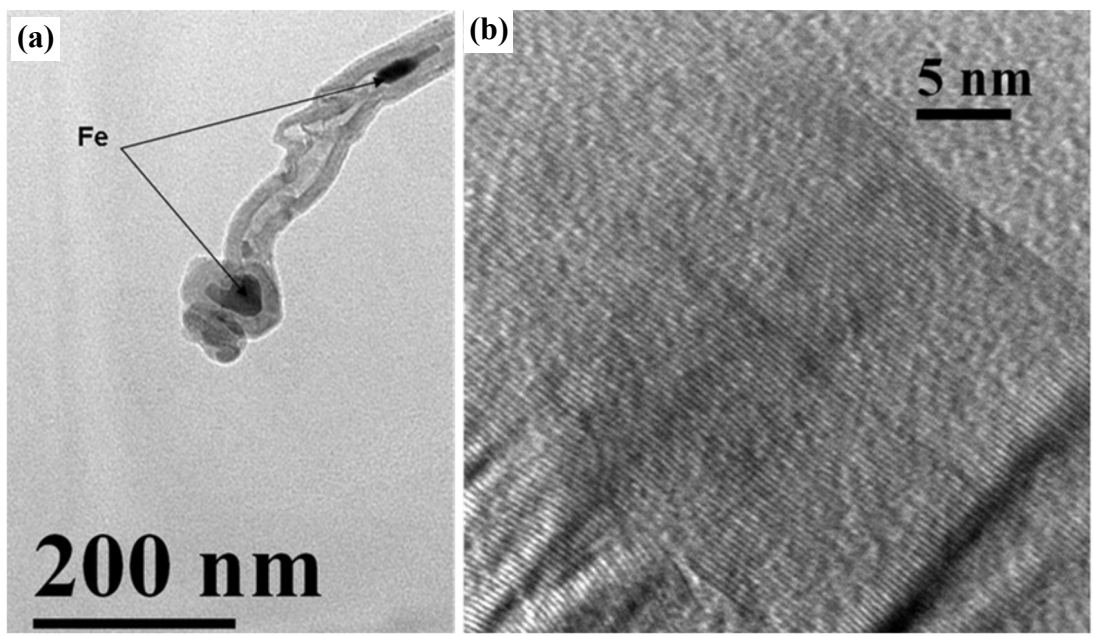

Figure 3.10: (a) High resolution transmission electron microscopy image of the Fe encapsulated carbon nanotube and (b) The lattice image of the nanotube grown after pre-heating of Fe at 850 °C

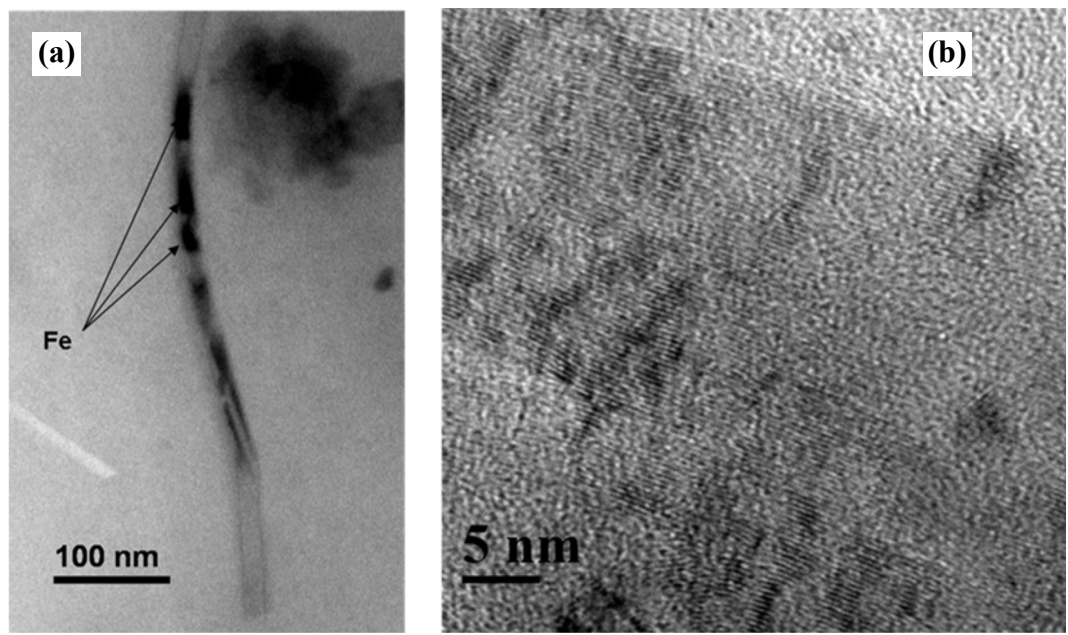

Figure 3.11: (a) High resolution transmission electron microscopy image of the Fe encapsulated carbon nanotube and (b) The lattice image of the nanotube grown after pre-heating of Fe at 900 °C

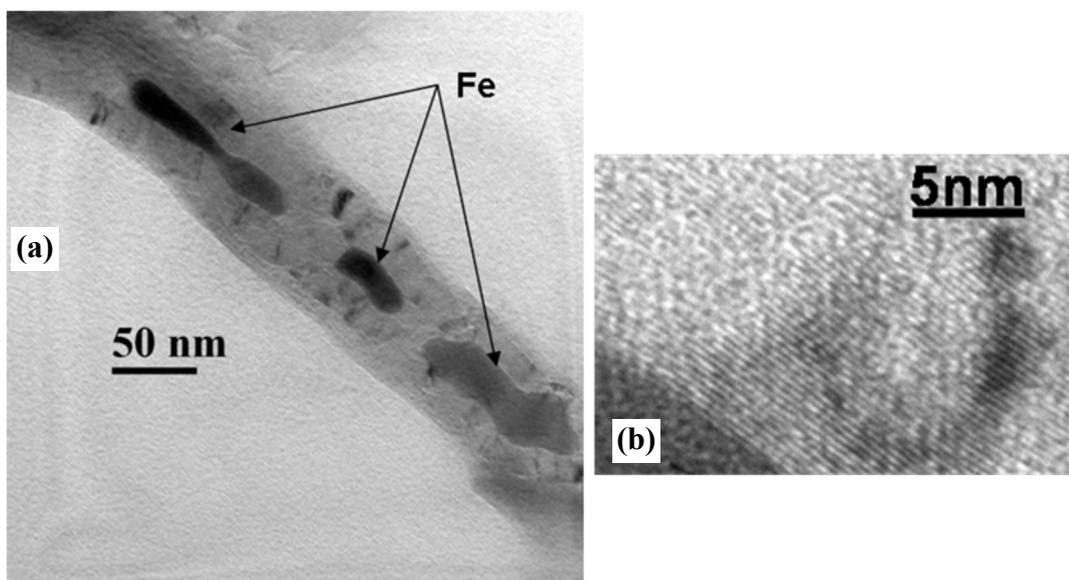

Figure 3.12: (a) High resolution transmission electron microscopy image of the Fe encapsulated carbon nanotube and (b) The lattice image of the nanotube grown after pre-heating of Fe at 1000 °C

nm size is encapsulated at the top of the a-CNT implying tip growth mechanism. Careful observations reveal that a few elongated particles are embedded in the core of the nanotubes, which establishes the hollow nature of the nanotubes deposited in this study.

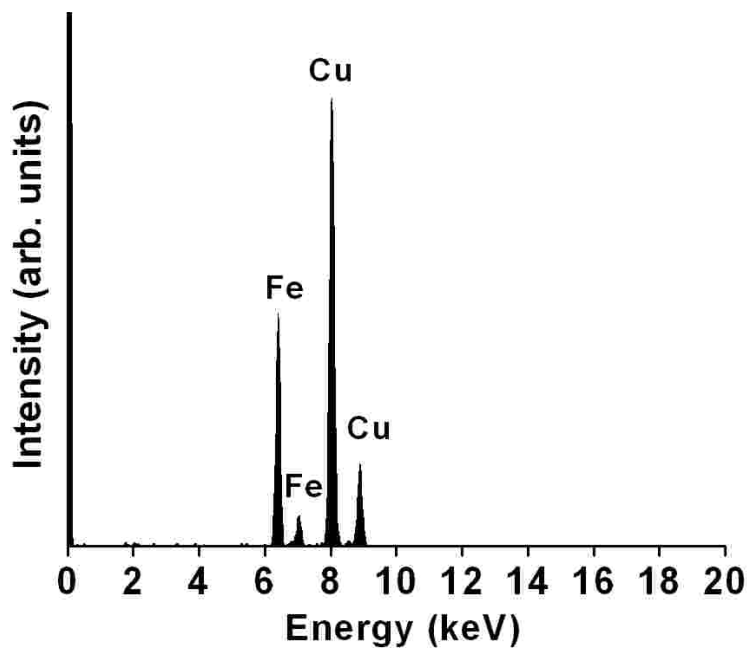

Figure 3.13: Energy dispersive X-ray spectrum obtained from a nanotube encapsulated metal nanoparticle

Chemical composition analysis (EDX) confirms that the elongated particles inside the tubes are Fe particles (Figure 3.13).

XRD measurements were performed to examine the structure of the CNTs and the resulting θ-2θ scan is shown in Figure 3.14 which confirms the presence of MWCNTs and Fe in all the samples. Intensities of all the CNT related peaks are nearly equal in case of a-CNT and b-CNT but significantly change in the c-CNT sample. The change in intensities of the XRD peaks can be explained from the SEM observations. All other conditions remaining the same the XRD peak intensity depends upon the number of CNTs present in the sample. As the number density of c-CNT is the lowest so the CNT peaks corresponding to c-CNT have lower intensity compared to other samples.

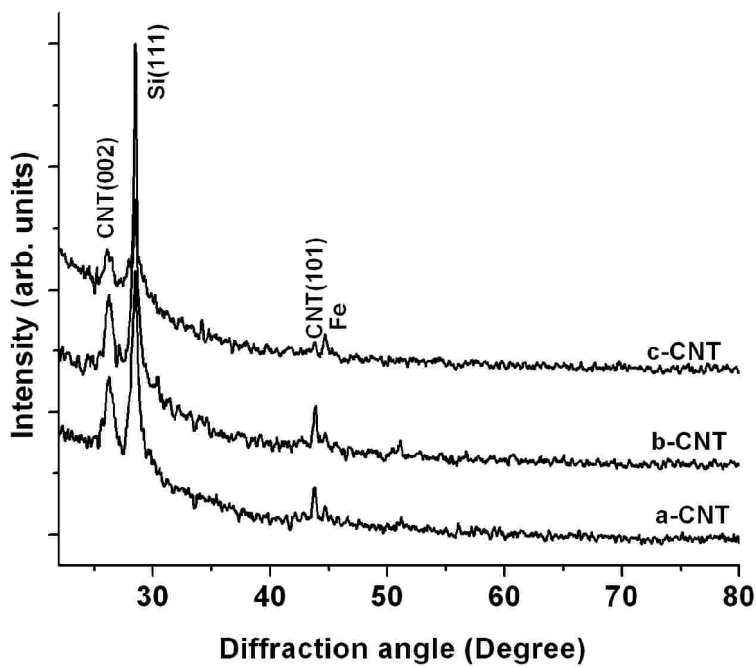

Figure 3.14: X-ray diffraction spectra of multi-walled carbon nanotubes grown on Si(111) substrate using Fe catalyst after pre-heating at different temperatures (a-CNT) 850 °C; (b-CNT) 900 °C; (c-CNT) 1000 °C

Figure 3.15 shows the room temperature Raman spectra of the MWCNT material taken with a laser excitation wavelength of 514.5 nm. The two main peaks are the D and G bands as before. The G band appears near 1575 cm$^{-1}$, which is very close to the value observed by Hiura et al. (1993), for CNTs and nanoparticles. The D band is located at around 1348 cm$^{-1}$. The position of the D band for MWCNT can be expressed as

$$\omega_D = 1285 + 26.5 E_{laser} \quad \quad \ldots(3.1)$$

where $\omega_D$ is the wave number and $E_{laser}$ is the laser energy in units of cm$^{-1}$ and eV (Wei et al., 2003). In the present case $E_{laser} = 2.41$ eV resulting in $\omega_D \approx 1349$ cm$^{-1}$, which is in agreement with the observed peak position. The intensity ratio of the G band with respect to the D band varies in the three samples. The values of ($I_D/I_G$) for the a-CNT, b-CNT and c-CNT are 0.23, 0.18 and 0.20 respectively. Although the shape and positions of peaks in the spectra are almost similar, the variation in the intensity ratio of the D to G peaks (R = $I_D/I_G$) are indicative of differences in the degree of graphitization. The decrease in the relative intensity of the disordered mode for b-CNT and c-CNT samples can be attributed to the decreased number of structural defects.

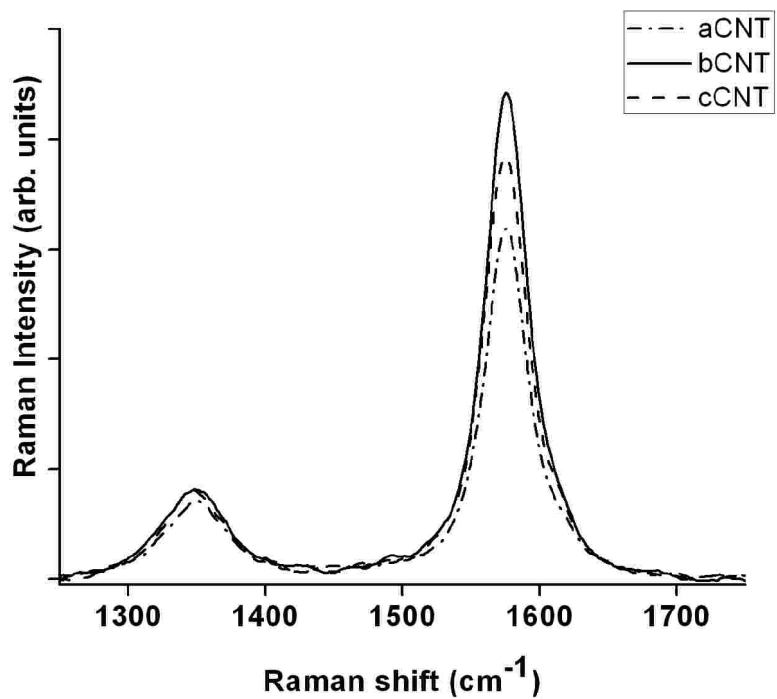

Figure 3.15: Raman spectra (514.5 nm excitation) of multi-walled carbon nanotubes grown after pre-heating at different temperatures (a-CNT) 850 °C; (b-CNT) 900 °C; (c-CNT) 1000 °C

**3.1.2.3 Summary**

The effect of pre-heating of catalysts on CNT growth has been studied. The study reveals that the pre-heating strongly affects not only degree of graphitization but also the number density of the partially Fe filled MWCNTs. The CNTs grown on Fe catalyst pre-heated at 900 °C reveal the best result in terms of number density as well as the degree of graphitization. Heat treatment at 1000 °C results in low number density and the CNTs grown after pre-heating at 850 °C exhibit the poorest degree of graphitization. CNT growth with Fe catalyst occurs primarily by the tip growth mechanism and HRTEM studies confirm that the grown materials are MWCNTs with partial catalyst filling.

## 3.2 Synthesis of carbon nanotubes using thermally evaporated elemental Ni metal catalyst

Here, we have discussed the growth morphology of the as-grown CNTs synthesized using APCVD over a pre-heated Ni catalyst on Si(111) substrate using propane as a source of carbon.

### 3.2.1 Experimental details

APCVD of CNTs was carried out at 850 °C with pre-heating at 900 °C, by catalytic decomposition of propane on Si(111) wafers with a pre-treated Ni overlayer in a hot-wall horizontal reactor. The resultant products were analyzed using AFM, XRD, FESEM, HRTEM, EDX and Raman spectroscopy.

### 3.2.2 Results and discussion

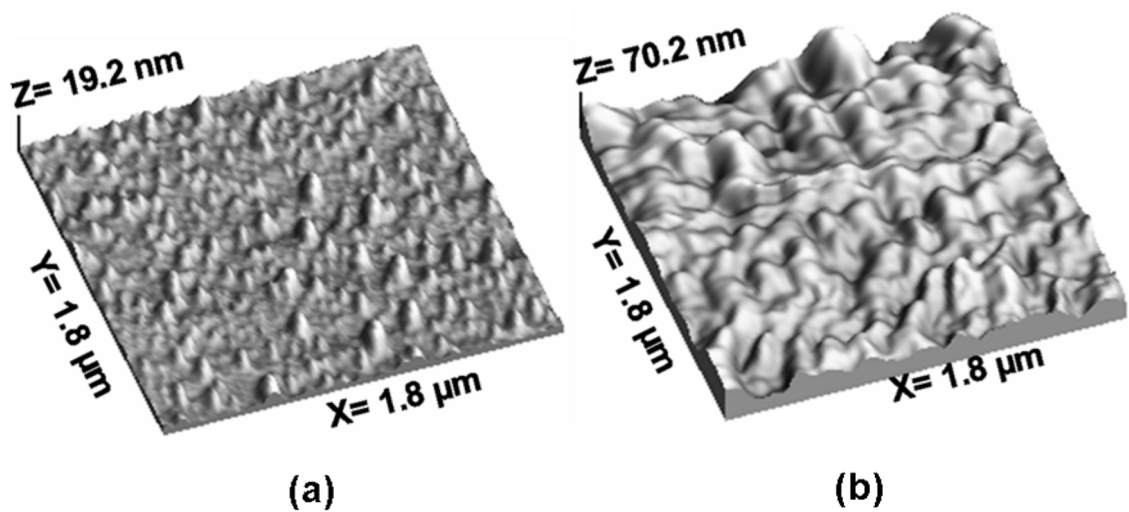

Figure 3.16: Atomic force microscopy images of the surface of the Ni catalyst on the Si substrates (a) The as-deposited catalyst surface; (b) Catalyst surface after the thermal treatment at 900 °C

Before the synthesis of CNTs, the samples were annealed in hydrogen at the growth temperature. Figure 3.16 show AFM images of Ni catalyst before and after the thermal annealing which reveal that heating above a certain temperature causes Ni clusters to coalesce and form macroscopic islands.

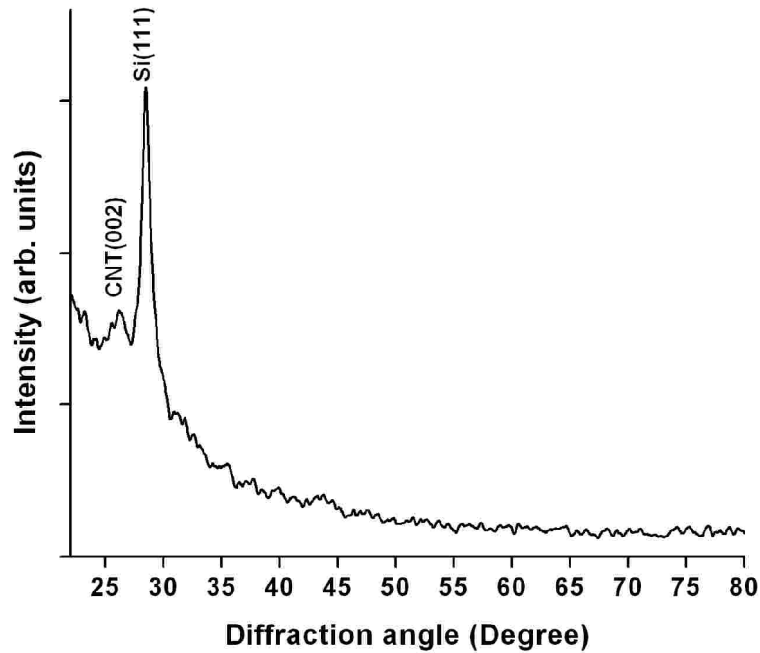

Figure 3.17: X-ray diffraction spectrum of multi-walled carbon nanotubes grown on a Si(111) substrate

The θ-2θ XRD scan is shown in Figure 3.17. This shows two peaks- one due to the presence of MWCNTs and the other, from the Si substrate.

Figure 3.18 is the FESEM micrograph showing the surface morphology of Ni-coated Si wafers after CNT growth. High aspect ratio CNTs are observed on the Ni film surfaces; there is no observable growth of these structures on bare Si substrates. The number density of the deposited MWCNTs was high with randomly oriented morphology.

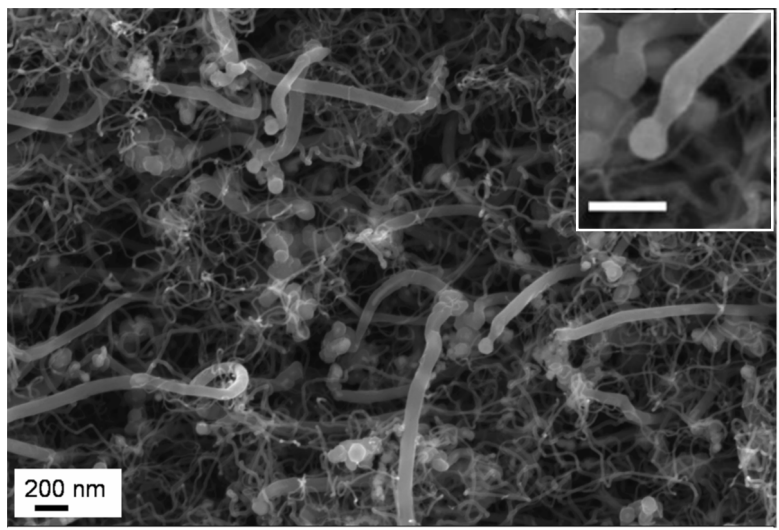

Figure 3.18: Field emission scanning electron micrograph of the as-grown carbon nanotubes; (inset) Small bright catalyst particles at the tip of nanotube and the scale bar length is 200 nm

It can be noted that in Figure 3.18 the number of small tubes is much higher than the big ones and the tube diameter is in the range of 15 to 100 nm. The formation mechanism of the MWCNTs on the Ni catalyst is a tip growth mode as evidenced by nanoscale catalyst particles encapsulated at the top end of the CNT (inset of Figure 3.18).

Figure 3.19 shows the HRTEM image of a MWCNT grown from the Ni catalyst. It can be clearly seen that the CNT is well graphitized with an inner diameter of about 14-17 nm and outer diameter 38-42 nm. The thickness of the tube wall lies in the range of 12-14 nm, which suggests that the tube wall is composed of approximately 30-40 graphitic layers. In the same figure, various defects and bamboo-like structures (diaphragms) inside the CNT can also be observed. The distance between two neighboring graphitic walls is about 0.34 nm (inset of Figure 3.19).

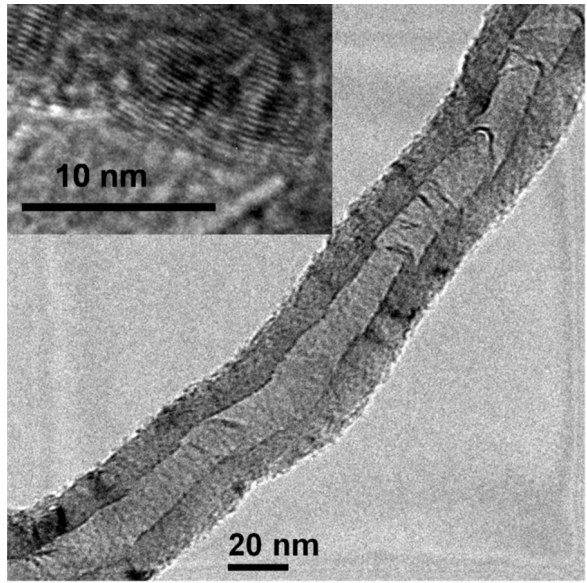

Figure 3.19: High resolution transmission electron micrograph of a multi-walled carbon nanotube grown using Ni; (inset) The fringe spacing between two carbon layers is about 0.34 nm

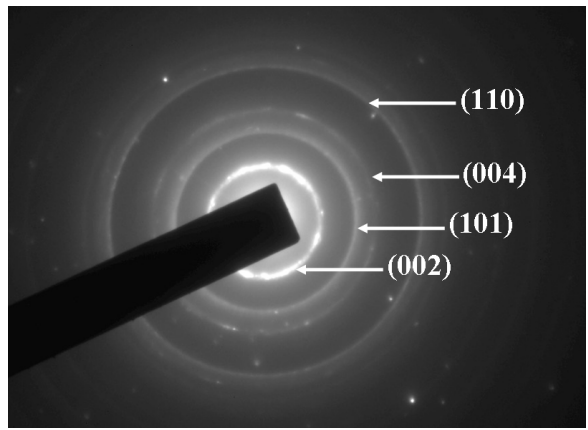

Figure 3.20: The selected area electron diffraction pattern from multi-walled carbon nanotube grown using Ni

Figure 3.20 shows the SAED pattern from the MWCNT. There are diffuse halos due to the amorphous carbon film on the copper grid and sharp rings due to the MWCNT. The diffraction rings, corresponding to layer spacings of 0.34 and 0.17 nm, are from the graphite (002) and (004) planes, respectively. The diffraction rings corresponding to layer spacings of 0.20 and 0.12 nm can also be seen in Figure 3.20. The former appears from the (101) and the latter from the (110) planes of graphite.

The room temperature Raman spectrum of the MWCNT material was obtained using a laser excitation wavelength of 488 nm. The spectrum is divided into two main zones: the low frequency region from 150-800 cm$^{-1}$ (Figure 3.21) and the high frequency zone from 1200–1750 cm$^{-1}$ (Figure 3.22). The vibrations of CNT originate from the curvature induced strain due to misalignment of the π-orbitals of adjacent coupled carbon atoms. These vibrations are reflected in the Raman peaks.

Evidence for the presence of radial breathing mode (RBM) vibrations in this sample is obtained from the low wave number range of the spectrum (Figure 3.21). The peaks located at 220 cm$^{-1}$ and 286 cm$^{-1}$ are due to the RBM of MWCNT (Jinno et al., 2004). This mode has A$_{1g}$ symmetry and all the carbon atoms move in phase in the radial direction creating breathing-like vibration of the entire tube (Zhao et al., 2002). The frequency of RBM is directly linked to the innermost tube diameter by the relation

$$\omega_{RBM} = 223.75/d \qquad \ldots(3.2)$$

where d is the innermost tube diameter and $\omega_{RBM}$ is the wave number in units of nm and cm$^{-1}$, respectively (Jinno et al., 2004). The peaks at 220 cm$^{-1}$ and 286 cm$^{-1}$ originate due to the RBM of MWCNT bundles with inner-core diameters of 1 nm and 0.78 nm respectively.

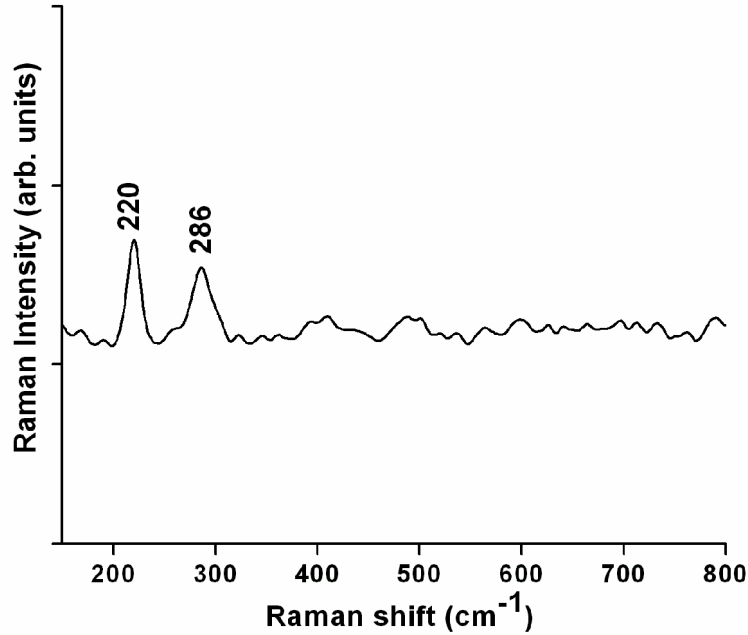

Figure 3.21: Low frequency Raman spectrum (488 nm excitation) of a multi-walled carbon nanotube film grown on Si showing the presence of radial breathing modes

The two main peaks observed in the high frequency zone (Figure 3.22) are the D and G band. The G band appears at 1567 cm$^{-1}$ (without any splitting) and the D band, at 1350 cm$^{-1}$. The position of the D band for MWCNT can be expressed by equation (3.1). In the present case $E_{laser}$ = 2.54eV resulting in $\omega_D \approx$ 1352 cm$^{-1}$, which is in agreement with the observed peak position. The observed full width at half maximum (FWHM) is 43 cm$^{-1}$. In contrast, Dillon et al. (2004) have reported that the FWHM of the D band of amorphous carbon and nano-crystalline graphite at 488 nm excitation as ~ 57 and 86 cm$^{-1}$. On the basis of these two characteristics (frequency and linewidth), it may be concluded that the presence of the D band in the sample is due to the MWCNT. The intensity ratio derived from Figure 3.22 is R = $I_D/I_G$ = 0.23, indicating that the grown MWCNT is highly crystalline in nature.

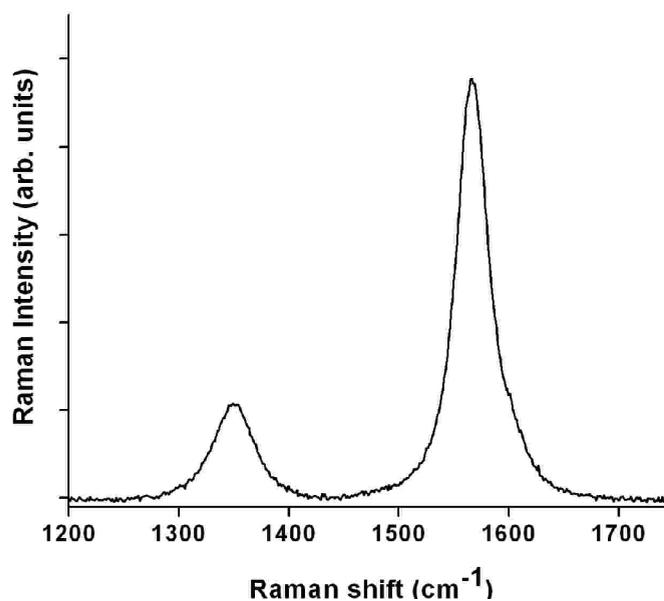

Figure 3.22: High frequency Raman spectrum (488 nm excitation) of a multi-walled carbon nanotube film grown by atmospheric pressure chemical vapor deposition on Si using elemental Ni catalyst

In the related literature (Arcos et al., 2004; Kwok and Chiu, 2005), there are several instances where the RBM has been found while there is no splitting of the G band (a split G band coupled with the presence of RBMs is a major indication for the presence of SWCNT) or any TEM images of SWCNTs. Those authors concluded that their CNT samples may be composed of a very small number of SWCNTs. However, the current study indicates that in those cases there might be no SWCNTs actually but some MWCNTs with very narrow diameter were present, which were responsible for the RBMs.

An understanding of the formation of bamboo-like nanotubes could be obtained by considering the published literature. The in-situ observation of the growth of bamboo-like nanotubes has been reported recently by Lin et al. (2007). Following this paper, the model can be described as schematically shown in Figure 3.23. After the deposition of Ni layer, the catalyst layer becomes fragmented into nanoparticles. The decomposition of propane on the surface of the Ni nanoparticles results in the formation of carbon, and the growth of CNTs occurs via diffusion of carbon through the Ni particle. The dissolved carbon diffuses towards the bottom of the Ni particle and the carbon segregates as graphite at the bottom and the sides of the Ni, thus forming a hemispherical cap within the nanotube and sealing the existing tube internally. This results in the formation of the knot in the CNT, which fully encapsulates the lower part of the catalyst particle. The graphitic layers formed at the bottom surface of the catalyst particle leads to elevation of the Ni particle to the tip. The motive force of pushing out the Ni particle may be due to the stress accumulated in the graphitic sheath due to the segregation of carbon from the inside of the sheath. If the carbon species are supplied in a steady-state manner, the growth process will repeat and a complete CNT with multiple knots along its length will appear (Figure 3.23).

The reason for the different growth morphologies in case of Fe and Ni is not presently clear. However a recent study of the diffusion-controlled growth of MWCNTs revealed that the catalyst particles are in liquid state during tubular growth and in solid state in case of bamboo-like growth (Bartsch et al., 2005). For the capillary action, the presence of liquid-like behavior of the catalyst material at the growth temperature is essential and therefore, the probability of catalyst incorporation within the bamboo-like CNTs is less than in the tubular CNTs.

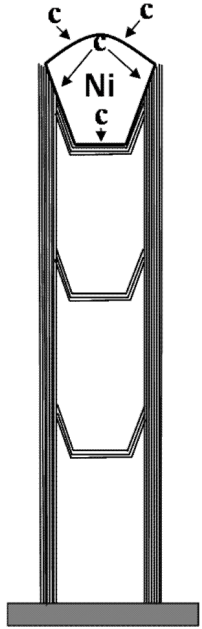

Figure 3.23: The proposed growth model for the formation of bamboo-like multi-walled carbon nanotube

### 3.2.3 Summary

MWCNTs have been grown at 850 °C using CVD by catalytic decomposition of propane on Si with a pre-treated Ni overlayer. At the growth temperature of 850 °C, the catalyst particles coalesce on the substrate as confirmed by AFM studies. The XRD results suggest that MWCNTs with good graphitization have been grown. Raman analysis shows that radial breathing modes are present in the spectrum, generated from small diameter MWCNTs. The FESEM images lead to the conclusion that, in this experiment, MWCNTs were grown over the Ni catalyst primarily by the tip growth mechanism. Lastly, HRTEM studies prove that the grown materials are CNTs indeed with a layer spacing of 0.34 nm.

# Chapter 4

# Growth and characterization of carbon nanotubes by chemical vapor deposition using a metal complex as catalyst

The CVD synthesis of CNTs on plain substrates generally requires metal deposition over the substrate, which is time consuming and the deposited area is also finite. These limitations not only increase the cost but also introduce a constriction in the mass production process of CNT by CVD. This can be easily overcome by employing the spin coating of the catalyst material on the desired substrate. The growth of CNT using spin coated Fe and Ni catalysts is discussed in this chapter.

## 4.1 Synthesis of carbon nanotubes using spin coated Fe-Mod-PR

### 4.1.1 Effect of growth temperature on the carbon nanotubes grown using Fe-Mod-PR

A method for the large scale synthesis of CNTs employing APCVD of propane on Si using a simple mixture of positive photoresist (PR) and a metalorganic molecular precursor, $Fe(acac)_3$ is discussed here. The mixture i.e. photoresist modified with $Fe(acac)_3$ was called Fe-Mod-PR. The effect of growth temperature, in the range of 550 to 950 ºC, on the morphology of as-grown CNTs was also investigated.

#### 4.1.1.1 Experimental details

Synthesis of CNTs was carried out at 550, 650, 850 and 950 ºC by catalytic decomposition of propane on Si(111) substrate with Fe-Mod-PR of 0.2M $Fe(acac)_3$ concentrations in a hot-wall horizontal APCVD reactor. The as-grown materials were analyzed by X-ray diffraction (XRD), field emission scanning electron microscopy (FESEM), high resolution transmission electron microscopy (HRTEM), energy dispersive X-ray (EDX) and Raman spectroscopy.

**4.1.1.2 Results and discussion**

XRD measurement was performed to investigate the structure of the precursor (Fe(acac)$_3$) prior to the mixing with photoresist and the resulting θ-2θ scan is shown in Figure 4.1a. All the diffraction peaks can be assigned to the orthorhombic structure of Fe(acac)$_3$ in agreement with the JCPDS card no. 30-1763. FESEM image (Figure 4.1b) shows that Fe(acac)$_3$ is comprised of thin platelets with a wide range of crystallite sizes.

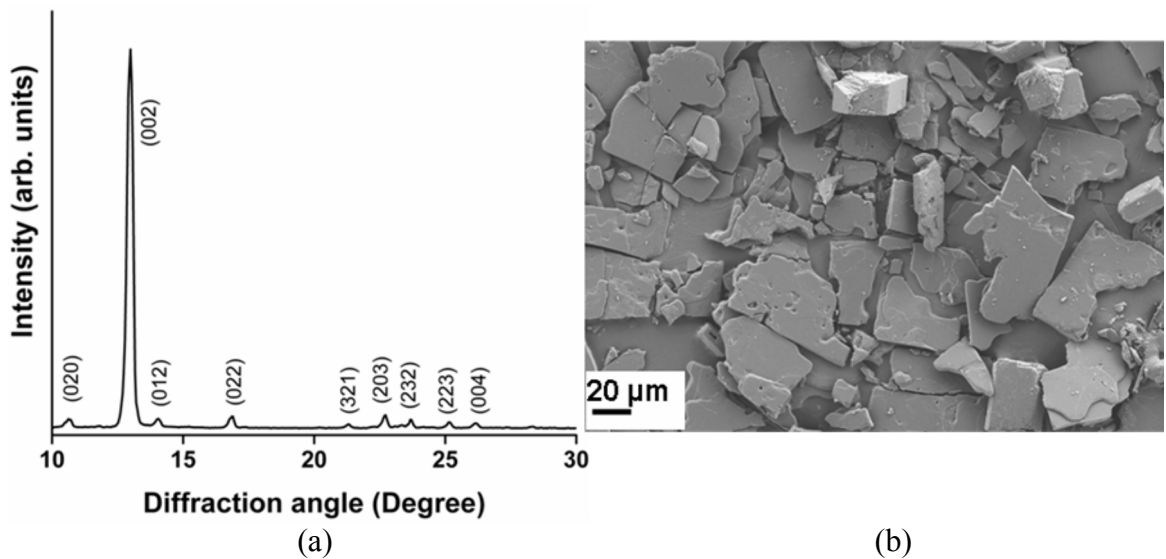

(a) (b)

Figure 4.1: (a) X-ray diffraction spectrum and (b) Field emission scanning electron microscopy image of Fe(acac)$_3$

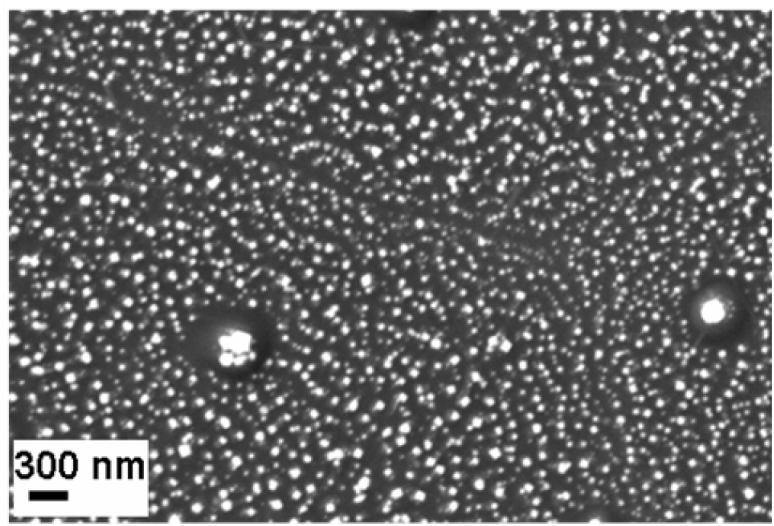

Figure 4.2: Field emission scanning electron micrograph of the catalytic nanoparticles after annealing Fe-Mod-PR at 850 ºC

Figure 4.2 shows the SEM image of the Fe catalytic nanoparticles formed on the Si substrate after annealing Fe-Mod-PR at 850 °C. It can be noted that after high temperature annealing, particles of nanometer order have been formed while the Fe(acac)$_3$ particles themselves were several microns in size.

After the growth of CNTs using Fe-Mod-PR, the analysis of the morphology and number density of the as-grown CNTs was performed using FESEM. Figure 4.3(a-d) are the FESEM images of nanotubes synthesized at 550, 650, 850 and 950 ºC on the Si(111) substrates. The micrographs reveal that the synthesized CNTs are randomly oriented and curved with a high aspect ratio. The observation of Fe nanoparticles at the tip of the CNTs supports the tip growth mechanism. SEM images suggest that the diameter of CNTs increases with the increase of growth temperature whereas the number density of the CNTs decreases. The surface diffusion of Fe nanoparticles over Si substrate is responsible for the change of CNT morphology which we have already discussed in section 3.1.1.2. EDX of the CNTs (Figure 4.3e) shows that the material contains only carbon and Fe with a Si peak (due to the substrate). It should be noted that even at 550 °C large numbers of CNTs have grown albeit with shorter length compared to the CNTs grown at higher temperatures. The increase in CNT length with temperature is due to the enhancement of carbon solubility and diffusivity (Cola et al., 2008).

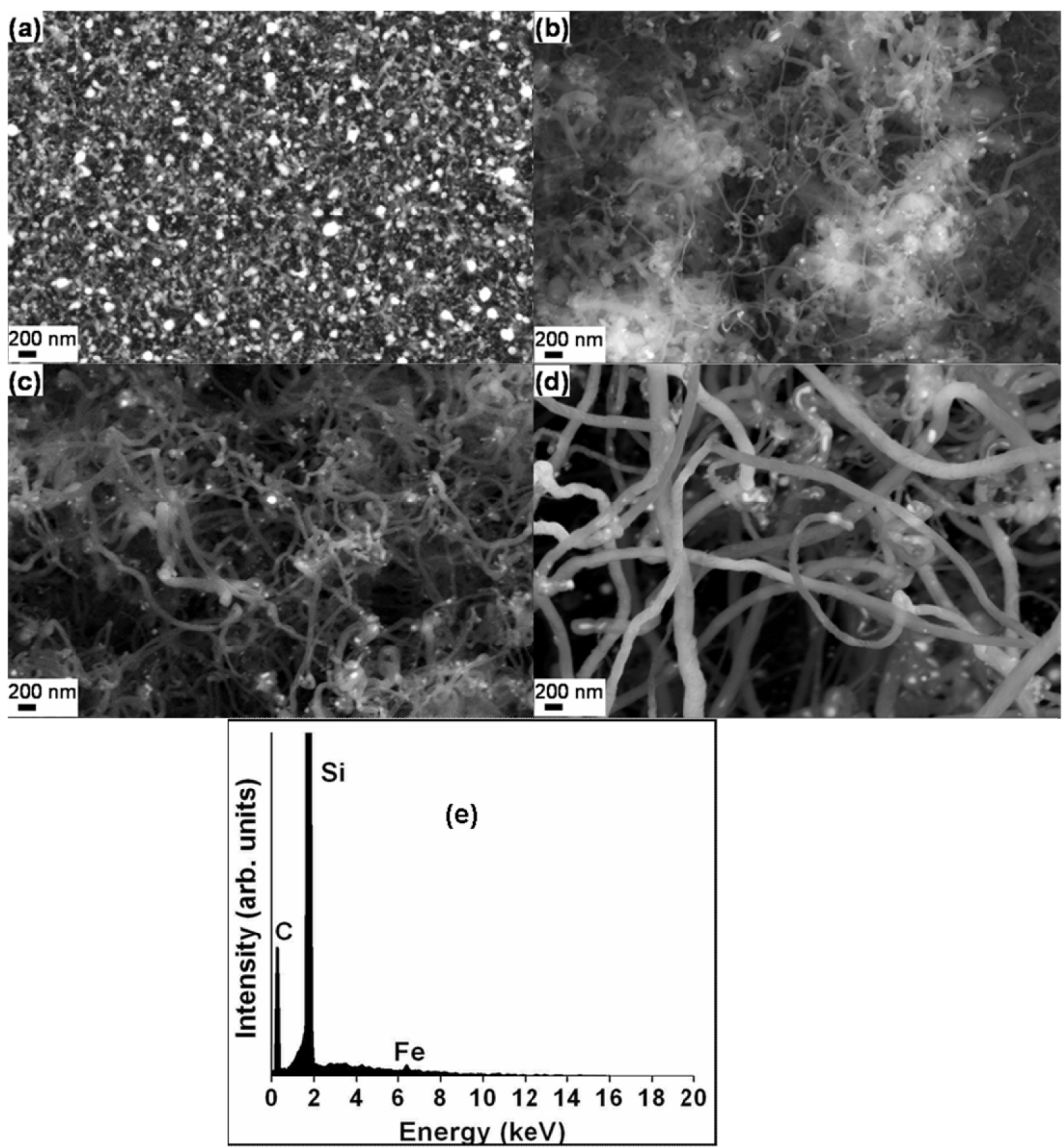

Figure 4.3: Field emission scanning electron microscopy images of carbon nanotubes grown using Fe-Mod-PR at different temperatures (a) 550 °C; (b) 650 °C; (c) 850 °C and (d) 950 °C; (e) Energy dispersive X-ray spectrum obtained from the sample grown at 850 °C

XRD measurement was performed using CuK$_\alpha$ radiation ($\lambda$=1.54059 Å) on the CNT samples grown at 850 °C and the resulting $\theta$-2$\theta$ scan is shown in Figure 4.4. The characteristic graphitic peak at 26.2° confirms the presence of MWCNTs in the sample. The peak near 44.7° is assigned to the Fe catalyst (JCPDS 06-0696). The peak near 28.5° is attributed to the (111) plane of the Si substrate (JCPDS 27-1402).

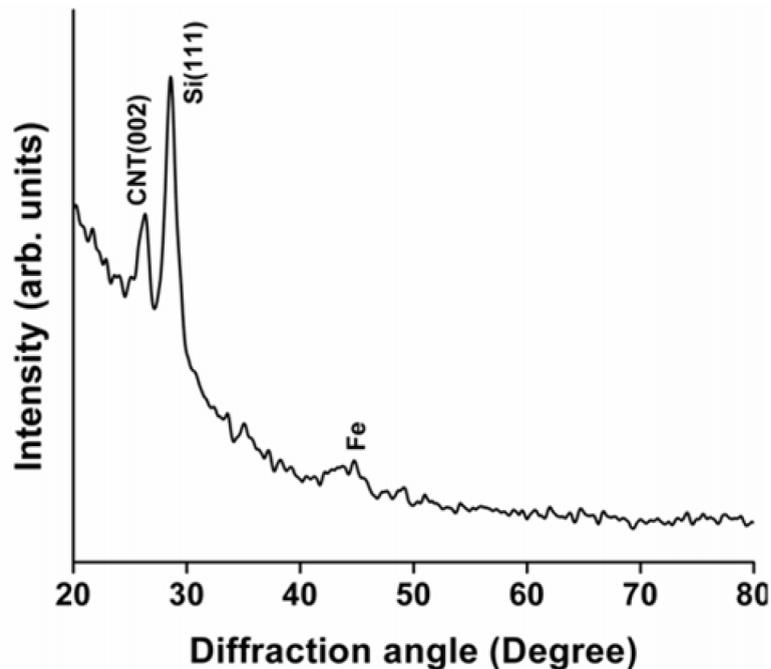

Figure 4.4: The X-ray diffraction spectrum of the carbon nanotubes prepared from Fe-Mod-PR at 850 ºC

The growth morphology and internal structure of the CNTs were further investigated as a function of temperature and at high resolution using TEM. Figure 4.5(a-d and the respective insets) are the HRTEM images of CNTs synthesized at 550, 650, 850 and

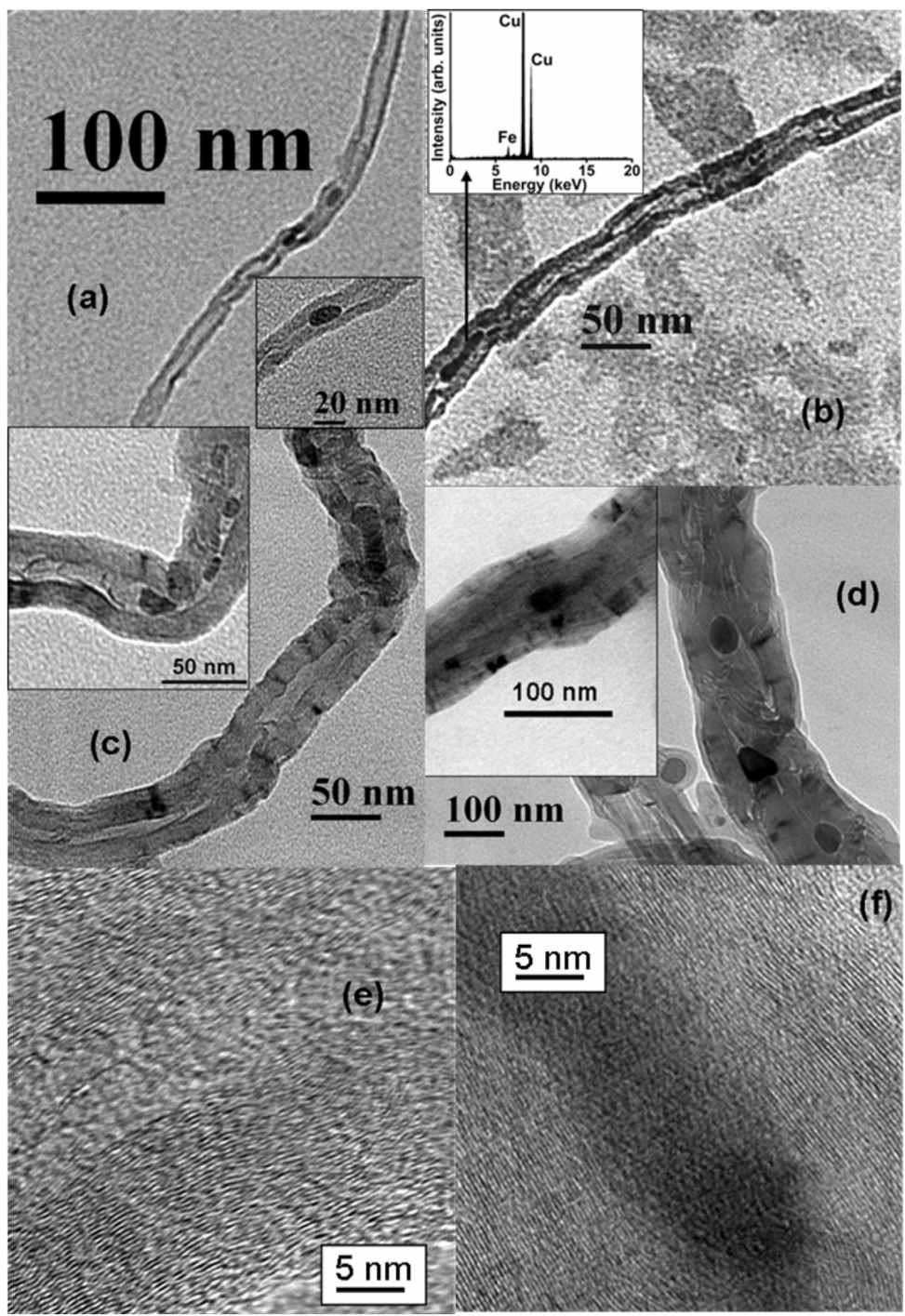

Figure 4.5: High resolution transmission electron microscopy images of carbon nanotubes grown at different temperatures using Fe-Mod-PR (4.5a and inset) 550 °C; (4.5b) 650 °C; (4.5c and inset) 850 °C; (4.5d and inset) 950 °C; (e) Lattice image from a carbon nanotube grown at 650 °C; (f) Lattice image from a nanotube grown at 850 °C; (inset of Figure 4.5b) The energy dispersive X-ray spectrum of a nanotube encapsulated metal nanoparticle

950 ºC, respectively, and illustrate that the nanotubes are always multi-walled irrespective of growth temperature. The TEM micrographs reveal that the diameter of the CNTs progressively increases with the increase of growth temperature. Further observations show that the CNTs are partially filled with metal nanoparticles and no nanoparticles are situated on the outer surface of the nanotubes. It indicates that the present growth method is a simple and effective one to introduce catalyst metals inside the CNT. Chemical composition analysis of an elongated particle (inset of Figure 4.5b) confirms that the particle is of Fe. The Cu signals come from the copper grid itself.

The HRTEM image of a CNT grown at 650 ºC (Figure 4.5e) shows the wavy structure of graphitic sheets due to the defects present in the graphite sheet. In contrast, the CNT grown at 850 ºC (Figure 4.5f) has fairly straight lattice fringes. HRTEM analyses suggest that the crystallinity of CNTs improves with the increase in growth temperature which is similar to the previous observation, discussed in section 3.1.1.2 for evaporated Fe metal catalyst.

Figure 4.6 shows the room temperature Raman spectra of the MWCNT material taken with a laser excitation wavelength of 514.5 nm. The two main peaks of the spectra are the D and G bands. The values of ($R=I_D/I_G$) for the CNT synthesized at 550, 650, 850, and 950 ºC are 1.23, 0.64, 0.35 and 0.24, respectively. The result indicates that high growth temperature is effective in reducing the number and severity of graphitic defects in the sample. This can be explained as follows: with the increase of the temperature, the carbon atoms achieve sufficient energy to rearrange themselves in a more ordered way, thereby eliminating the defects.

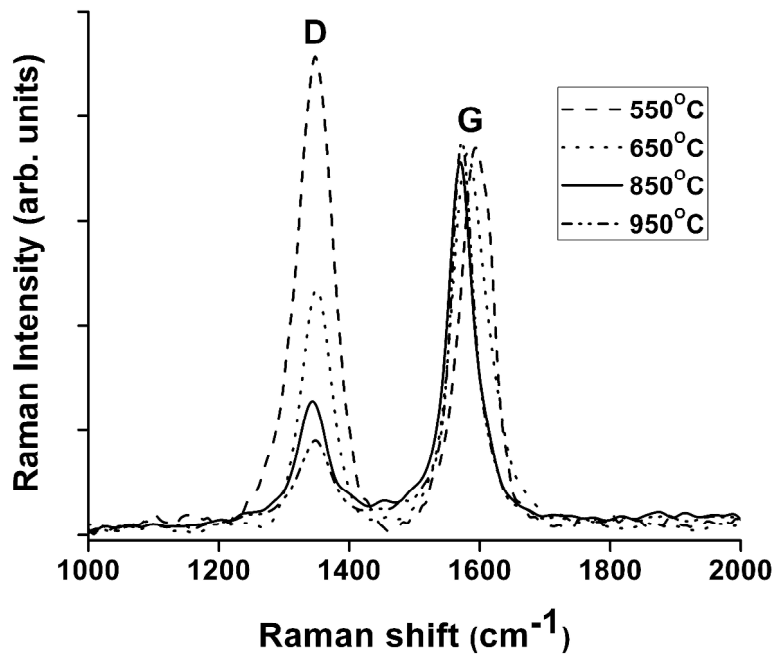

Figure 4.6 Raman spectra (514.5 nm excitation) of multi-walled carbon nanotubes grown at different temperatures using Fe-Mod-PR

**4.1.1.3 Summary**

A novel and scalable method has been demonstrated for the production of partially Fe filled CNTs using a mixture of Fe(acac)$_3$ with conventional photoresist followed by CVD growth. We observed that the average diameter and degree of graphitization of CNTs increases with growth temperature from 550 °C to 950 °C whereas the number density decreases.

## 4.1.2 Lithographically defined site selective growth of carbon nanotubes using Fe-Mod-PR

Here we report the growth of MWCNTs employing Fe-Mod-PR at selective regions on the Si substrate using a simple photolithographic route by CVD.

### 4.1.2.1 Experimental details

Synthesis of CNTs was carried out by catalytic decomposition of propane at 850 °C on Si(111) substrates with a pre-heating of Fe-Mod-PR (of 0.2M Fe(acac)$_3$ concentration) at 900 °C in hydrogen atmosphere in a hot-wall horizontal APCVD reactor. The as-grown materials were analyzed using XRD, SEM, HRTEM, EDX and Raman spectroscopy.

### 4.1.2.2 Results and discussion

We have used Fe(acac)$_3$ as a precursor and the structural characterization of this precursor has already been discussed in 4.1.1.2. Figure 4.7 shows the SEM micrograph of the catalyst surface form after annealing the Fe-Mod-PR film at 900 °C. The image reveals the non uniform distribution of catalyst nanoparticles over the Si substrate. For precise investigation of the effect of catalyst particle size on the CNT growth, the size distribution of the catalytic nanoparticles obtained from the measurements on plan view SEM image has been analyzed. The analysis (inset of Figure 4.7) shows that the catalyst nanoparticles range in size from 20 to 140 nm with a major proportion of particles in the interval 60 to 80 nm. The particles display a Gaussian profile with a mean of 70 nm.

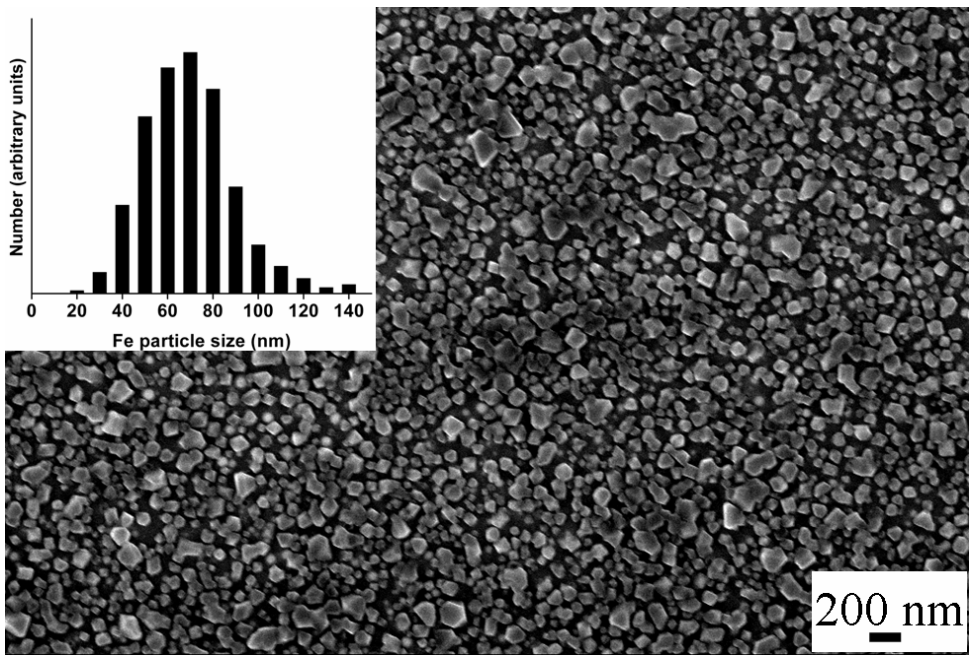

Figure 4.7: Scanning electron microscopy image of the catalytic nanoparticles prepared from Fe-Mod-PR film over the Si(111) substrate after annealing at 900 °C; (inset) The distribution histogram of the catalyst nanoparticles

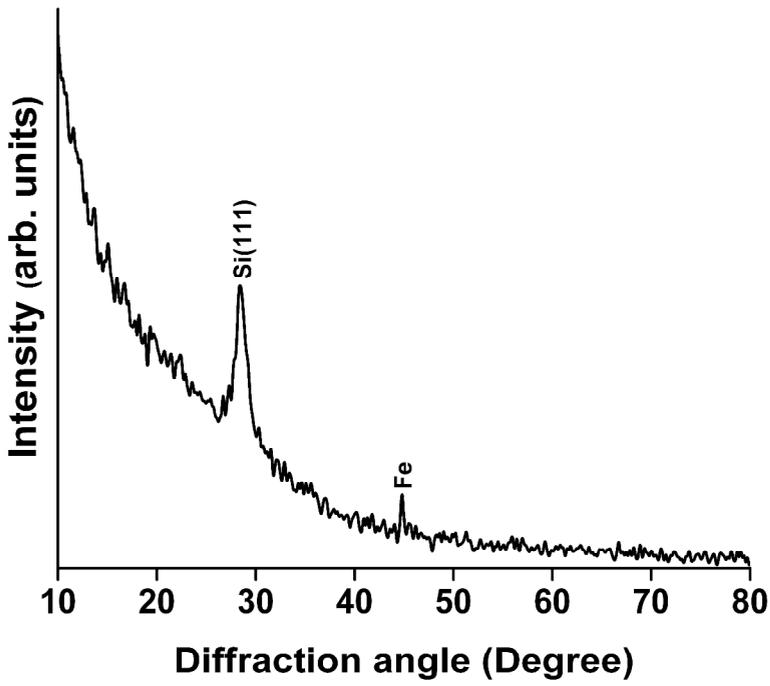

Figure 4.8: The X-ray diffraction spectrum of the Fe-Mod-PR thin film over the Si(111) substrate after annealing at 900 ºC

XRD measurement of the Fe-Mod-PR film annealed at 900 °C (Figure 4.8) shows only two peaks, the peak near 28.5º, which is attributed to the (111) plane of the Si substrate, and the peak near 44.7º, which is from the Fe catalyst.

The chemical composition analysis by EDX of the Fe-Mod-PR film annealed at 900 ºC in argon atmosphere (Figure 4.9) shows that the material mainly contains Fe, with a Si peak (due to substrate) and the presence of small amount of oxygen. As no oxide peak was detected in the XRD analysis of the same sample (Figure 4.8), the presence of oxygen may be due to absorbed oxygen from the surroundings. Absence of carbon peak in the XRD (Figure 4.8) and EDX analysis (Figure 4.9) of the Fe-Mod-PR film annealed at 900 ºC also confirm the complete removal of the organics from the Fe-Mod-PR pattern.

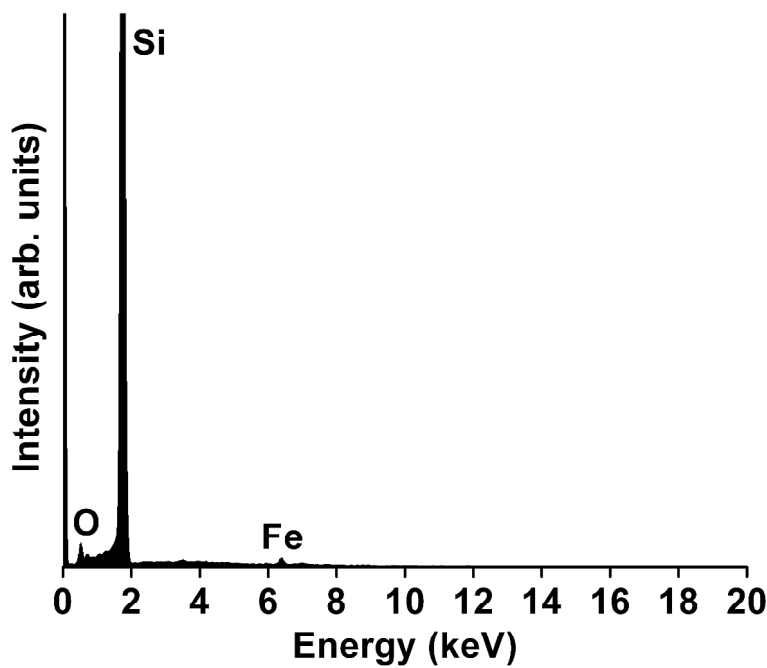

Figure 4.9: The energy dispersive X-ray spectrum of the Fe-Mod-PR thin film over the Si(111) substrate after annealing in argon at 900 ºC

The XRD pattern of the CNTs (Figure 4.10) shows peaks at 26.3°, 43.6°, 44.7° and 28.5°. The peaks at 26.3° and 43.6° are due to the (002) and (101) planes of CNT. The peak at 44.7° is assigned to Fe and 28.5° peak originates from the (111) plane of the Si substrate.

Figure 4.11 shows the SEM image of the CNTs synthesized on the Si(111) substrate using Fe-Mod-PR. The synthesized nanotubes had a randomly oriented web-like morphology with high number density and the measured yield was ~ 0.4 mg/(cm$^2$·h). It may be noted that in Figure 4.11, the tube diameters are in the range of 30 to 130 nm and their distribution profile (inset of Figure 4.11) is similar to the size distribution profile of the catalyst nanoparticles (inset of Figure 4.7). Thus, the size of the catalytic particles largely determines the diameters of the CNTs.

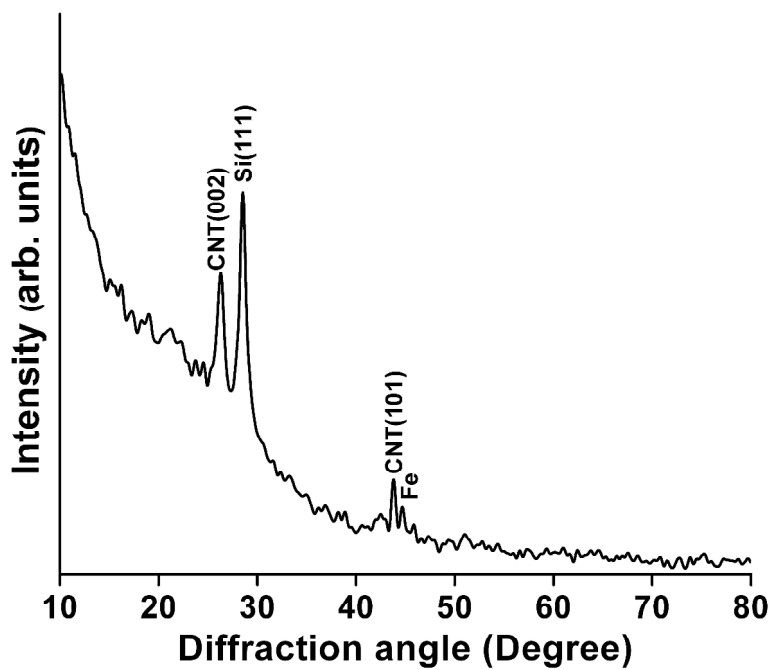

Figure 4.10: The X-ray diffraction spectrum of the multi-walled carbon nanotubes grown at 850 °C on a Si(111) substrate using Fe-Mod-PR

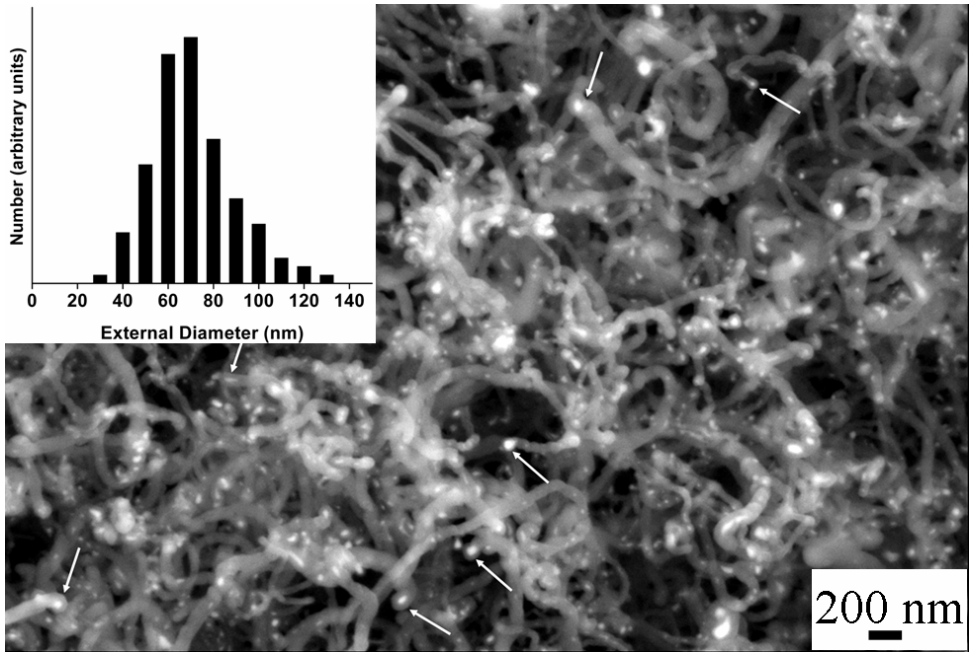

Figure 4.11: Scanning electron micrograph of the nanotubes synthesized using Fe-Mod-PR; (inset) The distribution histogram of the external nanotube diameter deposited by the chemical vapor deposition method

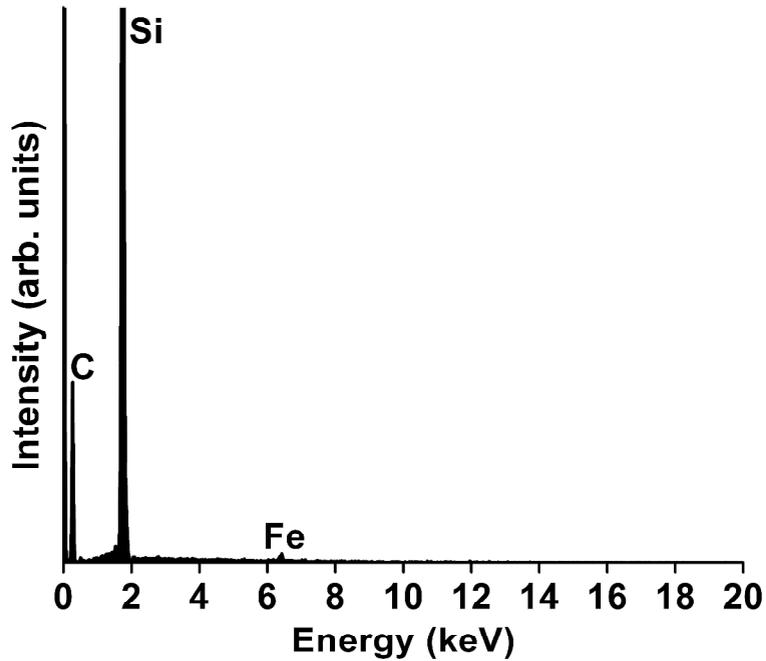

Figure 4.12: The energy dispersive X-ray spectrum of the multi-walled carbon nanotubes synthesized using Fe-Mod-PR over the Si(111) substrate

This is in agreement with earlier reports regarding CNT synthesis using CVD (Rümmeli et al., 2007). Catalyst particles were detected at the tip of the tubes (indicated in Figure 4.11 with white arrows) suggesting the tip growth mechanism. EDX of the CNTs (Figure 4.12) shows that the material contains only carbon and Fe with a Si peak (due to substrate) in agreement with XRD results (Figure 4.10).

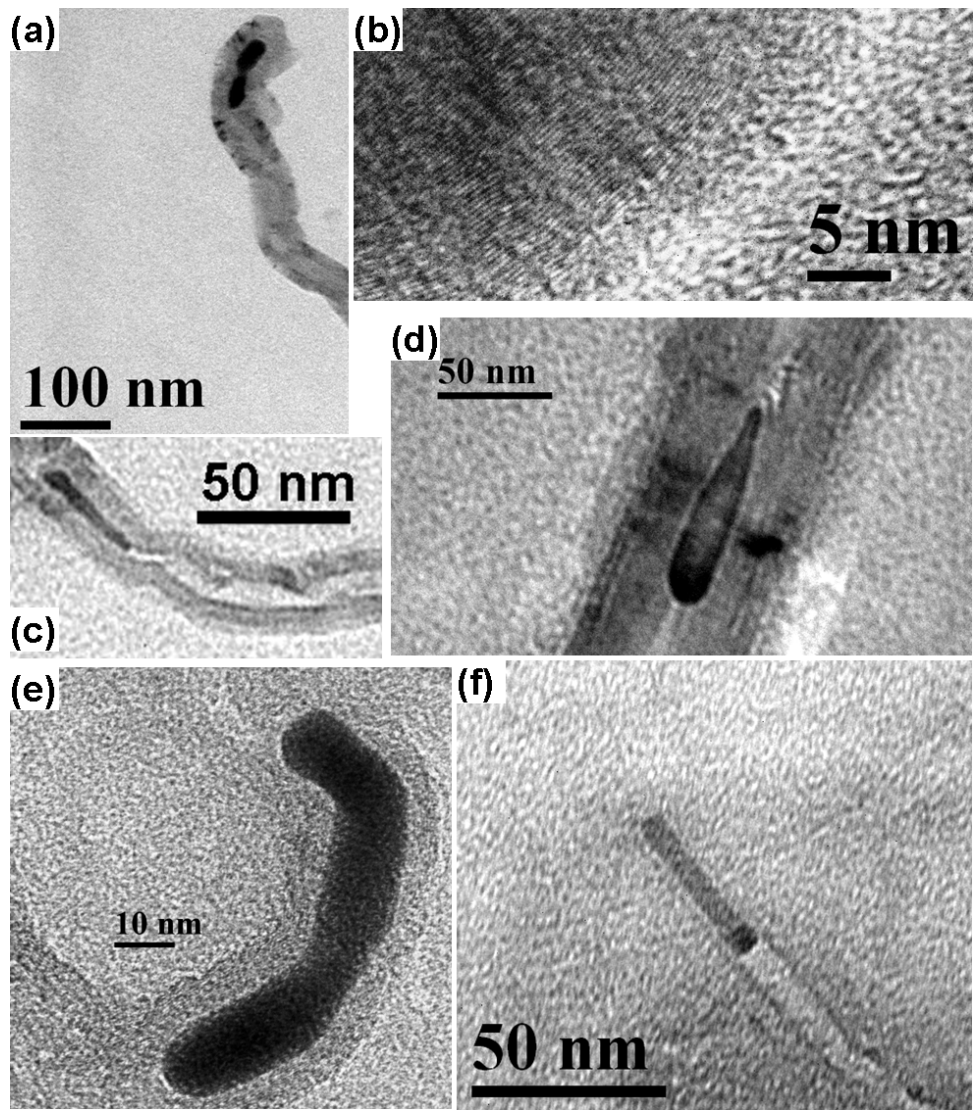

Figure 4.13: High resolution transmission electron microscopy images of the Fe encapsulated nanotubes grown using Fe-Mod-PR (a) A catalytic nanoparticle is encapsulated at the top of a nanotube implying tip growth mechanism; (b) High resolution image of interlayer spacing of graphitic carbon in nanotube; (c-f) Fe nanoparticles encapsulated in carbon nanotubes

Figure 4.13 show the HRTEM images of MWCNTs grown using Fe-Mod-PR. A catalytic nanoparticle of ~ 18 nm size is encapsulated at the tip of a CNT (Figure 4.13a) implying tip growth mechanism. It can be clearly seen that the CNT is well graphitized and has clear well-ordered lattice fringes of graphitic sheets (Figure 4.13b). Figure 4.13(c-f) demonstrate the presence of high aspect ratio nanoparticles encapsulated inside the CNTs deposited using Fe-Mod-PR. Chemical composition analysis of the CNT encapsulated nanoparticles (Figure 4.14) confirms that the elongated particles are of Fe.

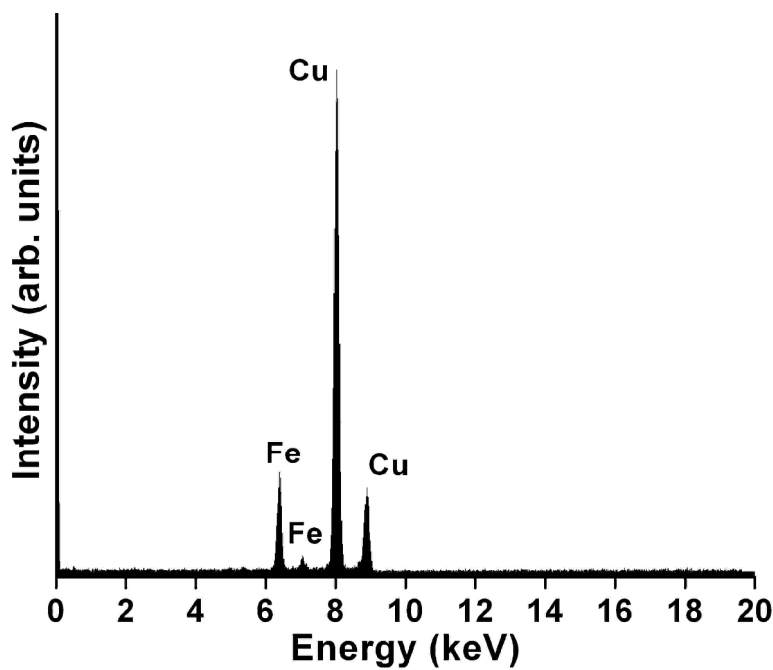

Figure 4.14: The energy dispersive X-ray spectrum of a nanoparticle encapsulated by a carbon nanotube grown at 850 °C

Figure 4.15 shows the room temperature Raman spectrum of the MWCNT material at a laser excitation wavelength of 514.5 nm. The two main peaks are the D and G bands. The G band appears near 1576 cm$^{-1}$ and the D band is around 1349 cm$^{-1}$. The intensity ratio derived from Figure 4.15 is $R = I_D/I_G = 0.28$, indicating that the grown CNTs are highly crystalline in nature.

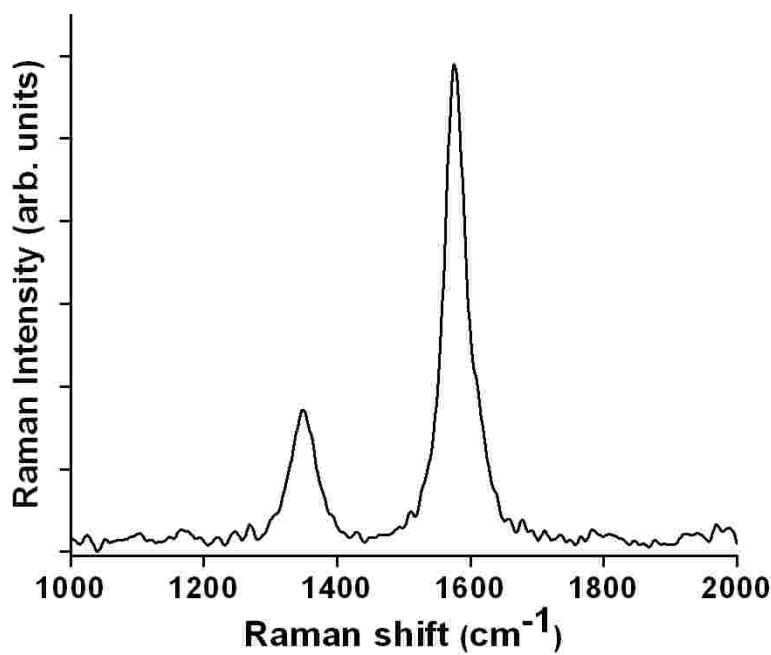

Figure 4.15: Raman spectrum (514.5 nm excitation) of multi-walled carbon nanotubes grown by chemical vapor deposition on Si using Fe-Mod-PR

Figure 4.16a is the SEM image of the Fe-containing patterns produced by the general photolithographic process using Fe-Mod-PR. Figure 4.16b is the SEM image of the CNTs grown from lithographically defined catalyst patterns. Figure 4.17 shows a photograph of the logo of Indian Institute of Technology, Kharagpur over which CNTs have been grown selectively. Thus, it has been demonstrated that catalyst locations can be defined in one simple lithographic step with the pattern size of microns to several centimeters.

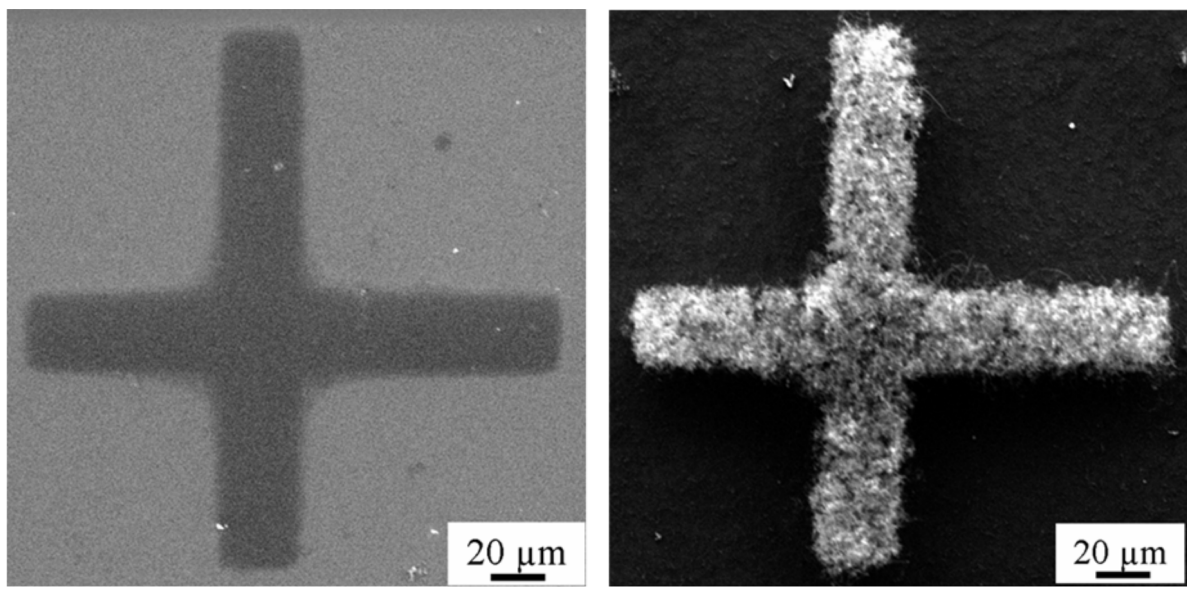

(a)                  (b)

Figure 4.16: Site selective growth of partially Fe filled multi-walled carbon nanotubes by direct photolithographic route using Fe-Mod-PR (a) Scanning electron microscopy image of Fe-Mod-PR pattern; (b) Scanning electron microscopy image after carbon nanotube growth

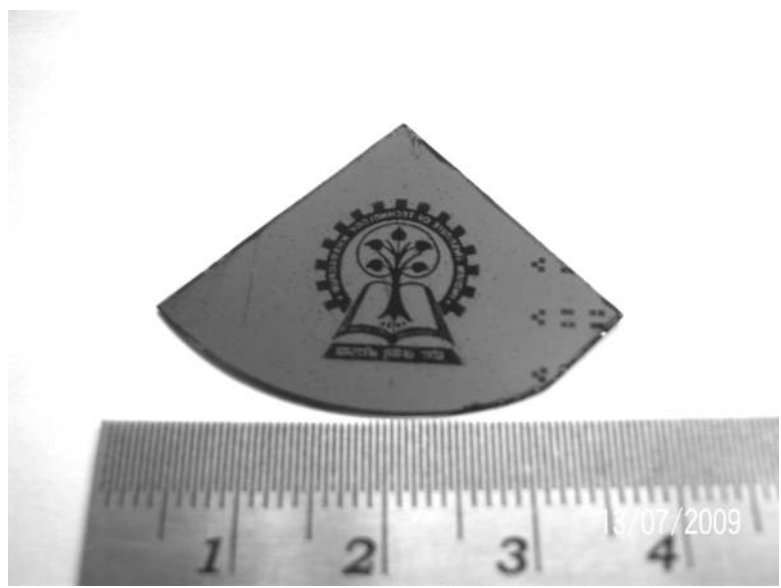

Figure 4.17: Large scale site selective growth of partially Fe filled multi-walled carbon nanotubes by direct photolithographic route using Fe-Mod-PR (The least count of the ruler is 0.5 mm)

This photolithography-based approach is a simple and yet robust process with excellent potential to produce partially Fe filled CNTs at lithographically defined locations in a reproducible and highly manufacturable manner. In other cases, like Kozhuharova et al. (2005), the process of catalyst pattern formation is quite complex as various steps are involved, such as photolithography, argon ion milling, sputter deposition and lift off. In the present experiment, a simple photolithographic step is enough to develop catalyst pattern on the substrate.

### 4.1.2.3 Summary

A simple and effective method has been demonstrated that can be used for site selective growth of partially Fe filled CNTs using Fe-Mod-PR. The experimental results show that the catalyst particle size effectively controls the diameter of the Fe filled CNTs grown through a tip growth mechanism. A feasible and inexpensive photolithographic route has been devised for the spatially selective synthesis of high quality Fe filled MWCNTs, which were grown only on a pre-defined surface in a manner compatible with current microfabrication processes. Since the overall process is suitable for conventional device fabrication, this method is a promising and practical pathway for large-scale industrial applications of various CNT-based electronic devices.

## 4.2 Synthesis of carbon nanotubes using spin coated Ni-Mod-PR

### 4.2.1 Lithographically defined site selective growth of carbon nanotubes using Ni-Mod-PR

In this section, the lithographically defined site selective growth of MWCNTs using a simple mixture of positive photoresist (PR) and Ni(salen) is reported. The mixture i.e. photoresist modified with Ni(salen) is named as Ni-Mod-PR. This is probably the first time where Ni(salen) has been used as a catalyst for CNT growth.

#### 4.2.1.1 Experimental details

The synthesis of MWCNTs was performed by APCVD with propane on Si(111) substrate at 850 °C using Ni-Mod-PR of 0.2M Ni(Salen) concentration. The Ni-Mod-PR film over Si substrate was heated at 900 °C in hydrogen atmosphere prior to CNT growth. The resultant as-grown products were analyzed using XRD, FESEM, HRTEM, EDX and Raman spectroscopy.

#### 4.2.1.2 Results and discussion

The SEM image (Figure 4.18) shows that the precursor i.e., Ni(salen), comprises of thin platelets with a broad range of crystallite sizes. The EDX analysis of Ni(salen) shows (inset of Figure 4.18) the presence C, N, O and Ni. Figure 4.19 shows the SEM image of the surface of Ni-Mod-PR film over the Si(111) substrate after being annealed at 900 °C. The catalyst nanoparticles are not uniformly distributed on the substrate and their sizes range from tens of nanometers to hundreds of nanometers.

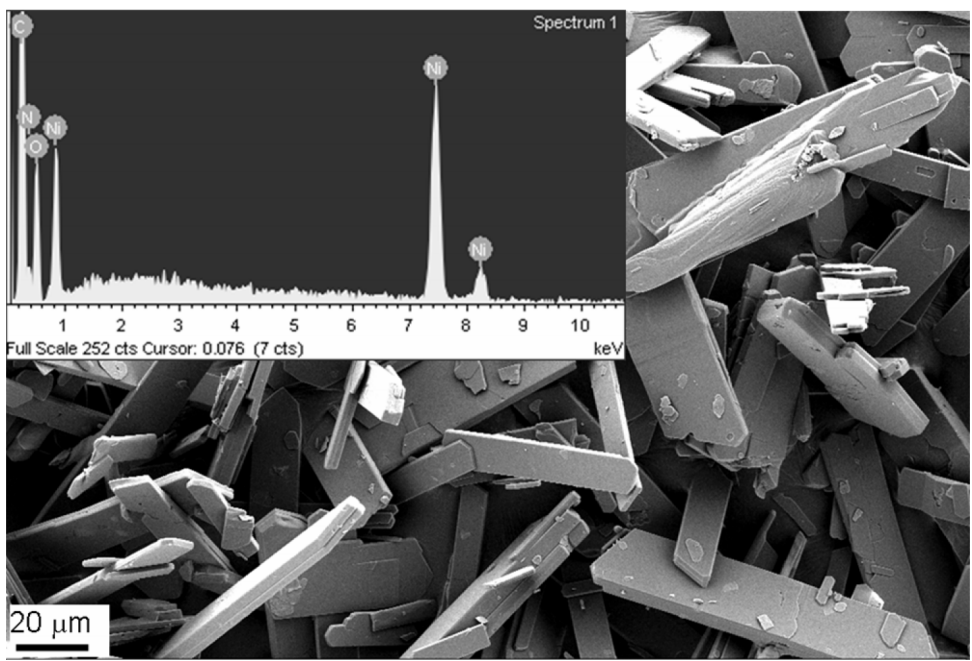

Figure 4.18: Scanning electron microscopy image of the Ni(salen); (inset) The energy dispersive X-ray spectrum of the precursor

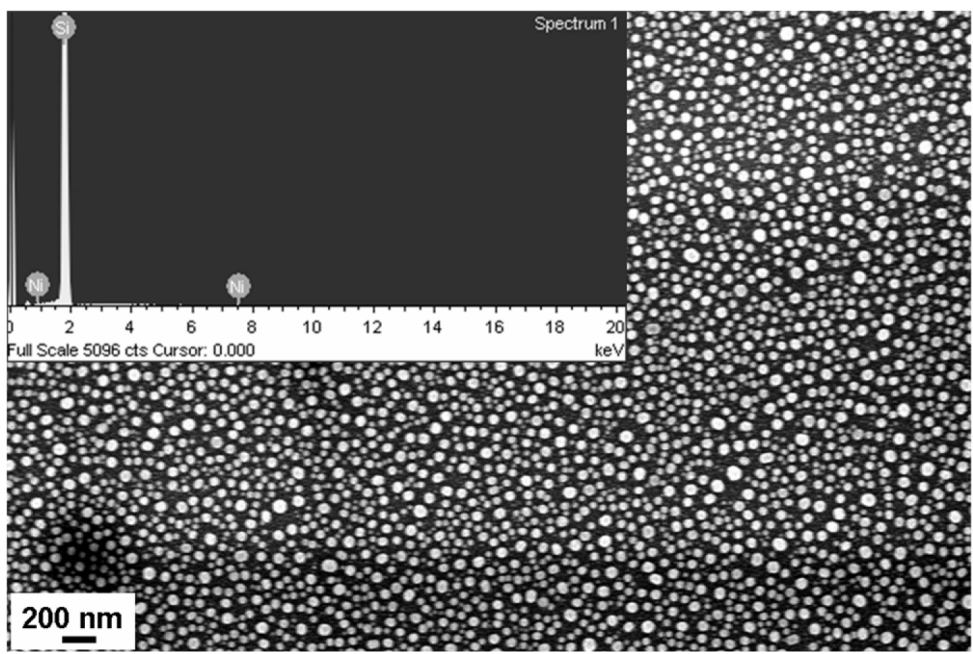

Figure 4.19: Scanning electron micrograph of the catalytic nanoparticles prepared from Ni-Mod-PR after annealing at 900 °C; (inset) The Energy dispersive X-ray spectrum of the Ni-Mod-PR over the Si(111) substrate after annealing in argon at 900 ºC

So, at the high temperatures, Ni(salen) particles also break down and form small Ni particles. The chemical composition analysis by EDX of the Ni-Mod-PR film annealed at 900 ºC in argon atmosphere (inset of Figure 4.19) shows that the material mainly contains Ni, with a Si peak (from the substrate). As there is no carbon peak in the EDX spectrum, it can be inferred that there are no residual organic species (within detection limit) in the catalyst particles.

The morphology of the as-grown carbon nanotubes was investigated by SEM. The SEM image (Figure 4.20) shows the presence of a homogeneous distribution of CNTs covering the catalyst surface in a web-like network. From this image, it can be seen that most of CNTs are nearly straight and long with a high number density. Furthermore, the wall surfaces of the CNTs appear relatively clean and smooth. EDX analysis of the as-grown CNTs (inset of Figure 4.20) shows that the material contains only carbon and Ni with a Si peak (due to the substrate). The high magnification image of the CNTs (Figure 4.21) shows metal particles at the tip of the nanotubes (marked in Figure 4.21 with white arrows), indicating the tip growth mechanism. It can be noted that the CNTs demonstrate a fair degree of alignment inspite of the lack of a bias or other mechanisms to align the tubes.

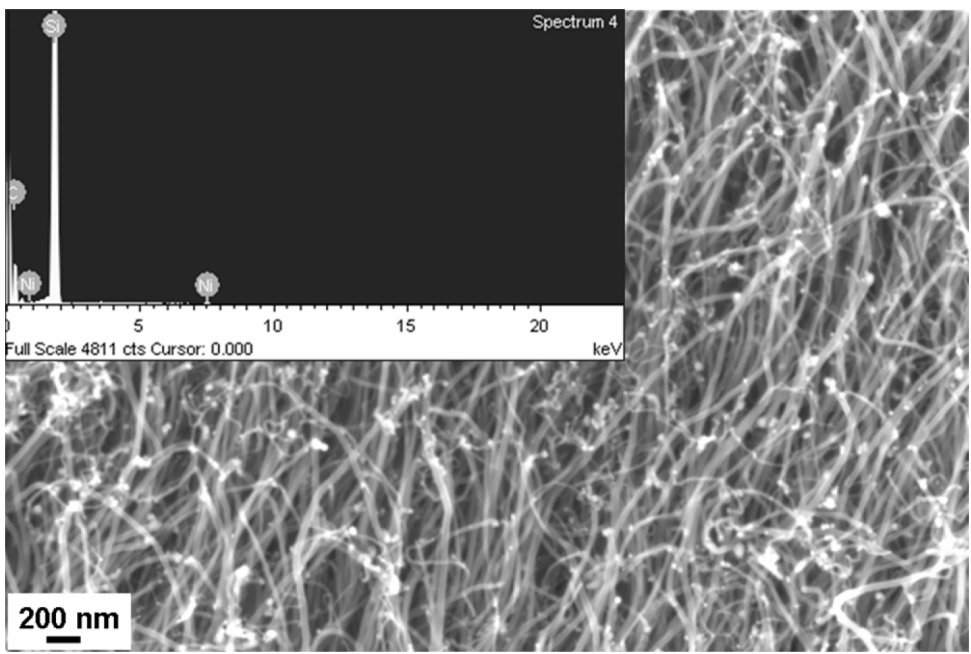

Figure 4.20: Scanning electron micrograph of the multi-walled carbon nanotubes synthesized using Ni-Mod-PR; (inset) The energy dispersive X-ray spectrum of the multi-walled carbon nanotubes

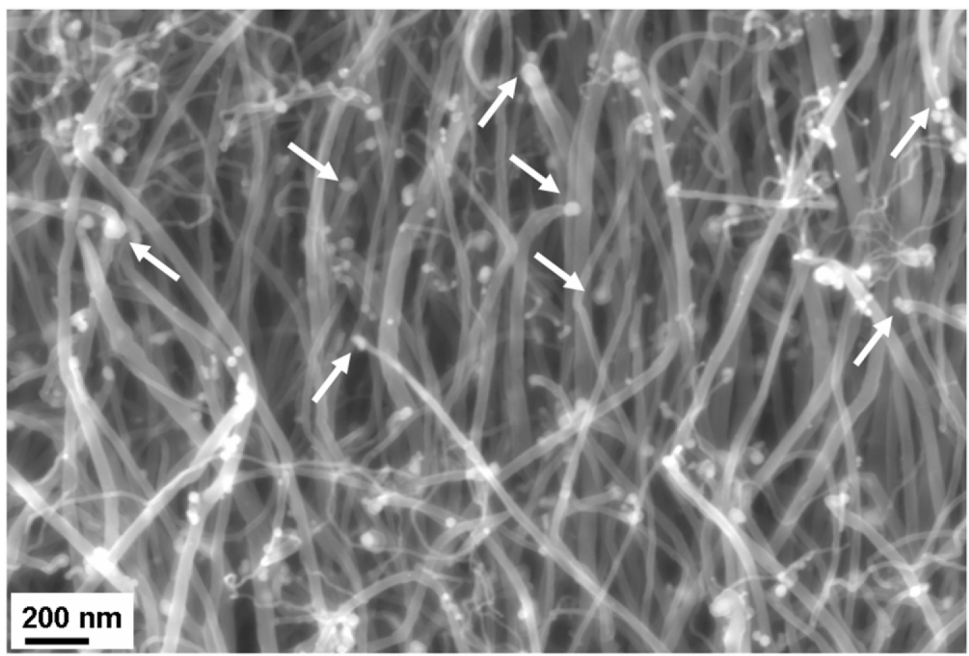

Figure 4.21: High magnification scanning electron micrograph of the multi-walled carbon nanotubes synthesized using Ni-Mod-PR indicating the tip growth mechanism

In order to analyze the internal structure of the nanotubes more closely, HRTEM studies were performed. The low magnification TEM images (Figure 4.22) show that the CNTs are the only product obtained and most of them are partially filled with metal nanowires. There is almost no trace of metal nanoparticles on the outer surface of the nanotubes according to the TEM observations which indicate the high selectivity of the synthesis method in favor of metal filled CNT formation.

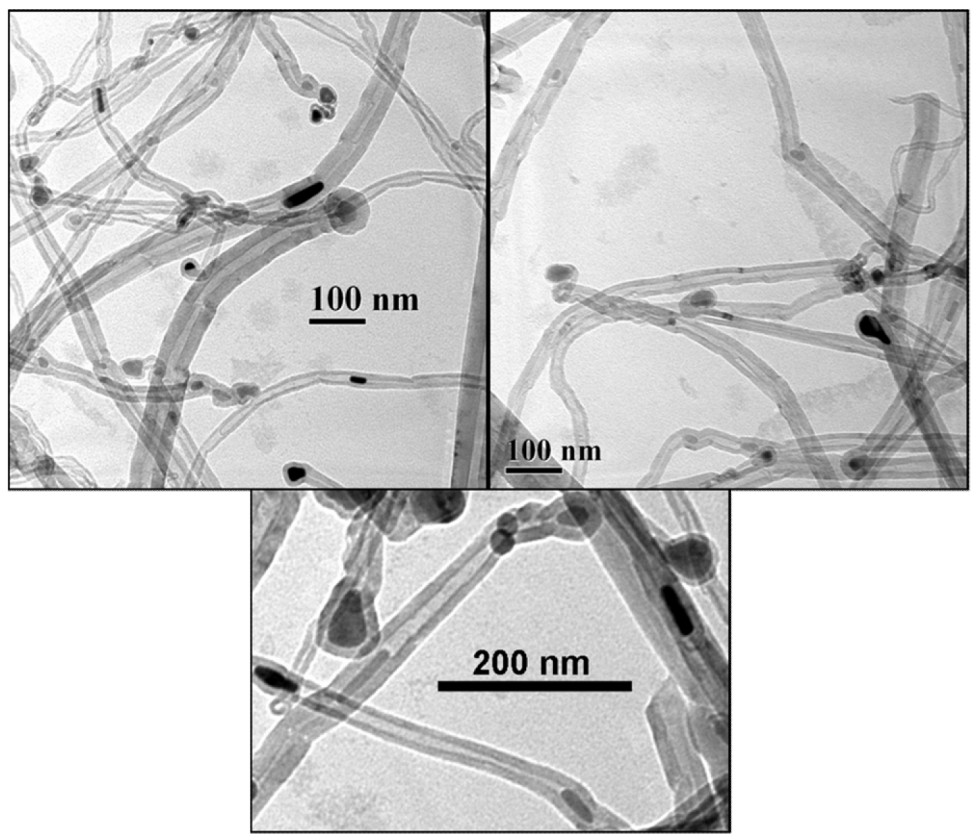

Figure 4.22: Low magnification transmission electron microscopy images of as-grown multi-walled carbon nanotubes synthesized using Ni-Mod-PR

TEM observations also demonstrate that some tubes are bent where defects occur during the growth, leading to a change in the growth direction. Further TEM examinations reveal that there are large numbers of long column-like metal nanowires in the cavities of CNTs (Figure 4.23). The length of these nanowires varies and is in the range of several nanometers to hundreds of nanometers. All the nanowires are tightly wrapped by the nanotube wall, and their diameters vary with the inner diameter of the corresponding CNT. Moreover, it can be observed that the catalyst nanoparticles are situated on the tip of the CNTs.

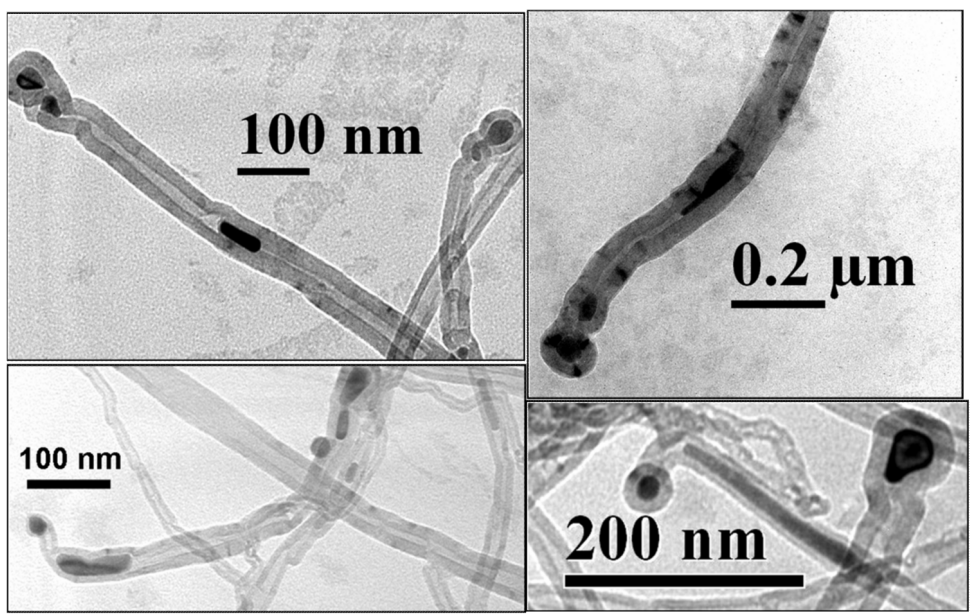

Figure 4.23: Transmission electron microscopy images of as-grown multi-walled carbon nanotubes synthesized using Ni-Mod-PR

Figure 4.24 is a TEM image of the CNTs filled with long nanowires. It can be seen that the CNT (indicated in Figure 4.24a with an arrow) is nearly straight and with outer diameter of about 65 nm. In particular, a long and uniform metal nanowire about 60 nm in length is encapsulated by the CNT. Figure 4.24b and 4.24c are high magnification images of the metal nanowire and the nanotube wall, respectively. The metal nanowire inside this nanotube exhibits clear well-ordered and straight lattice fringes (Figure 4.24b). The d- spacings of the lattice fringes are 0.21 nm, which corresponds to the (111) crystal

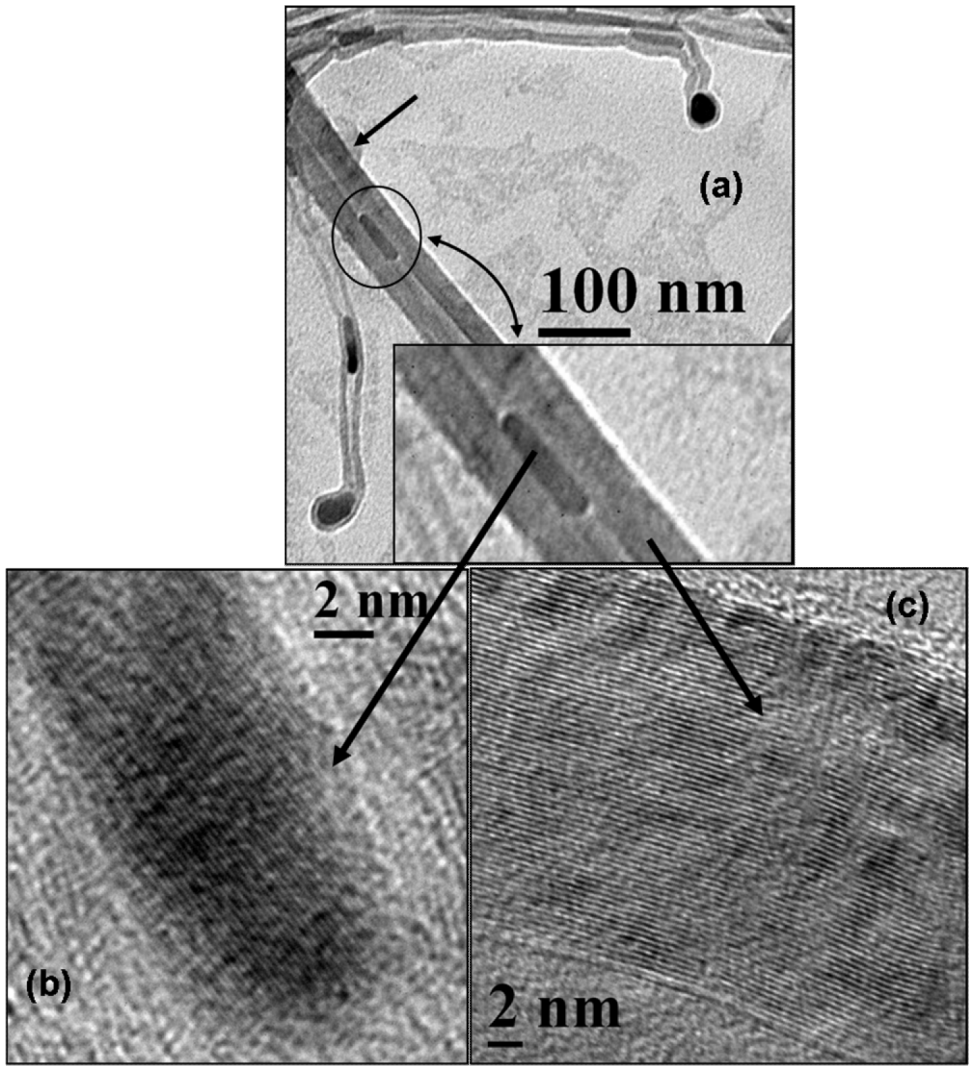

Figure 4.24: (a) Transmission electron microscopy image of as-grown multi-walled carbon nanotubes synthesized using Ni-Mod-PR (b) High resolution transmission electron microscopy image of the carbon nanotube encapsulated nanoparticle as indicated in the Figure 4.24a; (c) High resolution transmission electron microscopy image of the nanotube wall as indicated in the Figure 4.24a planes of Ni. Moreover, from Figure 4.24c, it is seen that the CNT is also well-graphitized and the interplanar distance between two adjacent graphene planes is about 0.34 nm, which is close to that of the (002) inter planar distance in graphite, i.e. 0.335

nm. The higher interplanar distance observed in MWCNTs is attributed to the curved shape of their structures, which induces strain inside the graphene planes stacking (Ajayan, 1999).

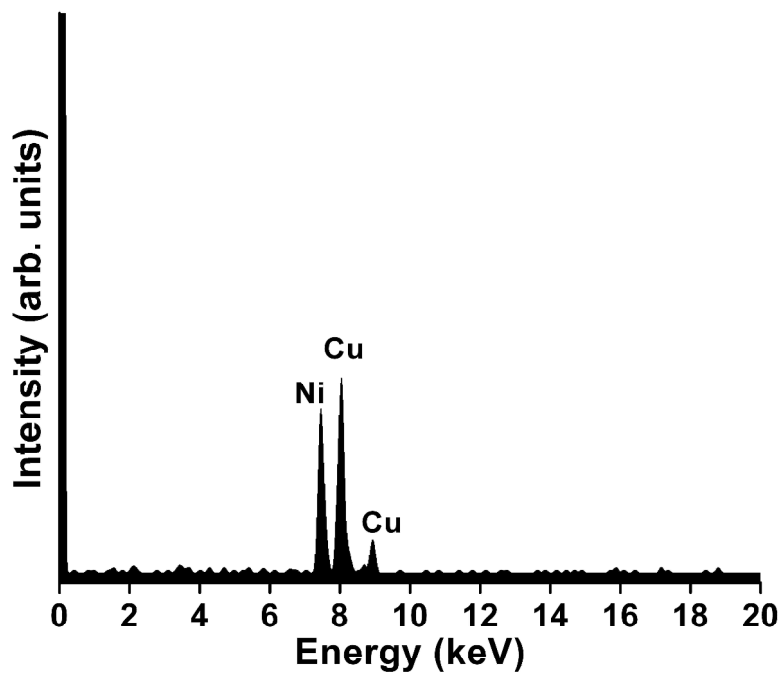

Figure 4.25: The energy dispersive X-ray spectrum of the catalytic nanoparticle encapsulated within the carbon nanotube as indicated in Figure 4.24a

A small amount of amorphous carbon is attached to the outer surface of this particular CNT. The amorphous carbon originates from the pyrolysis of the hydrocarbon species during the cooling of the system; this has been frequently reported in CVD synthesis (Dai et al., 2009). The EDX analysis of specimen (Figure 4.25) also demonstrates that the metallic nanowire inside the CNT indicated in Figure 4.24a is of Ni, which is in agreement with the lattice spacing obtained from the HRTEM image (Figure 4.24b). The Cu signals are due to the copper grid, supporting the sample.

Figure 4.26 shows the Raman spectrum for the as-grown CNTs at a laser excitation wavelength of 514.5 nm. The two main peaks are the D (1346 cm$^{-1}$) and G (1572 cm$^{-1}$) bands.

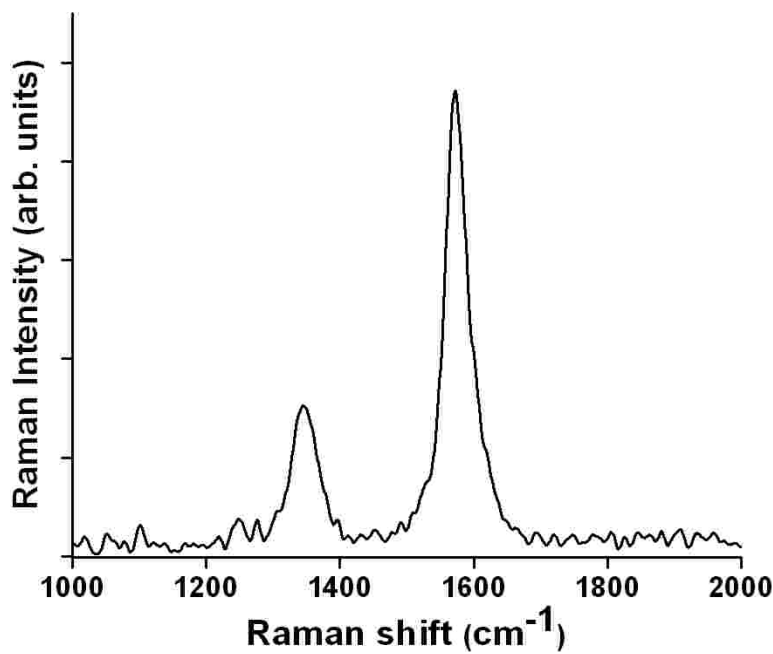

Figure 4.26: Raman spectrum (514.5 nm excitation) of multi-walled carbon nanotubes grown by chemical vapor deposition on Si using Ni-Mod-PR

The intensity ratio of the D peak to the G peak, which measures quality, as derived from Figure 4.26 is R = $I_D/I_G$ = 0.32, indicating that the grown CNTs are highly crystalline in nature.

Figure 4.27a is the SEM image of the Ni-containing patterns produced by the general photolithographic process using Ni-Mod-PR. Figure 4.27b is the SEM image of the grown CNTs on the catalyst pattern corresponding to Figure 4.27a and showing the high selectivity of the process.

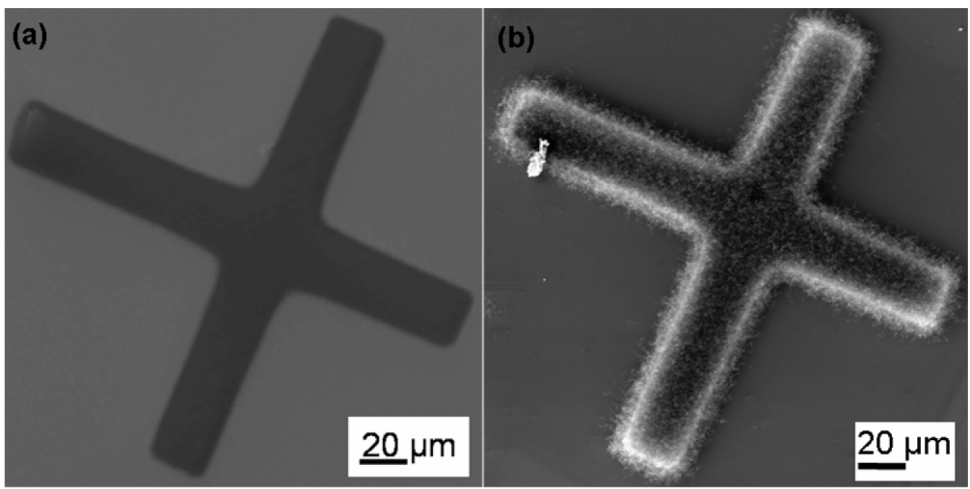

Figure 4.27: Site selective growth of partially Ni filled multi-walled carbon nanotubes by direct photolithographic route using Ni-Mod-PR (a) Scanning electron microscopy image of Ni-Mod-PR pattern; (b) Scanning electron microscopy image after nanotube growth

**4.2.1.3 Summary**

A novel method has been developed that can be used for economic mass production as well as for the selective growth of partially Ni filled CNTs using a mixture of Ni(salen) and conventional photoresist. The experimental results show that the Ni filled CNTs are multi-walled, well graphitized and grown through a tip growth mechanism.

# Chapter 5

## Summary, conclusions and scope for future work

This chapter presents a brief summary of the work performed in this thesis, draws a few conclusions and finally recommends some directions for future work.

### 5.1 Summary

CNTs represent an interesting class of nanomaterials which are used for various applications starting from the bulk scale to the nanometer regime. However, control over the synthesis of CNTs is currently minimal. The multiplicity of the catalyst systems and synthesis methods for CNTs makes it difficult to decide the best route for CNT production. The initial goals of this research were to analyze the effect of process temperatures and role of the metal catalysts on resulting carbon nanotubes for better understanding of their growth mechanism. The final objective was to develop a simple method for the site selective growth of catalyst filled carbon nanotubes over a substrate.

Over the last few years, CVD has been the preferred method among different CNT synthesis methods because of its potential advantage in producing a large amount of CNTs grown directly on a desired substrate with high purity. Accordingly in the present case, MWCNT synthesis was first studied using different elemental metal catalysts (Fe and Ni) via APCVD process by varying growth parameters such as synthesis temperature or pre-heating temperature. Secondly, synthesis of MWCNTs using different metal complex catalysts (Fe(acac)$_3$ and Ni(salen)) was performed on bare Si substrates as well as directly onto a selective region over Si substrates where the catalyst pattern has been developed via photolithography. MWCNTs were grown at atmospheric pressure with propane as the carbon feedstock. Catalysts were delivered to the silicon substrates either by evaporation (in case of elemental metal catalyst) or by spin coating (for metal complex catalysts). The effects of the synthesis conditions such as catalyst and temperature on CNT morphology were analyzed.

In the subsequent study on the morphology of the MWCNTs grown by CVD, a significant influence of the process temperature and the catalyst material on the quality of the CNTs was observed. The diameter, number density and degree of graphitization of the nanotubes can be adjusted by choosing the appropriate process temperature. It is worth noticing that a morphology transition of CNTs with variation of elemental metal catalyst was observed. For the elemental Ni catalyst, the grown nanotubes were bamboo-like whereas the CNTs grew as straight tubes partially filled with metal when elemental Fe catalyst was used. However, in both the cases, the nanotubes followed a tip growth mechanism. Models for the growth of carbon nanotubes have been suggested for partially catalyst filled as well as for bamboo-like nanotubes produced by the experiments performed. It was proposed that the state of the catalyst material plays an important role in determining the resultant CNT morphology. Therefore, it is possible to pre-select the characteristic of the CNT with high probability by using the appropriate process. In case of elemental metal catalysts, under the studied conditions, Fe is the better catalyst in terms of degree of graphitization and uniformity of diameter of CNTs. Elemental Fe catalysts show the best results when the CNTs are grown at 850 °C after pre-heating at 900 °C. However, the metal deposition over the substrate is time consuming and the deposited area is also finite. This has been overcome to scale up CNT production by using spin coating of metal complex catalyst materials after mixing with suitable photoresist (Mod-PR catalyst system).

The Mod-PR catalyst system represents a novel way to produce a large quantity of CNTs at relatively low temperatures with high yield. This also makes it possible to synthesize partially catalyst filled CNTs on substrates with site selectivity using simple photolithography and is compatible with current microelectronics fabrication technology. Using Fe-Mod-PR, it was possible to synthesize a measurable quantity of nanotubes even at 550 °C, which was not feasible with the elemental Fe catalyst. Additionally the yield has been increased from ~ 0.1 mg/(cm$^2$·h) (elemental Fe) to ~ 0.4 mg/(cm$^2$·h) (Fe-Mod-PR) when the CNTs are grown at 850 °C after pre-heating at 900 °C from Fe-Mod-PR. With this experiment, it has also been demonstrated that for metal complex Fe catalyst, the diameter of the nanotubes is strongly related to the diameter of the catalyst particles prior to nanotube growth. It can also be noted that the CNTs synthesized using Ni-Mod-PR demonstrate a fair degree of vertical alignment inspite of the lack of a bias or other mechanisms to align the tubes.

Like the elemental metal catalysts, in the Mod-PR catalysts also, the nanotube diameter, number density and degree of graphitization can be controlled by varying the growth temperature. Moreover, in both the cases (elemental metal catalysts and metal complex catalysts), the carbonaceous material content was quite low, reflecting the purity of the grown nanotubes.

Among the different Mod-PR catalysts used, CNTs grown using Fe-Mod-PR gives the best result in terms of degree of graphitization though the quality is inferior to the CNTs synthesized using elemental Fe catalyst. However, in contrast to the elemental metal catalyst systems, CNTs synthesized using Mod-PR approach always produces catalyst filled nanotubes.

## 5.2 Conclusions

1. The diameter, number density and degree of graphitization of the MWCNTs can be controlled by varying the growth temperature, irrespective of the type of catalyst used for the CNT synthesis.

2. Pre-heating temperature can also be used to control nanotube growth morphology in case of elemental metal catalyst systems.

3. Internal structure of the MWCNTs depends on the catalytic element in case of elemental metal catalyst whereas for metal complex catalyst, metal filled nanotubes are always produced.

4. The quality and yield of MWCNTs are affected by the physical form of the catalyst (elemental metal or metal complex).

5. Metal complex catalysts are promising candidates for scalable low temperature site selective growth of magnetic material filled MWCNTs.

## 5.3 Scope for future work

The results presented in this thesis show that the nanotubes are not entirely filled with magnetic material. Therefore, more control over filling and extensive magnetic characterization of those tubes is necessary for further improvement. One big challenge would be to grow aligned filled carbon nanotubes at low temperature with site selectivity. Another task would be to generate SWCNTs using both elemental metal and metal complex catalyst and it should be tested if the growth morphology of those nanotubes and their structure depends on the process parameters. Irrespective of the type of nanotubes, the yield and purity issues must be addressed in order to improve the quality of nanotube production. The selectively grown magnetic material filled CNTs can be used in sensors or in magnetic recording media.

**5.4 Author's contribution**

In the present work, MWCNTs have been grown using a relatively uncommon source of hydrocarbon species, viz. propane. A scalable, site selective growth method for magnetic material filled MWCNTs has been demonstrated which can potentially be performed at relatively low temperatures. The ability to tailor the growth of nanotubes by tailoring the chemistry of the catalyst has been demonstrated. Under optimized conditions, relatively high purity (low graphitic contamination) CNTs could be grown. The diameter, number density and degree of graphitization of the selectively grown MWCNTs can also be controlled by varying the growth temperature.

# List of research publications

## (A) Journal Articles

1. **Sengupta J.**, Jana A., Singh N. D. P. and Jacob C., (2010), Effect of growth temperature on the CVD grown Fe filled multi-walled carbon nanotubes using a modified photoresist, Materials Research Bulletin**,** Vol. 45, pp. 1189-1193.

2. **Sengupta J.**, Jana A., Singh N. D. P. and Jacob C., (2010), Lithographically defined site-selective growth of Fe filled multi-walled carbon nanotubes using a modified photoresist, Carbon**,** Vol. 48, pp. 2371-2375.

3. **Sengupta J.** and Jacob C., (2009), Growth temperature dependence of partially Fe filled MWCNT using chemical vapor deposition, Journal of Crystal Growth, Vol. 311, pp. 4692-4697.(TOP 25 Hottest Articles, October to December, 2009)

4. **Sengupta J.** and Jacob C., (2010), Pre-heating effect on the catalytic growth of partially filled carbon nanotubes by chemical vapor deposition, Journal of Nanoscience and Nanotechnology, Vol. 10, pp. 3064-3071.

5. **Sengupta J.** and Jacob C., (2010), The effect of Fe and Ni catalysts on the growth of multiwalled carbon nanotubes using chemical vapor deposition, Journal of Nanoparticle Research, Vol. 12, pp. 457- 465.

6. **Sengupta J.**, Panda S. K. and Jacob C., (2009), A comparative study of the synthesis of carbon nanotubes using Ni and Fe as catalyst, Advanced Materials Research, Vol. 67, pp. 89-94.

7. **Sengupta J.**, Panda S. K. and Jacob C., (2009), Carbon nanotube synthesis from propane decomposition on a pre-treated Ni overlayer, Bulletin of Materials Science**,** Vol. 32, pp. 135-140. (Cover page article)

8. Panda S. K., **Sengupta J.** and Jacob C., (2010), Synthesis of β-SiC/SiO$_2$ core-sheath nanowires by CVD technique using Ni as catalyst, Journal of Nanoscience and Nanotechnology, Vol. 10, pp. 3046-3052.

9. Gupta A., **Sengupta J.** and Jacob C., (2008), An atomic force microscopy and optical microscopy study of various shaped void formation and reduction in 3C-SiC films grown on Si using chemical vapor deposition, Thin Solid Films, Vol. 516, pp.1669-1676.

## (B) Conference Presentations

1. **Sengupta J.** and Jacob C., Growth and characterisation of carbon nanotubes synthesized by propane decomposition using CVD, International Conference on Hi-Tech Materials, 2009, Indian Institute of Technology, Kharagpur, India.

2. Panda S. K., **Sengupta J.** and Jacob C., Hot wall and cold wall CVD grown polycrystalline β-SiC - A comparative study, International Conference on Hi-Tech Materials, 2009, Indian Institute of Technology, Kharagpur, India.

3. **Sengupta J.**, Panda S. K. and Jacob C, Effect of reconstruction of catalyst on the catalytic growth of partially filled carbon nanotubes by chemical vapor deposition, The Materials Research Society of India, AGM 2009, SINP, Kolkata, India.

4. Panda S. K., **Sengupta J.** and Jacob C., β-SiC/$SiO_2$ nanocables synthesized by APCVD technique, The Materials Research Society of India, AGM 2009, SINP, Kolkata, India.

5. **Sengupta J.** and Jacob C., To study the reduction behavior of iron oxide using hydrogen, IUMRS- International Conference on Advanced Materials 2007, Bangalore, India.

6. **Sengupta J.**, Gupta A. and Jacob C., Void formation in 3C-SiC films grown on Si using CVD, The Materials Research Society of India, AGM 2007, NPL, New Delhi, India.

# Author's Vitae

Joydip Sengupta was born in Kolkata, India on 28$^{th}$ December 1981. He graduated (B. Sc., Honours in Electronics) with 1$^{st}$ class in the year 2003 from University of Calcutta, India. In 2005, he received the M. Sc. degree in Electronic Science with 1$^{st}$ class from the same university. He enrolled for the Ph. D. program at Materials Science Centre, Indian Institute of Technology (IIT), Kharagpur, India in the year 2006. He was awarded a Senior Research Fellowship by CSIR (Govt. of India) in 2008. His research interests include growth, characterization and application of carbon nanotubes and related materials. He has published nine scientific papers so far in various international/national journals. He has also presented six papers so far in international/national conferences.